\documentclass[chicago]{emulateapj-rtx4}

\usepackage{amsmath,amssymb,latexsym, graphicx}
\usepackage{epsfig}
\usepackage{booktabs}
\usepackage{lscape}
\usepackage[dvips]{color}

\newcommand{\tn}[1]{\textnormal{#1}}

\newcommand{\kref}[1]{{Section} \ref{#1}}

\newcommand{\AV}{A_V}

\newcommand{\MB}{M_B}
\newcommand{\TNT}{T_{90}}

\slugcomment{Submitted to ApJ}

\shorttitle{Afterglows of \emph{Swift} Type I and Type II GRBs}
\shortauthors{Kann et al.}

\begin{document}

\title{The Afterglows of \emph{Swift}-era Gamma-Ray Bursts II.: Type I GRB versus Type II GRB Optical Afterglows
\footnote{Based in part on observations obtained with the Very Large Telescope under ESO program 075.D-0787 (PI Tagliaferri); 076.D-0747 (PI Campana), 079.D-0884  (PI D'Avanzo); 077.D-0805 (PI Tagliaferri). Also based partly on observations made with the Italian Telescopio Nazionale Galileo (TNG) operated on the island of La Palma by the Fundaci\'{o}n Galileo Galilei of the INAF (Istituto Nazionale di Astrofisica) at the Spanish Observatorio del Roque de los Muchachos of the Instituto de Astrofisica de Canarias under programs AOT12 TAC\_38 (PI Antonelli); AOT16 TAC\_19 (PI Maiorano).}
}
\author{
D.~A.~Kann,\altaffilmark{1}
S.~Klose,\altaffilmark{1}
B.~Zhang,\altaffilmark{2}
S.~Covino,\altaffilmark{3}
N.~R.~Butler,\altaffilmark{4,5}
D.~Malesani,\altaffilmark{6}
E.~Nakar,\altaffilmark{7,8}
A.~C.~Wilson,\altaffilmark{9}
L.~A.~Antonelli,\altaffilmark{10,11}
G.~Chincarini,\altaffilmark{3,12}
B.~E.~Cobb,\altaffilmark{13,14,15}
P.~D'Avanzo,\altaffilmark{3,16}
V.~D'Elia,\altaffilmark{10}
M.~Della Valle,\altaffilmark{17,18,19}
P.~Ferrero,\altaffilmark{1,20}
D.~Fugazza,\altaffilmark{3}
J.~Gorosabel,\altaffilmark{21}
G.~L.~Israel,\altaffilmark{10}
F.~Mannucci,\altaffilmark{22}
S.~Piranomonte,\altaffilmark{10}
S.~Schulze,\altaffilmark{1,23}
L.~Stella,\altaffilmark{10}
G.~Tagliaferri,\altaffilmark{3}
K.~Wiersema\altaffilmark{24,25}
}

\altaffiltext{1}{Th\"uringer Landessternwarte Tautenburg,  Sternwarte 5, 07778 Tautenburg, Germany}
\altaffiltext{2}{Department of Physics and Astronomy, University of Nevada, Las Vegas, NV 89154-4002, USA}
\altaffiltext{3}{INAF, Osservatorio Astronomico di Brera, via E. Bianchi 46, 23807 Merate (LC), Italy}
\altaffiltext{4}{Townes Fellow, Space Sciences Laboratory, University of California, Berkeley, CA 94720-7450, USA}
\altaffiltext{5}{Astronomy Department, University of California, 445 Campbell Hall, Berkeley, CA 94720-3411, USA}
\altaffiltext{6}{Dark Cosmology Centre, Juliane Maries Vej 30, DK-2100 K{\o}benhavn {\O}, Denmark}
\altaffiltext{7}{Division of Physics, Mathematics, and Astronomy, California Institute of Technology, Pasadena, CA 91125, USA}
\altaffiltext{8}{Raymond and Beverly Sackler School of Physics \& Astronomy, Tel Aviv University, Tel Aviv 69978, Israel}
\altaffiltext{9}{Department of Astronomy, University of Texas, Austin, TX 78712, USA}
\altaffiltext{10}{INAF, Osservatorio Astronomico di Roma, via Frascati 33, 00040, Monteporzio Catone (RM), Italy}
\altaffiltext{11}{INAF National Institute for Astrophysics, I-00136 Rome, Italy}
\altaffiltext{12}{Universit\`{a} degli studi di Milano-Bicocca, Dipartimento di Fisica, piazza delle Scienze 3, 20126 Milano, Italy}
\altaffiltext{13}{Department of Astronomy, Yale University, P.O. Box 208101, New Haven, CT 06520, USA}
\altaffiltext{14}{Department of Astronomy, 601 Campbell Hall, University of California, Berkeley, CA 94720--3411, USA}
\altaffiltext{15}{Department of Physics, The George Washington University, Corcoran 105, 725 21st St, NW, Washington, DC 20052, USA}
\altaffiltext{16}{Dipartimento di Fisica e Matematica, Universit\`{a} dell'Insubria, via Valleggio 11, 22100 Como, Italy}
\altaffiltext{17}{INAF, Osservatorio Astronomico di Capodimonte, Salita Moiariello, 16 80131, Napoli, Italy}
\altaffiltext{18}{European Southern Observatory, Karl Schwarschild Strasse 2, 85748 Garching bei M\"unchen, Germany}
\altaffiltext{19}{International Centre for Relativistic Astrophysics Network, Piazzale della Republica 2, Pescara, Abruzzo, Italy}
\altaffiltext{20}{Instituto de Astrof\'isica de Canarias, C/ V\'ia L\'actea, s/n E38205 -- La Laguna (Tenerife), Spain}
\altaffiltext{21}{Instituto de Astrof\'{\i}sica de Andaluc\'{\i}a (IAA-CSIC), Apartado de Correos, 3.004, E-18.080 Granada, Spain}
\altaffiltext{22}{INAF, Osservatorio Astrofisico di Arcetri, largo E. Fermi 5, I-50125 Firenze, Italy}
\altaffiltext{23}{Centre for Astrophysics and Cosmology, Science Institute, University of Iceland, Dunhagi 5, IS 107 Reykjavik, Iceland}
\altaffiltext{24}{Astronomical Institute ``Anton Pannekoek'', University of Amsterdam, Kruislaan 403, 1098 SJ Amsterdam, The Netherlands}
\altaffiltext{25}{Department of Physics and Astronomy, University of Leicester, University Road, Leicester LE1 7RH, UK}


\begin{abstract}
Gamma-Ray Bursts (GRBs) have been separated into two classes, originally along the lines of duration and spectral properties, called ``short/hard'' and ``long/soft''. The latter have been conclusively linked to the explosive deaths of massive stars, while the former are thought to result from the merger or collapse of compact objects. In recent years indications have been accumulating that the short/hard vs. long/soft division does not map directly onto what would be expected from the two classes of progenitors, leading to a new classification scheme called Type I and Type II which is based on multiple observational criteria. We use a large sample of GRB afterglow and prompt-emission data (adding further GRB afterglow observations in this work) to compare the optical afterglows (or the lack thereof) of Type I GRBs with those of Type II GRBs. In comparison to the afterglows of Type II GRBs, we find that those of Type I GRBs have a lower average luminosity and show an intrinsic spread of luminosities at least as wide. From late and deep upper limits on the optical transients, we establish limits on the maximum optical luminosity of any associated supernova, confirming older works and adding new results. We use deep upper limits on Type I GRB optical afterglows to constrain the parameter space of possible mini-SN emission associated with a compact-object merger. Using the prompt emission data, we search for correlations between the parameters of the prompt emission and the late optical afterglow luminosities. We find tentative correlations between the bolometric isotropic energy release and the optical afterglow luminosity at a fixed time after trigger (positive), and between the host offset and the luminosity (negative), but no significant correlation between the isotropic energy release and the duration of the GRBs. We also discuss three anomalous GRBs, GRB 060505, GRB 060614, and GRB 060121, in the light of their optical afterglow luminosities.
\end{abstract}


\section{INTRODUCTION}

It has long been known that Gamma-Ray-Bursts (GRBs) come in (at least\footnote{There is statistical evidence from the duration distribution and the duration/hardness plots for a third class of GRBs, ``intermediate GRBs'', which has been found to be detector- and method-independent \citep[][and references therein]{Horvath2008, Horvath2010, Ripa2009, Chattopadhyay2007}. An analysis of probable intermediate GRBs from the \emph{Swift} sample has shown, though, that they are a subclass of temporally shorter, underluminous (both in prompt as well as afterglow emission) long/soft GRBs \citep{deUgartePostigoIntermediate} linked to the deaths of massive stars.}) two classes: Those typically lasting 2 s or less and having hard prompt-emission spectra (short/hard GRBs) and those lasting typically longer than 2 s and having softer prompt spectra which often show strong hard-to-soft spectral evolution (long/soft GRBs) \citep{Kouveliotou1993, Ford1995}\footnote{Note that there is evidence, though, that the first few seconds of long/soft GRBs temporally and spectrally resemble short/hard GRBs, with a harder spectral slope and higher peak energy than the integrated spectra of the complete long/soft GRBs (hence, the spectral evolution; \citealt{Ghirlanda2009}, and references therein). This has been interpreted as a universal central engine working in both classes of GRBs \citep[see also][]{Eichler2009}. In terms of integrated spectra, though, short/hard GRBs clearly exhibit both higher peak energies as well as harder spectral slopes $\alpha$, this is especially true for those detected by the Gamma-ray Burst Monitor (GBM) instrument of \emph{Fermi}, which has a broader spectral response than \emph{CGRO} BATSE \citep{Nava2010, GhirlandaGBM}. The ``Kouveliotou-Plot'' has also been successfully constructed with \emph{Swift} GRBs \citep{ZhangChoi, Gomboc2010}, but the difference in hardness between short/hard and long/soft GRBs disappears almost completely in the rest frame.}. Some long/soft GRBs have been related spectroscopically \citep{Galama1998, Hjorth030329, Stanek030329, Malesani2004, Pian060218, Chornock100316D} to the deaths of massive stars, the so-called collapsar model \citep[][for a review, see \citealt{WoosleyBloom}]{Woosley1993}, and it was proposed that all long GRBs are accompanied by photometric supernova (SN) signatures \citep{Zeh2004}. The short/hard GRBs have been more enigmatic, as they are rarer and harder to localize. It was already long suspected that they were of cosmological nature too due to their isotropic distribution in the sky, which shows no alignement with the supergalactic plane \citep[][and references therein]{Bernui2008}. The most favored model is the merger of two compact objects, i.e., a neutron star (NS) and a black hole (BH) or two neutron stars \citep{Blinnikov1984a, Blinnikov1984b, Paczynski1986, Goodman1986, Eichler1989}\footnote{Note that we do not consider the ``superflares'' of Soft Gamma-Repeaters (SGRs) in this work, which have also been suggested as a source for part of the short GRB population \citep[e.g.,][]{Hurley1806, Palmer1806}. While \cite{TanvirLocal} showed that a significant number of BATSE short GRBs originate in the local universe \citep[see also][but see \citealt{Tikhomirova2010}]{Chapman2009}, only two well-localized short GRBs, GRB 051103 and GRB 070201, are considered to be SGR flares in nearby galaxies \citep{Ofek051103, Frederiks051103, Hurley051103, PerleyGCN070201, GolenetskiiGCN070201, HurleyGCN070201, Abbott070201, Ofek070201}.}.

While there have also been extremely intense short/hard GRBs, such as GRB 841215 \citep{Laros1985}, GRB 930131 \citep[e.g.,][]{Kouveliotou1994}, and GRB 031214 \citep{Hurley031214}, the true breakthrough in the observations only came with the \emph{Swift} satellite \citep{Gehrels2004} and its discovery of well-localized afterglows in the X-ray band \citep{Gehrels050509B}, as well as the \emph{HETE-2} localization of GRB 050709 \citep{Villasenor050709}. These lead to identifications in the optical \citep{Hjorth050709, Fox050709, Covino050709} and radio \citep{Berger050724} bands\footnote{More recently, detections at GeV energies have also been reported \citep{Abdo090510, Abdo081024B, Ackermann090510}, whereas the TeV regime has not yielded a secure detection yet \citep[e.g.,][]{Abdo2007}, and neither have gravity waves \citep{AbadieLIGO}, which would yield insight into the central engine \citep[e.g.,][]{Kiuchi2010}.}, thus allowing for the first time an association between some short/hard GRBs and galaxies at moderate redshift that show no evidence of recent star-formation \citep{Gehrels050509B, Barthelmy050724, Berger050724, Gorosabel050724}. Thus, at least some short/hard GRBs must stem from a different progenitor class than long/soft GRBs\footnote{Recently, \cite{Lazzati2009} showed that under certain off-axis viewing angles, collapsars can produce prompt GRB emission which looks very similar to a short/hard GRB with extended emission. GRB 050724 is a counterexample, though, as it occurred in a galaxy without ongoing star-formation and did not exhibit any SN emission following it. Still, there seems to be evidence for some short/hard GRBs to have a different type of progenitor than a compact-object merger \citep[either collapsar or a third, unknown class,][, but see \citealt{LeiblerBerger}]{Virgili2010, Cui2010}.}, and several lines of evidence favor the compact object merger models \citep[e.g.,][]{Fox050709, Barthelmy050724, Berger050724, FongHST}. For reviews on short/hard GRBs and their progenitors, see \cite{NakarReview} and \cite{LeeReview}.

Further observations showed that the classic short/hard versus long/soft dichotomy could not hold. Observations of early, soft, extended emission bumps and late X-ray flares in the afterglows of short/hard GRBs \citep{Villasenor050709, Fox050709, Barthelmy050724, Campana050724} showed that even for compact object mergers, the central engine must be active for much longer time than previously thought, possibly through tidal tails refreshing the accretion disk at late times \citep[][and references therein, though it is unclear if matter fallback can actually create X-ray flares, \citealt{Rossi2009}]{Lee2010}. This leads to $\TNT$ measurements (in detectors which are sensitive to hard X-rays like \emph{HETE II} WXM/SXC and \emph{Swift} BAT) which far exceed the classic 2 s division line. In the ``other direction'', it has also been found that precursors can significantly precede the main prompt emission in some cases \citep{TrojaPrecursor}. The situation became even more complex when two temporally long GRBs at low redshift were discovered which showed no evidence for accompanying SNe, GRBs 060505 and 060614 \citep{DellavalleNature, FynboNature, GalyamNature, Ofek060505}. GRB 060614, with $\TNT=102$ s, was clearly a long GRB but showed \citep{GehrelsNature}, next to the missing SN, another clear sign of typical short/hard GRBs, negligible spectral lag between different energy bands of the prompt emission \citep{NorrisBonnell, Zhang2006, Guiriec090510} coupled with a low luminosity, making it a strong outlier of the lag-luminosity correlation of \cite{Norris2000}. \cite{Zhang060614} showed that it was possible to ``transform'' the prompt-emission light curve of GRB 060614 into a light curve strongly resembling that of GRB 050724 (short, hard spike, long, faint and soft emission tail) simply by reducing the luminosity of the GRB. This led to the idea of a new classification scheme for GRBs \citep{ZhangNature, GehrelsNature, Zhang060614}: In analogy to Type Ia and Type II SNe, GRBs can be classified as Type I events (induced by the catastrophic destruction of a compact star or stars, no associated SN, can be found in all types of galaxies) and Type II events (induced by the destruction of a massive star, associated with a SN -- most likely a broad-lined Type Ic SN -- and found only in galaxies with high specific star-formation). This definition is not based on a single observed quantity (such as the location in a $\TNT-$hardness diagram), and thus it is, in some cases, very difficult to place a GRB into the context of Type I or Type II. Still, we will adopt this terminology. \cite{Zhang080913} have extensively discussed the links between progenitors (collapsars versus mergers) and the expected observables, and have created a flowchart to help in the classification of Type I and Type II events (their Figure 8), which we shall employ. A second classification method which expands on \cite{Zhang080913} has been given by \cite{Lü2010}, we will compare our classification results with this classification also. For an even more detailed approach toward GRB classification based on physical progenitor models, see \cite{BloomPhysical}. 

In a companion paper \citep[][henceforth Paper I]{KannShortI}, we compiled optical light curves of \emph{Swift}-era Type II GRBs and compared them with those of a pre-\emph{Swift} sample \citep[][henceforth K06]{PaperIII}. We found that the two samples are very similar, and that, intrinsically, the afterglows had comparable luminosities. Therefore, it is justified to treat all afterglows as one large, coherent sample. In this work, we will compare the Type I GRB afterglows (which often only consist of non-detections and thus upper limits) with this large sample of Type II GRBs afterglows. The Type I GRB afterglow sample and the selection criteria are described in \kref{dataI}. We describe the construction of the Type I GRB afterglow light curves in \kref{LCs}, the method used for shifting them to a common redshift in \kref{Shift}, and background on the search for classical SN light and light from heavy-element- and neutron-decay driven ``mini-supernovae'' (mini-SNe, as proposed by \citealt{LiPaczynski1998a, LiPaczynski1998b} [henceforth LP98ab], and \citealt{Kulkarni2005} [henceforth K05]) in \kref{SNlimits} and \kref{Mini}, respectively. After discussing the observed Type I GRB afterglow light curves (\kref{Observed}), we shift them to a common redshift for a direct comparison between the afterglow luminosities of Type II and Type I GRBs. This allowed us to find the luminosity distribution of Type I GRB afterglows (\kref{Intrinsic}), place deep upper limits on accompanying SN emission (\kref{SNcon}), and put constraints on the parameter space of mini-SNe (\kref{Minicon}). We discuss our results in \kref{Disc}. Furthermore, using the derived bolometric energetics of our Type I and Type II GRB samples, we search for correlations between the parameters of the prompt emission and the afterglow luminosities derived at a fixed rest-frame time (\kref{EaC}). A summary and conclusions are given in \kref{End}. In the Appendix, we present additional Type I GRB afterglow observations along with a data table (\kref{Observations}), details on all the GRBs in our Type I sample as well as those that were not included (\kref{AppA}), the methods used to search for extra light in Type I GRB afterglows (\kref{AppC}), additional Type II GRBs that expand the sample of Paper I (\kref{PaperI}) and finally an Erratum to a table in Paper I (\kref{Erratum}).

In our calculations we assumed a flat universe with matter density $\Omega_M = 0.27$, cosmological constant $\Omega_\Lambda=0.73$, and Hubble constant $H_0=71$ km s$^{-1}$ Mpc$^{-1}$ \citep{Spergel2003}. Errors are given at the $1\sigma$, and upper limits at the $3\sigma$ level for a parameter of interest except where noted otherwise (e.g., host galaxy offsets that use XRT error circles at 90\% confidence).

\section{DATA GATHERING AND ANALYSIS}
\label{data}
\subsection{Selection Criteria for the Type I GRB Afterglow Sample}
\label{dataI}

As has been stated in the introduction, the ``Type I and Type II'' nomenclature is based on different observational quantities, not all of which are as easily accessible as $\TNT$, the duration (in the observer frame) during which the GRB emits 90\% of its fluence, starting after the first 5\% and extending to 95\% \citep{Kouveliotou1993}. A first classification scheme based on multiple observables for deciding whether a GRB belongs to the ``merger population'' or the ``collapsar population'' was given by \cite{DonaghyHETE}. A more advanced classification scheme has been proposed by \cite{Zhang080913}, and we will employ their flowchart in this work to split samples. Taking into account our complete sample\footnote{Note that we use only \emph{Swift}-era Type I GRBs. \cite{Gal-Yam2008} present observations of small IPN error boxes of BATSE-era Type I GRBs, and associate GRB 000607 with a bright galaxy at $z=0.1405$. We caution that the error box is still much larger than the X-ray error circes of \emph{Swift} Type I GRBs without optical afterglows which have been associated with nearby galaxies \citep{Bloom060502B, Stratta061201, BergerHighzSGRB}, and recent cases make it unlikely that this is the host galaxy, such as the detection of an extremely faint (27th mag) host galaxy for GRB 070707 \citep{Piranomonte070707} or studies on halo retention \citep{Zemp2009}. Also, based on IPN and BATSE data, \cite{Schaefer2006} and \cite{Tikhomirova2010} show that the pre-\emph{Swift} Type I GRBs they study are not associated with bright, nearby galaxies.}, we focused on deciding which GRBs are Type I or Type I candidates. Any remaining GRBs are then part of the Type II sample, without any controversial evidence that they might belong to the merger-population\footnote{We refer to Paper I and references therein on the possibility of further subclasses within the collapsar population, like intermediate GRBs and local low-luminosity GRBs.}.

Most GRBs in this paper that are classified as Type I GRBs also fulfill the ``classical'' $\TNT$ definition out of the BATSE era \citep{Kouveliotou1993} (25 of 38, 66\%). While some of these cases have so-called Extended Soft Emission Components (ESECs), these are so faint that they make up less than 5\% of the fluence at the end of the GRB. None of the GRBs in the Type II sample in Paper I have $\TNT\leq2$ s, although a few are just beyond the limit \citep[e.g., XRF 050416A, which is an X-Ray Flash and has an associated photometric SN,][]{Soderberg050416A}. Those that are longer than $\TNT=2$ s have other indicators that put them into the Type I class. \cite{NorrisBonnell} propose that a strong indicator for a merger-population event is a negligible spectral lag between different energy bands \citep[see also][]{Zhang2006}, i.e., Type I GRBs do not follow a lag-luminosity correlation \citep{Norris2000}\footnote{There are Type II GRBs with negligible spectral lag too, such as GRB 050717 \citep{Krimm050717}, but these are also extremely luminous, in accordance with the correlation.}. Furthermore, \cite{NorrisBonnell} propose that certain BATSE GRBs that had been classified as ``long'' are indeed Type I GRBs, consisting of a short (usually $\leq2$ s) single spike or series of spikes (the Initial Pulse Complex, IPC), followed by a long, usually soft and faint ESEC. This link was made due to several \emph{Swift} GRBs showing this characteristic light curve shape, with the classic example being GRB 050724, which was a clear Type I event due to its association with an old stellar population \citep{Barthelmy050724, Berger050724, Gorosabel050724}, and the most extreme \emph{Swift} event so far being GRB 080503 \citep[][note that \citealt{Xu060505060614} report a spectral lag for the uncharacteristically hard tail, though the detection is $2\sigma$ only]{Perley080503}. Of the remaining 13 events, 12 are described by light curves of the IPC+ESEC type (in which the bright ESEC contributes strongly to $\TNT$) the single exception being GRB 060505.

While we expect no or only very little light contribution from the decay of radioactive elements in the light of a Type I GRB afterglow \citep[][and references therein]{Hjorth050509B}, the lack of detected SN emission to very deep limits several weeks after the GRB would seem to be a strong criterion for a Type I origin (assuming it has otherwise been shown that the GRB afterglow is not overly reddened by line-of-sight extinction; this is not the case in either GRB 060505 [\citealt{Thoene060505, Xu060505060614}] or GRB 060614 [\citealt{DellavalleNature, Mangano060614, Xu060505060614}, this work]), especially for GRB 060505, where the SN limits are deeper than for any other GRB \citep[$M_R\gtrsim-10.5$ mag,][this work]{Ofek060505}. But it has been argued that the lack of radioactivity-powered emission may also be due to fallback black holes or extremely low energy deposition \citep[e.g.,][]{FynboNature, GalyamNature, DellavalleNature, Tominaga2007, Moriya2010}, so \cite{Zhang080913} use the presence of an SN as an incontrovertible Type II signal\footnote{Even here, it may be caution must be exercised, as \cite{TrojaSN} find a possible NS-NS or NS-BH merger channel in which the newly born NS after the second SN in a massive binary (the first star having already gone SN, leaving a NS or BH) can collide with the other compact object withing hours or days of the SN, which could create a Type I GRB coincident with a SN explosion. But they also point out that these SNe could be of any core-collapse type, and it seems quite unlikely a broad-lined Type Ic SN would create a NS which lives long enough to collide or merge with the other compact object \citep[but see, e.g.,][]{Ritchie2010}. Also see \cite{Dokuchaev2011}.}, the absence though is only one further step toward determining a Type I origin.

Almost all events in our paper are classified as Type I GRBs or Type I GRB candidates (e.g., any that do not have host associations cannot fulfill the definite Type I criteria of \citealt{Zhang080913}) without controversy. The steps leading to our classification of the GRBs in our sample are detailed in Table \ref{tabTypeISelection}. Three events, which we will later dedicate an entire chapter to, bear further discussion\footnote{Two recent events which are discussed as part of the Type II sample in Paper I and not here are the two highest-redshift GRBs, GRB 080913 \citep{Greiner080913} and GRB 090423 \citep{Tanvir090423, Salvaterra090423}. Both have $\TNT<2$ s in the rest frame, leading to the question if they might be Type I GRBs \citep{Zhang080913, Perez-Ramirez080913, Belczynski2010}. While a merger origin cannot be ruled out with high confidence, they are most likely due to collapsars. A very similar case which was recently reported is the $z\approx9.4$ GRB 090429B, which has $T_90\approx6$ s in the observer frame \citep{Cucchiara090429B}. A further intriguing case is GRB 090426, which has $\TNT<2$ s already in the observer frame, yet lies at $z=2.609$. Still, this is probably a Type II GRB \citep{Zhang080913, Antonelli090426, Levesque090426, Thöne090426, Nicuesa090426}. A Type I origin is supported by the duration and indications for a low circumburst environment density from the Ly$\alpha$ absorption line. Ambiguous indicators are the isotropic energy release (high for a Type I GRB, but comparable to GRB 090510), as well as the host galaxy, which has a very high luminosity and probably high metallicity (favoring a Type I origin), but the GRB occurred in a knot with higher SSFR and extinction (favoring a Type II origin). A Type II origin is favored by the afterglow luminosity, the redshift and the column density of highly ionized species which point to the environment being a star-forming region. This GRB is discussed in the light of its optical afterglow luminosity in \cite{Nicuesa090426}, where a luminosity typical of Type II GRB afterglows is found.}.

GRB 060614 has $\TNT\gg2$ s, but an IPC+ESEC light curve structure. A first indication that this may be a Type I event is the low host SSFR. The prompt emission also shows no spectral lag \citep[neither the IPC nor the ESEC,][]{GehrelsNature}, despite the isotropic energy release being on the low end for Type II events. It does fulfill the Amati relation using the integrated spectrum \citep{Amati060614}, but the spectrum is dominated by the extended emission, and the IPC (which is the part directly comparable to typical $\TNT\leq2$ s Type I GRBs) does not obey the Amati relation \citep{AmatiCONF}, as would be expected from a Type I event. All in all, either the low SSFR makes it a Type I event, or the low energy release makes it a Type I candidate.

GRB 060505 is a very special case. Its duration is in total $\approx10$ s including a faint precursor \citep{McBreen060505}, and it does not follow the IPC+ESEC prompt emission shape. As stated above, it has no associated SN down to very deep levels. It has been shown to have significant spectral lag \citep{McBreen060505}, but is far off the Amati relation \citep{KrimmBATWAM} as well as the lag-luminosity relation \citep{Zhang080913}. The host galaxy, a massive face-on spiral galaxy, is also ambiguous, with the GRB occurring at a large offset from the core but within a high SSFR star-forming region \citep{Thoene060505, Ofek060505}. Following the \cite{Zhang080913} flowchart, it is a Type I candidate due to the very low energy release.

According to the \cite{Zhang080913} flowchart, GRB 060121 should be classified as a ``Type II candidate''. It follows the Type I path well ($\TNT\leq2$ s, and also the IPC+ESEC light curve shape), but it obeys both the Amati relation and the lag-luminosity relation \citep{Zhang080913}, which brings it back to the Type II path, and then, at least for the $z=4.6$ solution \citep{Ugarte060121}, energetics arguments imply it is not of Type I. Still, it was seen as a traditional ``short/hard'' burst, so we will discuss it in the context of this work, as the \emph{observed} afterglow was extremely faint. It will therefore make an interesting test case as to how the luminosity of the afterglow light curve may add additional information which can help in the classification of such ambiguous events.

Furthermore, we have also compared our sample with the recent $\varepsilon$-classification criterion of \cite{Lü2010}. These authors define $\varepsilon$ as a ratio between the isotropic energy release and the rest-frame peak energy of the spectrum (this implies the redshift must be known), then plot $\log\varepsilon$ versus $\log T_{90,z}$, the rest-frame duration. They find that GRBs tend to cluster in three groups: The first group follows the Type II (and candidates) from \cite{Zhang080913}, the second the Type I GRBs (if only the IPC component is used in IPC+ESEC GRBs), and a third group (low $\varepsilon$ but large $T_{90,z}$) consist of nearby low-luminosity events which are often associated with SNe but show no classical afterglow emission. We determine $\varepsilon$ for all our candidates (using our assumed redshift, except for GRB 100117A, which has a secure redshift, \citealt{Fong100117A}; note that changing the redshift for cases with unknown redshift within reasonable bounds does not lead these GRBs to become high-$\varepsilon$ events) and find that they all belong to the low-$\varepsilon$, low-$T_{90,z}$ class (i.e., Type I GRBs), \emph{again} with the exception of GRB 060121, which is a marginal Type II GRB for the $z\approx4.6$ solution and a Type I GRB for the less likely $z\approx1.7$ solution.

\cite{GoldsteinBATSE} also recently presented a further distinction criterion, the energy ratio ($\mathrm{E_{peak}/fluence}$), to distinguish between the two different classes. We shall not discuss this here further, though.

Details on all GRBs in our sample can be found in the Appendix (\kref{AppA}), where we also list the sources of the data used in this study. In \kref{Observations} of the Appendix, we also gives details on additional original data on afterglows in our sample which we present in this work. We present data for eight GRBs in our sample, including the deepest published upper limits in some cases (GRB 050911, GRB 051210, GRB 060801) as well as several afterglow detections (GRB 060121 [afterglow discovery observations], GRB 060313, GRB 070714B [marginal]).

\subsection{Type I GRB Afterglow Light Curves}
\label{LCs}

In many cases (16 of 38, 42\%), no optical afterglows were discovered, so that only upper limits are available, either ground-based or by \emph{Swift} UVOT. In order to maximize the available light curve information for our study, we transformed the data of all filters to the $R_C$ band \citep[after correcting for the individual foreground extinction for each GRB and each filter,][]{SFD} by making the following assumptions: First, we assumed that the intrinsic spectral slope of the optical/NIR afterglow of each GRB is $\beta=0.6$, unless the data were sufficient to measure it. In the fireball model\footnote{While Type I GRBs derive from a different type of progenitor as Type II GRBs, most of the physics behind the GRB and the afterglow are expected to be identical \citep{NakarReview, Nakar2007, Nysewander2009}, i.e., a hyperaccreting accretion torus around a black hole which powers an ultrarelativistic fireball that propagates into the external medium \citep{Eichler2009, Lazzati2009}. The viability of both neutron star-neutron star and neutron star-black hole mergers to create Type I GRBs has been shown in numerical simulations \citep[e.g.,][]{Rosswog2003, Aloy2005, Rosswog2005, Oechslin2006, Lee2010, Pannarale2011, Rezzolla2011} and has also recently been supported by the measurement of a $2M_\odot$ mass for the pulsar PSR J1614$–$2230 \citep{Demorest2010, Özel2010}. BH-NS mergers, though, may account only for small numbers of Type I GRBs \citep{Belczynski2008, O'Shaughnessy2008}.}, if the cooling frequency $\nu_c$ lies blueward of the optical bands, it is $\beta=(p-1)/2$ \citep[e.g.,][and references therein]{SPN1998, ZhangMeszaros2004, Piran2005}, with the canonical value $p=2.2$ \citep{Kirk2000, Achterberg2001}, implying $\beta=0.6$. Observations of Type II GRB afterglows show that this situation has the highest probability (K06), and the mean and median values of the complete sample of Paper I are close to 0.6. While we caution that it has been shown that $p$ is not universal \citep[K06,][]{Shen2006, StarlingPaperII}, and that $\nu_c$ can also lie redward of the optical bands (e.g., the case of GRB 060505, \kref{AppA}), our assumption should be valid in the majority of cases. The influence of a different spectral slope on the shift $dRc$ (\kref{Shift}) is dependent on redshift. E.g., for $z=0.2$, $\Delta dRc=0.3$ mag between $\beta=0.5$ and $\beta=1.1$, for $z=0.8$, it is only $\Delta dRc=0.07$ mag. For the luminosity distribution, these small differences are not critical. Our second assumption is that the observed SED is unaffected by wavelength-dependent extinction through dust in the GRB host galaxies. As merger-induced events are typically expected to occur far from star-forming regions \citep[but see, e.g.,][]{Belczynski2006, Belczynski2007, Dewi2006, vandenHeuvel2007, D'AvanzoThree, TrojaSN}, this assumption is reasonable (see Table \ref{tabTypeISample})\footnote{We must make several notes of caution, however. At least one Type I GRB afterglow SED, that of GRB 050709, seems to show line-of-sight extinction even though the GRB is located in the outskirts of its host galaxy \citep{Ferrero050813}. While \cite{Gehrels2008} did not find any dark Type I GRBs, \cite{ZhengDark} show that the very red afterglow of GRB 070809 \citep{PerleyGCN070809B} is dark, and also suspect this could be the case for GRB 070724A, which has now been confirmed by the discovery of the very red afterglow of this event by \cite{Berger070724A}. In this work, we show that extinction along the line of sight to these two GRBs, if it is the source of the steep spectral slope, must be high ($\AV\approx0.9-1.5$, \kref{AppA}). Therefore, there must be cases where Type I GRB progenitors are surrounded by significant local extinction, and it is possible some cases where only upper limits were measured are affected additionally by dust. Note that \cite{Xu060505060614} also claim GRB 080503 is dark at 0.05 days after the GRB, but it is very unclear what the actual X-ray luminosity is as the X-ray afterglow is not detected any more.}. In those cases where no afterglow has been detected and we have upper limits only, we choose successively deeper limits, as the afterglows are not expected to rebrighten significantly and follow a typical monotonic decay (see Figure \ref{Bigfig1DetSec}).

Many Type I GRBs do not have measured redshifts. So far, no absorption spectroscopy of a Type I GRB afterglow has been successful \citep[see][Hjorth et al., in preparation, but see \citealt{TanvirGCN100816A, GorosabelGCN100816A} for the case of GRB 100816A]{Stratta061201, Piranomonte070707}, so that redshifts can only be determined from host galaxy spectroscopy. In some cases, no galaxies (or only extremely faint ones) are found in the \emph{Swift} XRT or optical afterglow error circles \citep[e.g.,][]{Piranomonte070707, Perley080503, FongHST, Rowlinson090515, Berger2010}, and the GRBs are instead assumed to be associated with bright nearby galaxies, such as in the case of GRB 050509B \citep[localized in the outskirts of a bright elliptical galaxy which itself is part of a cluster,][]{Gehrels050509B, BergerClusterSGRB}, GRB 060502B \citep{Bloom060502B}, GRB 061201 \citep{Stratta061201} and GRB 070809 \citep{PerleyGCN070809C, Berger2010}, or galaxy clusters, as for GRB 050813 \citep{ProchaskaSGRB, Ferrero050813}, GRB 050911 \citep{BergerClusterSGRB}, and GRB 090515 \citep{Berger2010}. Finally, if no association can be made at all, we choose a redshift $z=0.5$, which is the (rounded) median value of all measured redshifts we consider secure \citep[see also][]{Nysewander2009, deUgartePostigoIntermediate}.
In four cases (GRB 051227, GRB 060313, GRB 070707 and GRB 080503), we choose $z=1$, as the host galaxies of these GRBs (localized to subarcsecond precision through their optical afterglows) are exceedingly faint ($R\gtrsim26$, \citealt{D'AvanzoThree}, \citealt{BergerHighzSGRB}, \citealt{FongHST}, \citealt{Piranomonte070707}, \citealt{Perley080503}, Hjorth et al., in preparation) and thus resemble the hosts of Type II GRBs (although we caution that we have no detailed information on properties such as star-formation rates etc.).
We caution that while there is evidence that these GRBs do not lie much beyond $z=1$ \citep[e.g., the detection of the afterglow of GRB 060313 in all UVOT filters,][]{Roming060313}, they may lie significantly closer, with their host galaxies lying at the faint end of an as yet unknown luminosity distribution\footnote{Recent studies of the galaxy population hosting Type I GRBs \citep{Berger2009} show that they resemble the typical field galaxy population, having higher luminosities, higher metallicities and lower SSFR than Type II GRB host galaxies, and being larger as well \citep{FongHST, LeiblerBerger}. Many of them show exponential disk profiles, which are typical of spiral galaxies, whereas none shows disturbed morphology (typical of star-bursting Type II GRB host galaxies, \citealt{Conselice2005, FruchterNature, Wainwright2007}) and only GRB 050709 has an irregular host \citep{FongHST}. Such moderately star-forming galaxies are also predicted by population synthesis models to host many Type I GRBs \citep{O'Shaughnessy2008}. On the other hand, the cases of GRB 070707 \citep{Piranomonte070707} and GRB 080503 \citep{Perley080503} show that Type I GRB host galaxies can be extremely faint and probably low-mass.}. On the other hand, population synthesis models predict high rates of Type I GRBs at high redshifts from rapid merger channels in proto-elliptical galaxies \citep{O'Shaughnessy2008}, though it is likely that these GRBs cannot be detected by the current generation of detectors \citep{Belczynski2010}.

In the following, we will discuss the sample with secure and the sample with insecure redshifts separately, also aiming to compare the two to each other.

\subsection{Shifting Light Curves to a Common Redshift}
\label{Shift}
With knowledge of the redshift $z$, the extinction-corrected spectral slope $\beta$ and the host galaxy rest frame extinction $\AV$, we can use the method described in K06 to shift all Type I GRB afterglows to a common redshift of $z=1$, corrected for extinction along the line of sight (as well as $z=0.1$, see \kref{SNlimits}). As stated in \kref{dataI}, we do not have $\beta$ and $\AV$ for most Type I GRB afterglows, and in some cases, even $z$ is unknown. Thus, for many Type I GRB afterglows, we cannot derive results analogous to the Type II GRB afterglows (Paper I), but have to view them as an ensemble. With the exception of some special cases which we place at $z=1$ (see Appendix \ref{AppA}), as mentioned, we assume $z=0.5$ for GRBs with no known or assumed host galaxy. Compared with the true but unknown values for the parameters needed for the magnitude shift $dRc$, the magnitudes or upper limits of some GRBs may be fainter or brighter. The effect is stronger for low redshifts, for $z=0.2$ in comparison to $z=0.5$, it is $\Delta dRc=2.1$, for $z=0.8$ in comparison to $z=0.5$, it is $\Delta dRc=1.1$. Still, in a statistical sense, the effect will not be strong as we expect the true redshifts of the GRBs to be distributed relatively evenly around $z=0.5$.

\subsection{Determining Upper Limits on a SN Light Component}
\label{SNlimits}
Using the method described in \kref{Shift}, we have shifted Type I GRB afterglow light curves to a redshift of $z=0.1$. In most cases, the interval between the redshift of the GRB and $z=0.1$ is smaller in $z$-space than if it were shifted to $z=1$, implying a smaller uncertainty through the unknown $\beta$. Another reason for performing this analysis at $z=0.1$ and not at $z=1$ is that at the latter redshift, the $R_C$-band light curve of the SN template in the observer rest-frame may provide inaccurate flux measurements given the UV-deficiency exhibited by Type Ic SNe such as those which are found to be associated with (Type II) GRBs \citep{Filippenko1997}. Our sample consists of those Type I GRBs that have a known redshift (which, in some cases, is derived only from associating the GRB with a nearby bright galaxy or a cluster, we place less significance on these cases) and late detections/upper limits: GRB 050509B, GRB 050709, GRB 050724, GRB 050813 (less significant), GRB 050906 (less significant), GRB 051221A, GRB 060502B (less significant), GRB 060505, GRB 060614 (the latter two being the ``SN-less long GRBs''), GRB 061201 (less significant), and GRB 080905A. We then compare the detections/upper limits with the template light curve of SN 1998bw \citep{Galama1998}\footnote{For the world model used here, SN 1998bw was 0.19 mag less luminous than given in \cite{Galama1998}.}, see \cite{Zeh2004} for details of the method and descriptions of the parameters $k$ and $s$, which measure the GRB-SN luminosity in units of the luminosity of SN 1998bw at peak and the light curve stretching in comparison to the SN 1998bw light curve, respectively. In our comparison, we conservatively assume that the late optical emission from the Type I GRBs is due only to SN light and there is no contribution from afterglow emission. In the case of deep (host-galaxy subtracted) detections, we fit the template to pass through the brighter $1\sigma$ error bar of the faintest data point, and in the case of an upper limit, we fit the template to pass through the most restrictive upper limit. As we have no information at all about the stretch factor $s$, we assumed $s=1$ in all cases. If the stretch factor is smaller than SN 1998bw, such as XRF 060218/SN 2006aj \citep{Ferrero060218} or the photometric SN bump of XRF 050824 \citep{Sollerman050824} the luminosity limit typically would be slightly less constraining. Our fitting then results in a value of the luminosity factor $k$, e.g., $k=0.1$ implies a SN that has 0.1 times the peak luminosity of SN 1998bw in the same band at the same redshift. As there have been no signs of SN bumps in the light curves of Type I GRB afterglows, our $k$ values can be seen as conservative upper limits on any SN contribution.

\subsection{The Mini-SN/Macronova Model}
\label{Mini}

The mini-SN model was introduced by LP98ab as a potential observational consequence following the merger of two compact objects (NS + NS or  NS + BH). During the merger, neutron star matter at nuclear densities can be ejected at subrelativistic velocities, condensing into neutron-rich nuclei which rapidly decay, yielding a similar heating source as for classical radioactivity-driven SNe.

From the computational point of view the analytical solutions given in LP98ab are easy to handle and, therefore, we used them in our study. Basically, the LP98ab model depends on three free parameters, the ejected mass $M_{\rm ej}$ which is assumed to be identical to the radiating mass, the expansion velocity $v$ of the ejected matter, which is assumed to be independent of time, and the fraction $f$ of rest mass energy that is transformed into internal heat of the ejecta. A more detailed description can be found in \kref{AppC} of the Appendix.

LP98ab considered two cases of decay laws for the heating source, an exponential-law decay and a power-law decay. They found that in both cases for, e.g., $f=0.001$ bright mini-SNe are predicted with bolometric peak luminosities up to $10^{44}$ erg/s, or even higher. Thirteen years later, in the \emph{Swift} era, it has become clear however that such bright mini-SNe following short bursts are not seen \citep{Fox050709, Hjorth050509B, Perley080503, Kocevski070724A}.

\section{RESULTS}

\begin{figure*}[!t]
\epsfig{file=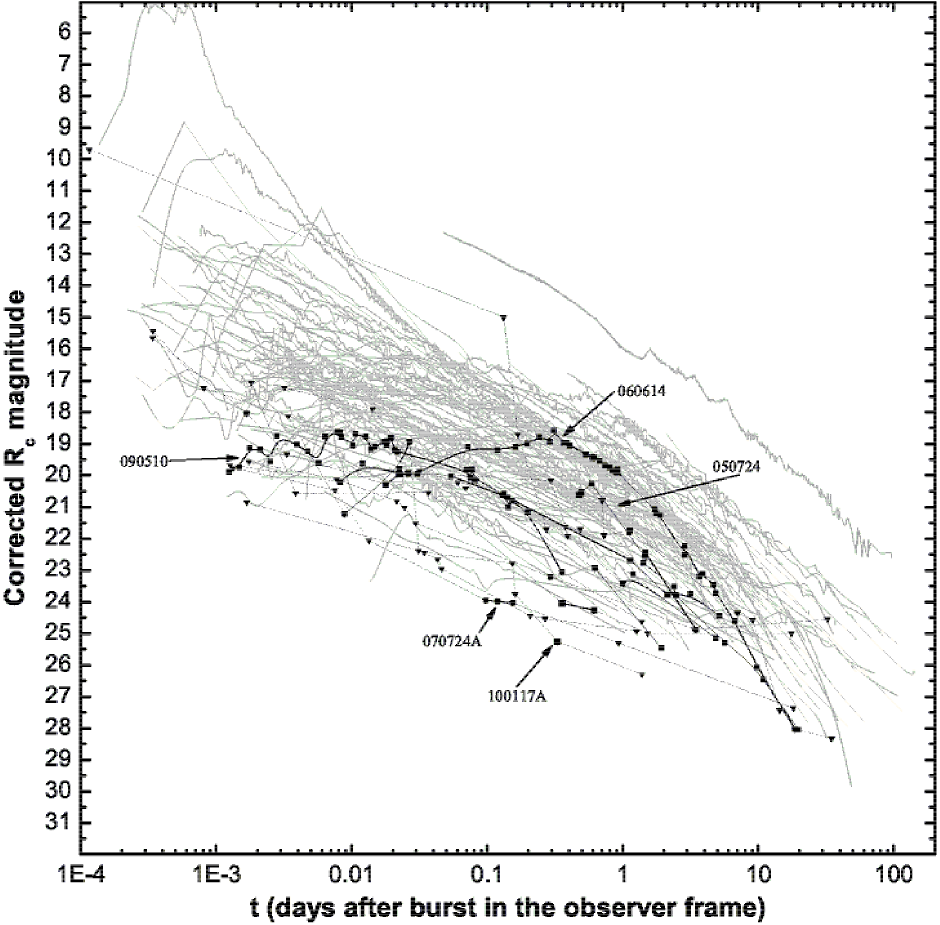,width=1\textwidth}
\caption[]{Afterglows of Type I and Type II GRBs in the observer frame. All data have been corrected for Galactic extinction and, where possible, the contribution of the host galaxy has been subtracted. Thin gray lines are Type II GRB afterglows, taken from K06, Paper I as well as this work. Black lines with data points are upper limits (thin straight dashed lines, downward pointing triangles) or detections (splines, squares) of Type I GRB afterglows, in this case for Type I GRBs that have both at least one afterglow detection as well as a redshift that we consider secure. In general, the detected afterglows are comparable with the fainter part of the observed Type II GRB afterglow sample, though several cases (such as the detections of GRB 070724A and GRB 100117A) are fainter than any Type II GRB afterglow in our sample. The single detected Type I GRB afterglow that is comparable in brightness to the brighter Type II GRB afterglows is that of GRB 060614. Several other exceptional GRB afterglows mentioned in the text are labeled.}
\label{Bigfig1DetSec}
\end{figure*}

\begin{figure*}[!t]
\epsfig{file=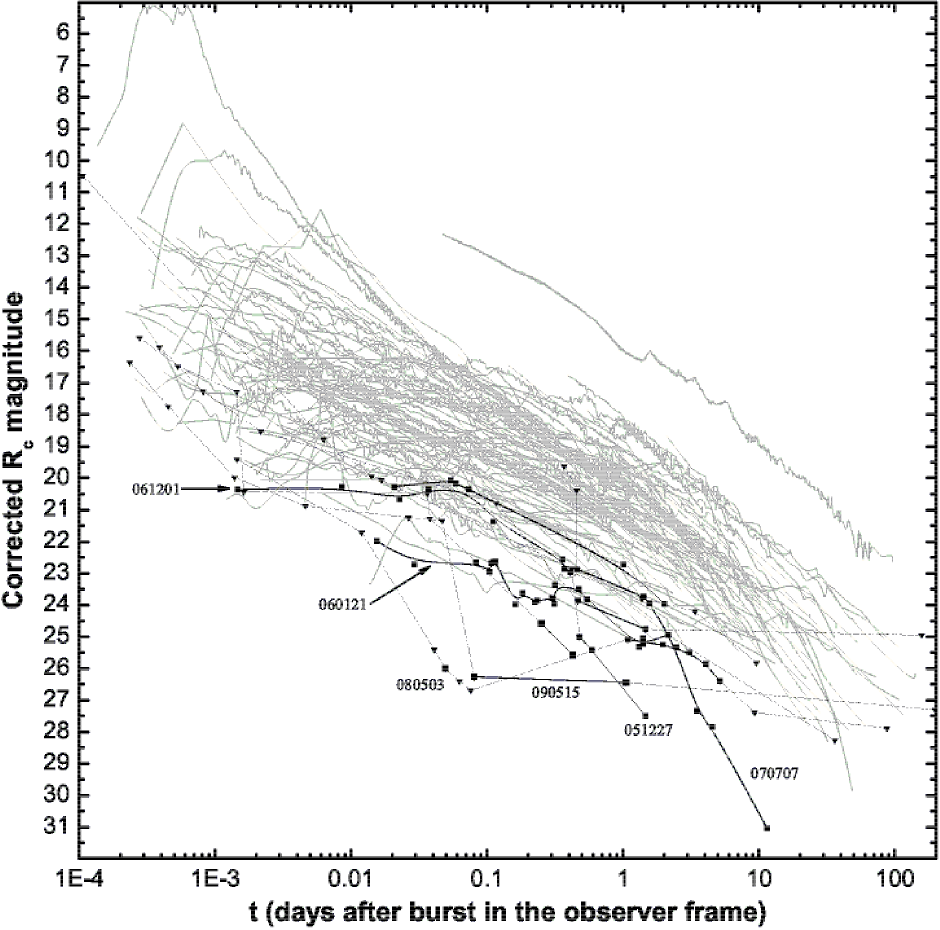,width=1\textwidth}
\caption[]{The same as Figure \ref{Bigfig1DetSec}, but with Type I GRB afterglows with at least one detection and insecure (or even unknown) redshifts. These afterglows present a quite strong contrast to those shown in Figure \ref{Bigfig1DetSec}, as most of them are significantly fainter than any Type II GRB afterglow in our sample. The early detections of GRB 080503 and GRB 090515 are the faintest-ever detected afterglows, and GRB 061201 is one of the faintest-ever at very early times. The final detections of GRB 051227 and GRB 070707 are derived after subtracting the host galaxy magnitudes and are associated with large errors.}
\label{Bigfig1DetInSec}
\end{figure*}

\begin{figure*}[!t]
\epsfig{file=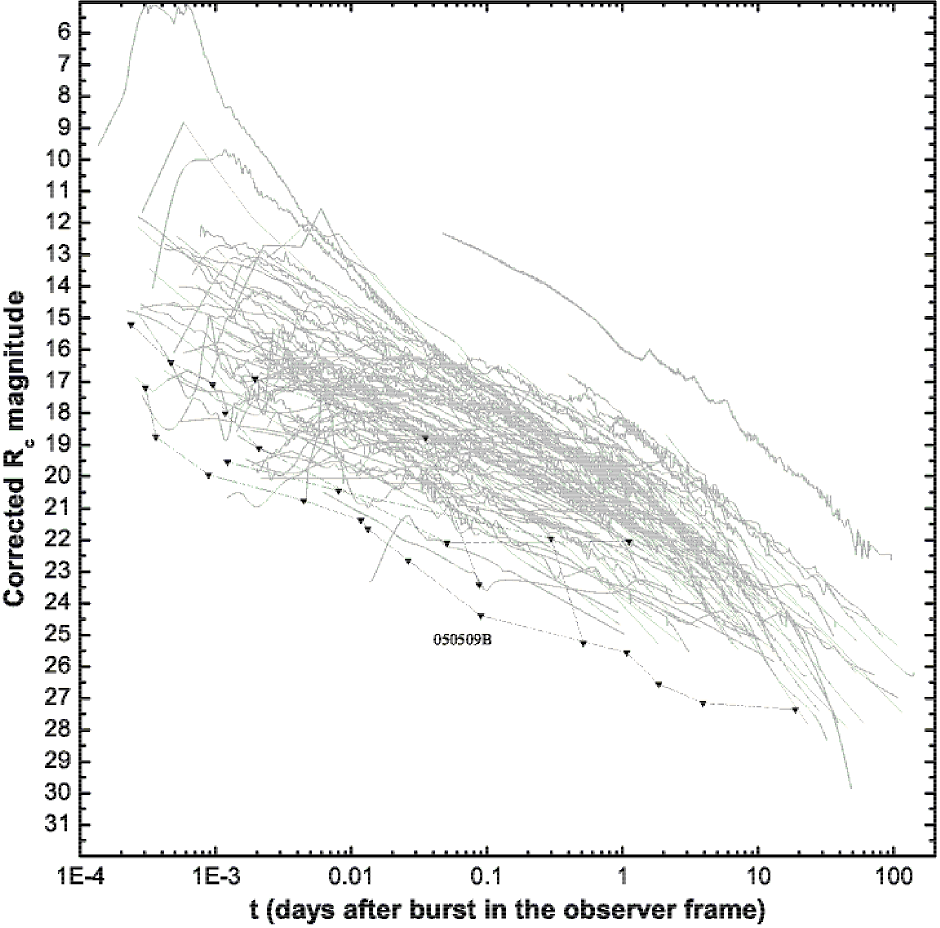,width=1\textwidth}
\caption[]{The same as Figure \ref{Bigfig1DetSec}, but with Type I GRBs for which only upper limits could be derived on their afterglows, but for which host galaxies have been securely (so we consider) identified and redshifts measured. Only four GRBs fulfill these criteria, GRB 050509B is exceptional by being fainter than any Type II GRB afterglow in our sample at almost any given time when measurements were taken. The single even fainter Type II GRB afterglow (at $\approx0.01$ days) is that of GRB 070802, this afterglow suffered from very high (and uncorrected in this figure) extinction \citep{Kruehler070802, Eliasdottir070802} and this $R_C$ band light curve was created by shifting the much brighter NIR data to the observed level of the very faint $R_C$ afterglow (Paper I).}
\label{Bigfig1ULSec}
\end{figure*}

\begin{figure*}[!t]
\epsfig{file=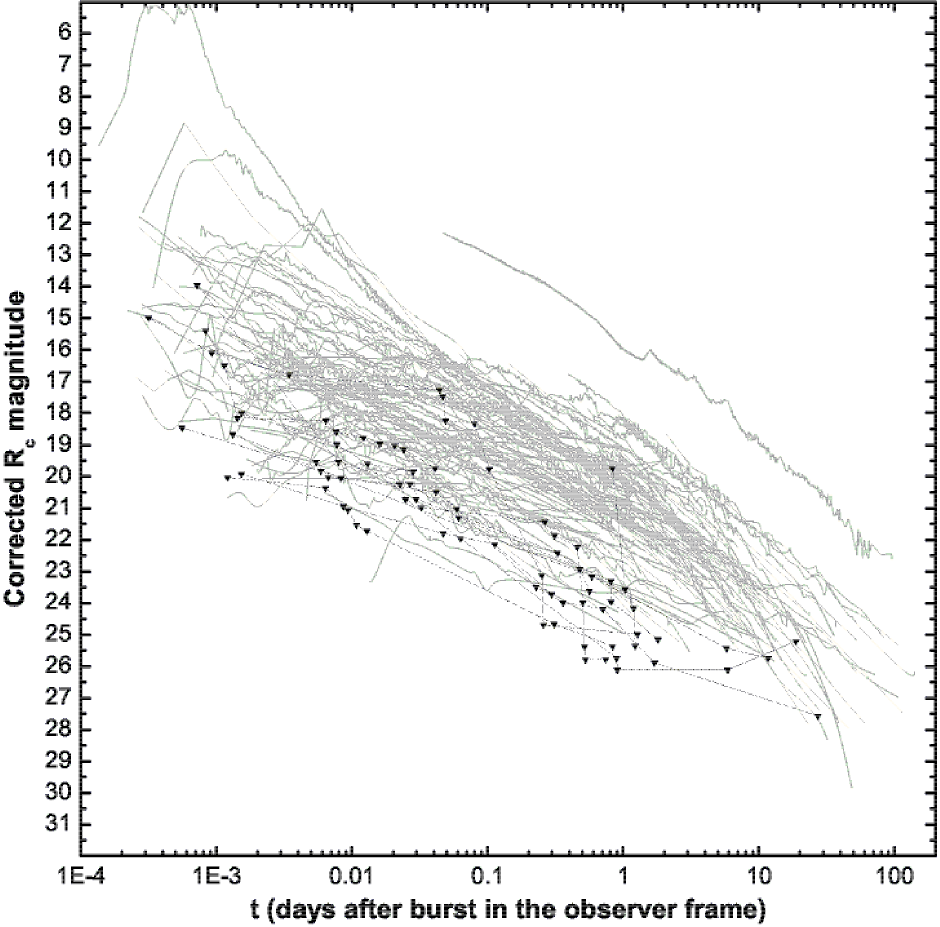,width=1\textwidth}
\caption[]{The same as Figure \ref{Bigfig1DetSec}, but with Type I GRBs for which only upper limits could be derived on their afterglows, and for which redshifts are insecure or even unknown. At early times, almost all upper limits are less deep than the faintest detected Type II GRB afterglows (mostly due to this being the detection limit of today's telescope technology), but starting at $\approx0.01$ days, there is a whole cluster of upper limits which would have detected any Type II GRB afterglow in our sample.}
\label{Bigfig1ULInSec}
\end{figure*}

\subsection{Observed Type I GRB Afterglows}
\label{Observed}
The observed light curves of the afterglows of our Type I GRB sample are presented in comparison with the pre-\emph{Swift} and \emph{Swift}-era Type II GRB afterglow light curves (K06, Paper I, this work) in Figures \ref{Bigfig1DetSec}, \ref{Bigfig1DetInSec}, \ref{Bigfig1ULSec}, and \ref{Bigfig1ULInSec}. Upper limits are marked with downward pointing triangles connected by thin straight dashed lines, while detections are squares connected with splines. All the afterglow data have been corrected for Galactic extinction (which is often small) and in some cases, the contribution of the host galaxy was subtracted (see Appendix \kref{AppA} for more details on each single GRB). We have separated the Type I GRB afterglow sample into four subsamples: Figure \ref{Bigfig1DetSec} shows all afterglows of Type I GRBs with detections as well as redshifts we consider secure (note that in some cases, most data points are upper limits only). Figure \ref{Bigfig1DetInSec} shows Type I GRBs with detected afterglows (again, partly also upper limits) but insecure or unknown redshifts. Figure \ref{Bigfig1ULSec} shows the cases where the redshift is considered secure, but only upper limits have been found (a total of only four cases), and Figure \ref{Bigfig1ULInSec} shows all afterglows with only upper limits and with insecure or unknown redshifts. We have labeled only a few special afterglows in each figure, as this would otherwise decrease legibility.

It is visible immediately that observationally, the optical afterglows of Type I GRBs are typically fainter than those of Type II GRBs \citep[see also][]{Gehrels2008, Nysewander2009}. Many optical afterglows are not detected at all, to upper limits that would have clearly detected almost all Type II GRB afterglows in this sample\footnote{We caution that the Type II sample of Paper I is biased toward (observationally) bright afterglows due to the sample selection criteria. There have been dark Type II GRBs in the \emph{Swift}-era that are also undetected optically to limits similar to the Type I GRB afterglow limits. While this makes the relative faintness of Type I GRB afterglows a less significant result from an observational point of view, a more significant distinction will be made in the intrinsic luminosities as we show in \kref{Intrinsic}.}. This is especially the case for early times ($<0.01$ days), where only a few Type II GRB afterglows (e.g., GRB 050820A, XRF 050416A, GRB 070110, GRB 070419A, GRB 070802, GRB 080603A, see Paper I and Appendix \kref{PaperI}) are fainter than most limits.

The most constraining upper limits at early times are on GRB 050509B (Figure \ref{Bigfig1ULSec}), which was observed rapidly by ROTSE \citep{RykoffGCN050509B} and RAPTOR \citep{WozniakGCN050509B}, an upper limit of $R_C>18.75$ is found after just 30 s. Furthermore, \cite{Bloom050509B} give an upper limit $R_C>24.4$ at only 0.09 days after the GRB, over 1 mag deeper than needed to detect any Type II GRB in the sample of Paper I\footnote{Note that the two faintest GRBs at this time are GRB 080913, which was at an extremely high redshift \citep{Greiner080913}, and GRB 070802, which was highly extincted \citep{Kruehler070802, Eliasdottir070802}. GRB 080913 was completely undetected in the $R_C$ band, this composite light curve was created by shifting the NIR data via the derived spectral slope to the magnitude that would be expected in the $R_C$ in a completely ionized universe \citep[see][]{Kann050904}. The light curve of GRB 070802 is also a composite made up of mostly NIR detections shifted to the $R_C$ band, it was completely undetected in $R_C$ at early times. Therefore, both of these light curves do not represent the actual observational capabilities.}. At about 0.05 days after the GRB, one GRB, 080503, sticks out (Figure \ref{Bigfig1DetInSec}), with both upper limits and a single detection at $\approx26$th mag, these are the deepest early detections and non-detections achieved for an afterglow so far \citep{Perley080503}. At $\approx0.1$ days, the afterglow of GRB 090515 (Figure \ref{Bigfig1DetInSec}) is also found to be exceedingly faint \citep{Rowlinson090515}. In both cases, detections were achieved only due to the combination of 8 m-class telescopes and excellent observing conditions. The faintest Type I afterglows in our sample around one day are those of GRB 051227 (Figure \ref{Bigfig1DetInSec}), discovered by the VLT \citep{MalesaniGCN051227a} and seen to decay very rapidly, possibly due to post-jet-break decay \citep[][see Appendix \kref{AppA}]{BergerHighzSGRB, D'AvanzoThree}, as well as GRB 090515, even though this afterglow decays very slowly \citep[note that a very late and even deeper observation shows that there is no underlying host galaxy contributing to this slow decay,][]{Rowlinson090515}. At later times, another extremely faint and rapidly decaying afterglow is that of GRB 070707 (Figure \ref{Bigfig1DetInSec}), which also had a very faint host galaxy \citep{Piranomonte070707}. Note that in both cases, the host galaxy magnitude has been subtracted, the deepest data points have large errors. The only afterglow of a Type I GRB (and a controversial one at that) that is comparable to the brighter Type II GRB afterglows is that of GRB 060614 \citep[Figure \ref{Bigfig1DetSec},][]{DellavalleNature, FynboNature, GalyamNature, Mangano060614, Xu060505060614}. This afterglow starts out faint but rises to a peak at about 0.25 days \citep{GalyamNature}, followed by a typical afterglow decay that includes a jet break \citep{Mangano060614, Xu060505060614}. The afterglow of GRB 050724, which experiences a flare at the time of its earliest detection \citep[Figure \ref{Bigfig1DetSec},][]{Malesani050724}, is comparable to the mean magnitude of the Type II GRB afterglow sample at this time but decays rapidly.

\subsection{The Luminosity Distribution of Type I GRB Afterglows}
\label{Intrinsic}

\begin{figure*}[!t]
\epsfig{file=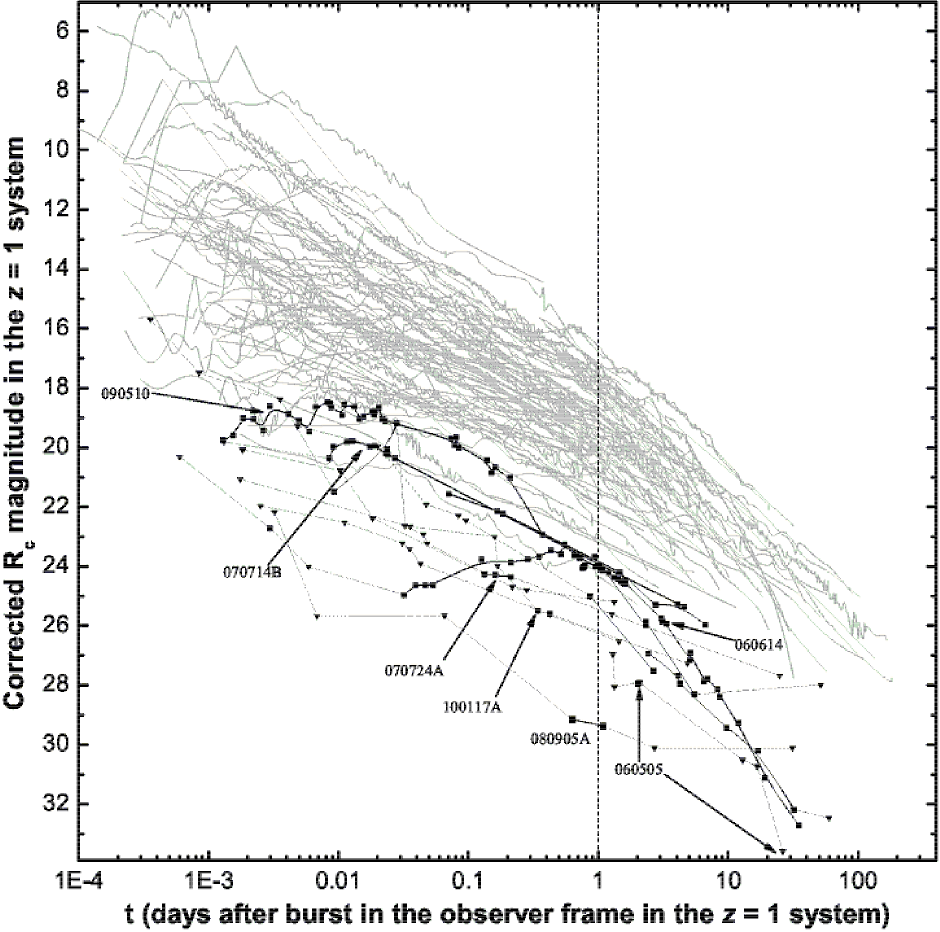,width=1\textwidth}
\caption[]{Afterglows of Type I and Type II GRBs in the observer frame after transforming all afterglows to $z=1$. The additional \emph{Swift}-era Type II afterglows expand the luminosity distribution in comparison the the pre-\emph{Swift} distribution, weakening the clustering reported before (see Paper I). The selection of Type I GRB afterglows in this figure is identical to that of Figure \ref{Bigfig1DetSec}, i.e., afterglows that have both at least one detection and a redshift we consider secure. It is evident that afterglows of Type I GRBs, including that of GRB 060614, are fainter than those of Type II GRBs at one day in general, with the brightest being only as luminous as the faintest Type II GRB afterglow in our sample, that of XRF 050416A (Paper I). At early times, the afterglow of GRB 090510 is comparable to faint Type II GRB afterglows. The afterglow of GRB 060505, which is a unique, unclear case (\kref{Hybrid}), is extremely faint. GRB 080905A, which occurred at a low redshift \citep{Rowlinson080905A}, has an extremely underluminous afterglow.}
\label{Bigfig2DetSec}
\end{figure*}

\begin{figure*}[!t]
\epsfig{file=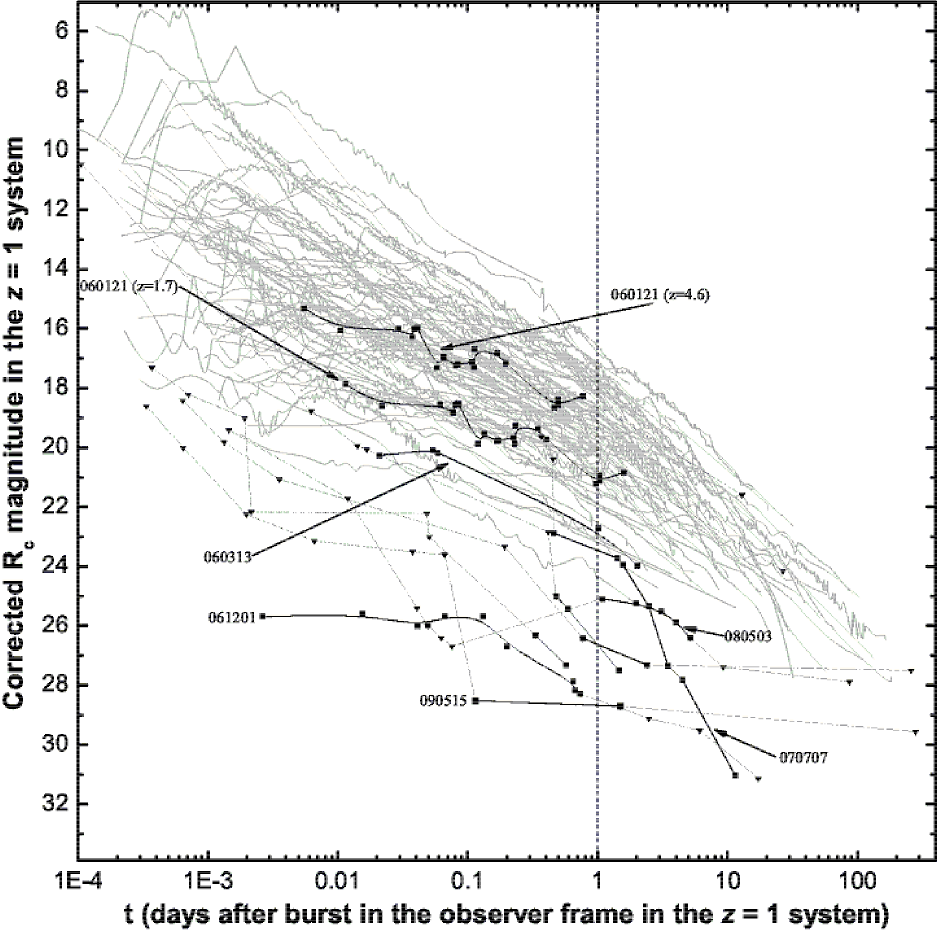,width=1\textwidth}
\caption[]{The same as Figure \ref{Bigfig2DetSec}, but with the same selection as in Figure \ref{Bigfig1DetInSec}. The faintest early afterglow is that of GRB 061201, assuming $z=0.111$ \citep{Stratta061201}. This is about 11 mag fainter than typical Type II GRB afterglows detected at this time, and still 6 mag fainter than the faintest ones. The afterglow of GRB 060313, assuming $z=1$, is comparable to faint Type II GRB afterglows, as is that of GRB 070707 at one day (also, $z=1$ is assumed), before the steep decay sets in \citep{Piranomonte070707}. The afterglow of GRB 090515 is extremely subluminous, comparable to that of GRB 080905A in Figure \ref{Bigfig2DetSec}. Assuming $z=4.6$, the afterglow of GRB 060121 is comparable to typical Type II GRB afterglows, and comparable to moderately faint ones if $z=1.7$, in stark contrast to all other Type I GRB afterglows.}
\label{Bigfig2DetInSec}
\end{figure*}

\begin{figure*}[!t]
\epsfig{file=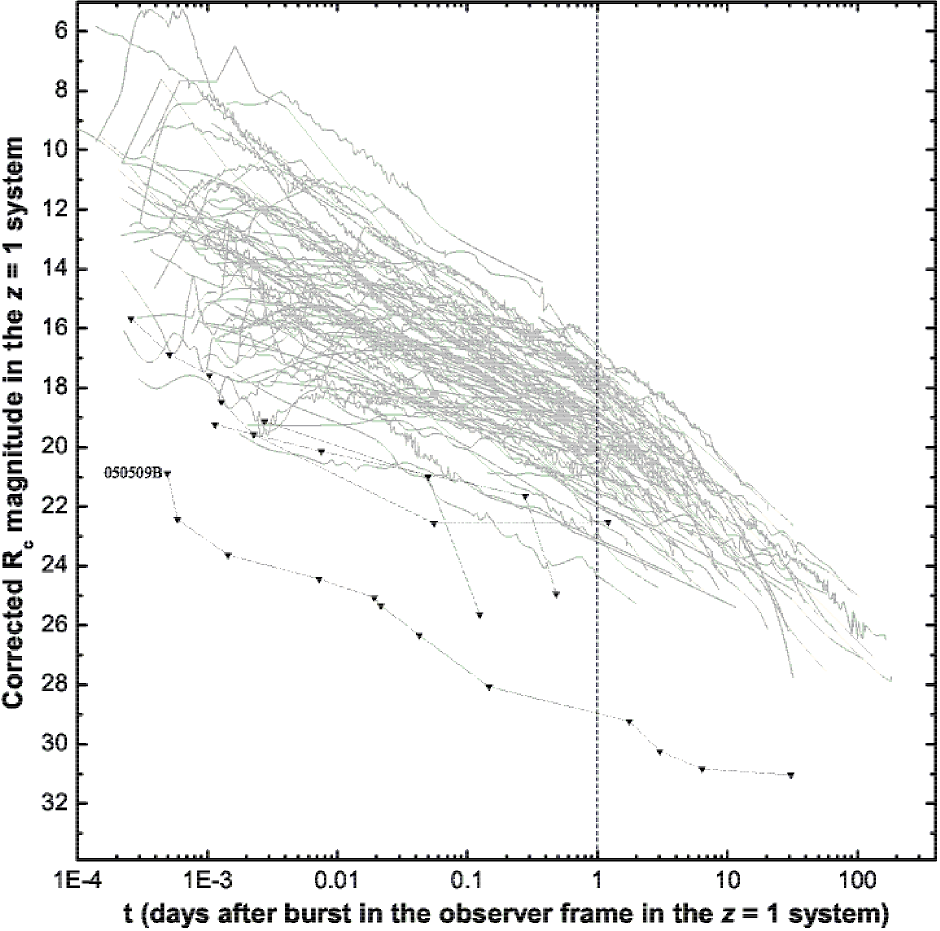,width=1\textwidth}
\caption[]{The same as Figure \ref{Bigfig2DetSec}, but with the same selection as in Figure \ref{Bigfig1ULSec}. The upper limits on the afterglow of GRB 050509B are much lower than any Type II GRB afterglow, and the three other afterglows in this plot are also fainter than any Type II GRB afterglow at least during certain epochs.}
\label{Bigfig2ULSec}
\end{figure*}

\begin{figure*}[!t]
\epsfig{file=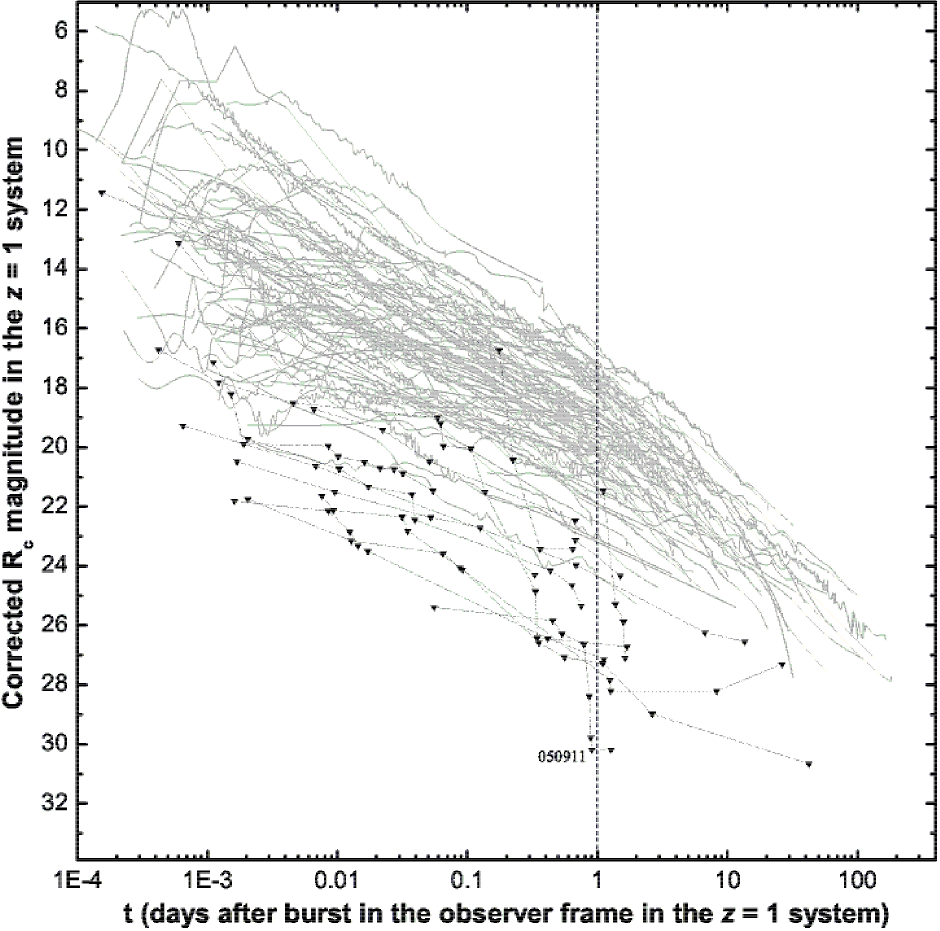,width=1\textwidth}
\caption[]{The same as Figure \ref{Bigfig2DetSec}, but with the same selection as in Figure \ref{Bigfig1ULInSec}. Most of the upper limits at $\approx1$ d are lower than any Type II GRB afterglow in our sample, but here we caution that the uncertain/unknown redshifts make this finding less significant than in the cases with secure redshifts. If the assumed cluster redshift \citep{BergerClusterSGRB} for GRB 050911 is correct, it represents the deepest upper limits obtained so far on a GRB afterglow at one day in the $z=1$ frame.}
\label{Bigfig2ULInSec}
\end{figure*}

After shifting all afterglows to $z=1$ (\kref{Shift}), we can compare the afterglows of Type I and Type II GRBs. The results are shown in Figures \ref{Bigfig2DetSec}, \ref{Bigfig2DetInSec}, \ref{Bigfig2ULSec}, and \ref{Bigfig2ULInSec}. The labeling and splitting into different subsamples is is identical to that in Figures \ref{Bigfig1DetSec}, \ref{Bigfig1DetInSec}, \ref{Bigfig1ULSec}, and \ref{Bigfig1ULInSec}, respectively. Several afterglows (partly different ones from Figures \ref{Bigfig1DetSec} to \ref{Bigfig1ULInSec}) have been labeled. Magnitude shifts $dRc$ and absolute magnitudes $M_B$ at one day after the burst are given in Table \ref{tabTypeI}.

If one considers all afterglows independent of how secure their redshift is, it is apparent that the afterglows of Type I GRBs spread even further apart, whereas the distribution of Type II GRB afterglows retains about the same width (Paper I, the clustering claimed in K06 is not found anymore with the addition of the \emph{Swift}-era sample). At 0.1 days, the total span is greater than 11 mag, from GRB 060121 at 17th mag \citep[assuming $z=4.6$, though this is insecure,][]{Ugarte060121} to the afterglow of GRB 090515 at $R_C=28.5$ mag (also a case with insecure redshift, but an upper limit on the GRB 050509B afterglow is almost as deep). Assuming $z=1.7$ for GRB 060121, the spread is about 1.5 mag less. At the same time, the spread of Type II GRB afterglows is about 8 mag, from 13th (the insecure case of GRB 060210, see Paper I) to 21st mag (XRF 050416A), and these afterglows tend to cluster even more strongly at later times. If one uses only the Type I GRB afterglows with secure redshifts, the spread becomes similar to that of the Type II GRB afterglows, from 20th mag (GRB 090510) to $>28$th mag (GRB 050509B, as mentioned above, of course the actual afterglow may be significantly fainter, which would again imply an increased spread). At one day, the spread of Type I GRB afterglows with secure redshifts is smaller, 5.5 mag from 23.8 mag (GRB 051221A) to 29.3 mag (GRB 080905A). This minimum spread is very secure, as both afterglows are detected at this time. Using the insecure redshifts additionally, the spread is much larger, almost twelve mag from 18.5 mag (GRB 060121 at $z=4.6$) to $>30.2$ mag (GRB 050911), or about nine mag if GRB 060121 is at $z=1.7$. Discounting GRB 060121 completely due to it being classified as a Type II candidate, the brightest afterglow is GRB 060313 at one day assuming $z=1$, and the spread is very similar to that of the Type II GRB sample.

The variance of the complete Type II GRB afterglow Golden Sample of Paper I and this work (69 events, both pre-\emph{Swift} as well as \emph{Swift}-era) at one day is 2.9 mag, and it is 2.6 mag for the larger sample of all Type II GRBs (94 events). For the Type I GRB afterglow sample, we find the following values: For the detections with secure redshifts at one day (eleven GRBs), 2.7 mag. For the detections with insecure redshifts, it is 11.9 mag (with GRB 060121, assuming $z=4.6$, nine events), 8.2 mag (with GRB 060121, assuming $z=1.7$, nine events), and 5.8 mag (without GRB 060121, eight events). If we add all detections with secure redshifts to the latter three samples, we find variances of 6.4 mag, 4.9 mag and 4.0 mag, respectively. Giving the variance of the upper-limit only samples yields little valuable information, but if we, for completeness' sake, assume the actual afterglow magnitudes to be equal to the upper limits at one day, we find 3.6 mag and 4.7 mag for the upper limits with secure redshifts (seven GRBs) and those without (ten GRBs), respectively.

Furthermore, the Type I GRB afterglows are much less luminous than those of Type II GRBs, as has been predicted by \citet[][see also \citealt{Fan2005}]{Panaitescu2001}. The one clear exception is GRB 060121 (Figure \ref{Bigfig2DetInSec}), which probably lies at high redshift and is strongly collimated \citep{Ugarte060121, Levan060121}, and which we classify as a Type II GRB candidate. It is comparable in luminosity to typical Type II GRB afterglows if $z=4.6$, and comparable to faint Type II GRB afterglows if $z=1.7$. The afterglow of the extremely energetic GRB 060313 \citep{Roming060313}, assuming $z=1$, is also comparable to the faintest Type II GRB afterglows of the sample (Figure \ref{Bigfig2DetInSec}), the same is the case for the also extremely energetic \emph{Fermi}-LAT GRB 090510 (Figure \ref{Bigfig2DetSec}, for this GRB the redshift is secure), though it fades rapidly after 0.02 days \citep{DePasquale090510, McBreenLAT}. At about one day, the afterglow of GRB 060614, by far the brightest observed Type I GRB afterglow (\kref{Observed}), is just slightly brighter than the afterglow of XRF 050416A (the faintest afterglow in the sample of Paper I), and then it becomes even fainter rapidly. The late optical flare of GRB 050724 \citep{Malesani050724} is seen to peak at a similar magnitude, and the magnitude of the afterglow of GRB 051221A is also comparable (Figure \ref{Bigfig2DetSec}, all three redshifts are secure). Assuming the association with a galaxy at $z=0.111$ \citep{Stratta061201, Berger2010}, the afterglow of GRB 061201 has a magnitude of $R_C\approx25.5$ just a few minutes after the GRB, which is about 11 mag fainter than the typical early Type II GRB afterglows (Figure \ref{Bigfig2DetInSec}). The faintest detected afterglow at one day is that of GRB 080905A \citep{Rowlinson080905A}, at $R_C=29.3$ (Figure \ref{Bigfig2DetSec}); the redshift for this GRB is secure. Even fainter are the upper limits (derived in this work) on GRB 050911 (Figure \ref{Bigfig2ULInSec}), if one assumes an association with a galaxy cluster at $z=0.1646$ \citep{BergerClusterSGRB}. The afterglow of GRB 060505, for which it is unclear if it is a Type I GRB \citep[][as well as our own classification according to \citealt{Zhang080913} as well as \citealt{Lü2010}]{Ofek060505} or a Type II GRB \citep{FynboNature, Thoene060505, McBreen060505}, is here seen to be about 3 mag fainter than the faintest Type II GRB afterglows, but well comparable to the other Type I GRB afterglows or upper limits thereon. It is thus clearly not a classical Type II GRB, but also not of the subluminous Type II family, such as GRB 980425 \citep{Galama1998}, GRB 031203 \citep{Sazonov031203, Soderberg031203, Malesani2004} and XRF 060218 \citep{Campana060218, Pian060218, Soderberg060218}, as these GRBs, while possessing very faint afterglows, were also accompanied by energetic SNe. We refer to \kref{Hybrid} for a deeper discussion on this GRB. Three afterglows which were seen to be exceptional observationally (\kref{Observed}), namely GRBs 051227, 070707 and 080503, are all not remarkable any more. In all three cases we caution that we do not know a redshift, but have assumed $z=1$ due to the fact that all three have very faint host galaxies -- and in this case, their exceptional observational faintness would mostly be due to a distance effect (though they are all still much fainter than Type II GRB afterglows).

\begin{figure}[!t]
\epsfig{file=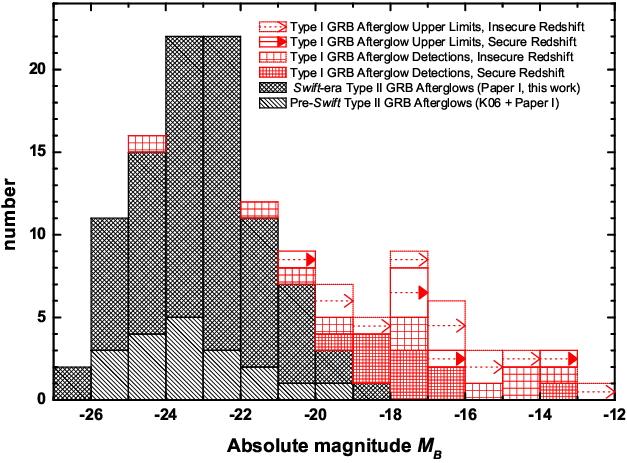,width=0.47\textwidth}
\caption[]{Absolute $B$-band magnitudes of Type II and Type I GRB afterglows or upper limits thereon. They are measured at one day after the burst in the observer frame after shifting the afterglows to $z=1$. For the Type I GRB sample, we distinguish between detected afterglows with redshifts we consider secure, detected afterglows with insecure or estimated redshifts, upper limits with secure redshifts (in some cases these have detected afterglows at earlier times) and finally upper limits with insecure or estimated redshifts. The Type I GRB afterglows with detections and secure redshifts are almost 6 mag fainter than the Type II GRB afterglow Golden Sample (mean mag $\overline{\MB}=-17.34\pm0.50$ versus $\overline{\MB}=-23.17\pm0.21$, respectively; we use both pre-\emph{Swift} and \emph{Swift}-era Golden Sample GRBs, K06, Paper I, this work).}
\label{Histo}
\end{figure}

A histogram of the absolute magnitudes $M_B$ (at one day after the burst assuming $z=1$) is shown in Figure \ref{Histo}. The luminosity distribution of \emph{Swift}-era Type II GRB optical afterglows are very similar to the sample of K06 (Paper I). For the complete Type II GRB afterglow Golden Sample (for which the extinction corrections are well-defined), we find $\overline{\MB}=-23.17\pm0.21$ (FWHM 1.71 mag) if we also include the Type II GRBs presented in Appendix \kref{AppA} of this paper. If we use the entire Type II GRB sample (i.e., add the Silver and Bronze samples of Paper I), we find $\overline{\MB}=-23.14\pm0.17$ (FWHM 1.61 mag), which is essentially an identical value. In comparison to this value, we find the following mean absolute $B$ magnitudes for different Type I GRB afterglow samples: 
$\overline{\MB}=-17.34\pm0.50$ mag (FWHM 1.65 mag) for the sample with detections at one day in the $z=1$ frame and secure redshifts,
$\overline{\MB}=-17.33\pm1.15$ mag (FWHM 3.45 mag) for the sample with detections and insecure redshifts, with GRB 060121 lying at $z=4.6$,
$\overline{\MB}=-17.04\pm0.96$ mag (FWHM 2.87 mag) for the sample with detections and insecure redshifts, with GRB 060121 lying at $z=1.7$, and
$\overline{\MB}=-16.45\pm0.85$ mag (FWHM 2.40 mag) for the sample with detections and insecure redshifts, without GRB 060121. Due to its very strong outlier nature and possible Type II classification, we will not include it in the following considerations. We note that in the sample with detections, there are five GRBs with assumed redshifts (as well as several where the association with a nearby galaxy is not strongly significant, e.g., GRB 061201, GRB 070809, GRB 090515). But four of these, GRBs 051227, 060313, 070707, and 080503 are assumed to lie at $z=1$ (only GRB 091109B is assumed to lie at $z=0.5$). Almost all other Type I GRBs with redshifts are closer than this, so it is more likely that the true redshifts of these four GRBs will be $z<1$ than $z>1$, making their absolute magnitudes even fainter.
For the upper limits, the resulting mean absolute $B$ magnitude is also an upper limit (it basically assumes all afterglows lie just at the detection threshold). The FWHM of the luminosity distributions are just given for completeness, as they convey little information here. We find
$\overline{\MB}>-17.28\pm0.72$ mag (FWHM 1.90 mag) for the upper limits with secure redshifts (note that in the case of GRB 070429B, GRB 070724A and GRB 100117A, afterglows are detected at earlier times, but only an upper limit can be given at one day as the decay slopes are not constrained), and
$\overline{\MB}>-16.66\pm0.68$ mag (FWHM 2.16 mag) for the upper limits with insecure or estimated redshifts. Finally, if we join the samples of detections and upper limits, we find 
$\overline{\MB}=-16.97\pm0.46$ mag (FWHM 1.99 mag) for all detections without GRB 060121, and
$\overline{\MB}>-16.92\pm0.49$ mag (FWHM 2.02 mag) for all upper limits.
It is evident that the cases without secure redshifts are fainter (though not significantly) than those with secure redshifts. A possible explanation is an observational bias, more luminous afterglows will have smaller XRT error circles, a higher chance at having a detected optical afterglow (a yet again smaller positional uncertainty than an X-ray-only error circle), and the higher luminosity may be related to a higher circumburst medium density (see \kref{OFFSET}) at a smaller offset. All these factors combine to make it easier to identify the (very probably) correct host galaxy and make the association more secure. Alternately, it might indicate that the redshifts we use or assume for our insecure redshift sample are incorrect and generally too low.

The Type I GRB afterglows which are detected at one day at $z=1$ and have secure redshifts are found to be $5.8\pm0.5$ mag fainter in the mean compared to the Type II GRB Golden Sample, thus, a factor of $\approx210^{+130}_{-80}$ less luminous. The factors are $\approx470^{+580}_{-260}$ for the Type I GRB afterglow detections with insecure redshifts, and $\approx290^{+170}_{-110}$ for the complete detection sample (without GRB 060121 in each case). Similar values are also derived for a comparison between the upper limits (both with secure as well as insecure redshifts) and the Type II GRB afterglow sample. Concerning the comparison between samples of detected afterglows (Type II) and upper limits (part of the Type I sample), a note of caution. As is extensively discussed in Paper I, the Type II GRB sample is afflicted by several sample selection biases. The need for a measured redshift as well as, in most cases, good multi-color light curves to determine the SED and thus the dust-extinction correction biases the sample against both very dusty as well as intrinsically subluminous afterglows which could skew the luminosity distribution toward lower values (more in accordance with the Type I GRB sample). As such, using upper limits on Type I GRB afterglows in comparison to Type II GRB afterglows is statistically imbalanced as long as the Type II sample does not also include all possible upper limits. As has been discussed in Paper I, though, an inclusion of such limits is not feasible within the bounds of these works, as both the redshift insecurities as well as probably the dust extinction insecurities are much greater than for the Type I GRB sample, and no significant information could be gleaned by trying to transform the Type II GRB afterglow upper limits to $z=1$. Furthermore, in recent years, technological advances in GRB follow-up have allowed us to lessen the bias of the sample presented in Paper I compared to when the paper was first submitted, and especially compared to the pre-\emph{Swift} sample of K06. Large rapid-response telescopes/detectors such as the P60 \citep[e.g.,][]{CenkoDark} and GROND \citep[e.g.,][]{GreinerDARK} are obtaining more and more high-quality data on faint/reddened events, of which many have entered our sample (e.g., GRB 070802, GRB 080913), lessening the impact of Type II GRB non-detections. These cases (also see the extremely extinguished afterglow of GRB 080607, \citealt{Perley080607}, this paper) show that generally, after correcting for extinction, the intrinsic luminosity of these afterglows is comparable to the mean of the (observationally brighter) Type II GRB sample (or even far exceeds it in the case of the hyperluminous GRB 080607), intrinsically subluminous events like GRB 060805A \citep{PerleyHosts2009} are rare. For the Type I GRBs, while there is some evidence for a few cases with high reddening (see \kref{LCs}), reddening is negligible in most cases where it could be determined, and there is no evidence so far for the redshift distribution to extend much beyond $z\approx1$, and finally the more often detected X-ray afterglows indicate many Type I GRB afterglows are very subluminous. Therefore, the additional deep upper limits on Type I GRB afterglows add further evidence to the strong luminosity bimodality.

Further support for the significance of the luminosity bimodality comes if we examine if the samples could be drawn from a single luminosity distribution via the a Kolmogorov–Smirnov (K-S) test \citep{Press1992}. Comparing the two most secure samples, namely the Type II GRB Golden Sample and the Type I GRB sample with detections and secure redshifts, we find that $P=1.34\,\times\,10^{-8}$, which is strong evidence that the two samples are inconsistent with being drawn from the same distribution. A comparison between the Type II GRB Golden Sample and the Type I GRB afterglows with detections but insecure redshifts ($P=1.23\,\times\,10^{-6}$), those with upper limits and secure redshifts ($P<1.21\,\times\,10^{-5}$) and finally the upper limits with insecure/estimated redshifts ($P<6.81\,\times\,10^{-8}$) all also support the finding of two separate luminosity distributions.
Also, we can compare our Type I GRB samples with secure redshifts with those with insecure/estimated redshifts. For the detections, we find $P=0.38$, and for the upper limits, we find $P=0.58$. This implies in both cases that the two compared samples have been drawn from the same distribution, which is at least an indication that our choices for the redshifts for the insecure/unknown cases were in the right range. Comparing the detections with secure redshifts to the upper limits, also with secure redshifts, results in $P=1$. The most likely interpretation of this result is that of a detector threshold. While it could be possible that the upper limits are much fainter and thus $p$ would be much smaller, the telescopes available at this time are just not capable of taking such deep observations, conversely, many Type I GRB afterglow detections have already pushed modern detector technology to its limits, as clearly seen in cases such as GRB 080503 \citep{Perley080503} and GRB 090515 \citep{Rowlinson090515}. We consider it less likely that all non-detected GRBs had afterglow luminosities that lay just beneath the detection limits in all cases.
Finally, taking all Type II GRB afterglow absolute magnitudes (94 data points) and all Type I GRB afterglow absolute magnitudes (35 data points, not including GRB 060121), including the upper limits, we find $P=3.63\,\times\,10^{-20}$ (it is $P=1.1\,\times\,10^{-18}$ with GRB 060121 included, still an extremely low value).
As we do not expect the basic fundamental principles of afterglow emission to be different for Type II and Type I GRB afterglows (i.e., both are external forward shock emission from a relativistic fireball, \citealt{NakarReview, Nakar2007, Nysewander2009}), the reason for this bimodality must lie elsewhere, as will be discussed below.

\subsection{Constraints on SN 1998bw Light in Type I GRB Afterglows}
\label{SNcon}
The appearance of classical SN light, both photometrically and spectroscopically,  in a GRB afterglow is the main observational evidence for the origin of the burst being a collapsing massive star. Its non-detection within about the first 2 weeks down to deep luminosity limits is therefore usually considered as a strong argument in favor of the identification of the burst under consideration as a Type I GRB, especially if one considers Type I GRBs as those that do not originate from the deaths of massive stars. In the last years, theoretical work has indicated that other explosion channels of single stars that do not produce bright SNe may be realized, namely  stars that collapse more or less directly to a black hole \citep{Fryer2006, Fryer2007, Moriya2010}. This has been suggested as an explanation for the ``SN-less long GRBs'' GRB 060505 and GRB 060614 \citep{FynboNature, GalyamNature, DellavalleNature, Thoene060505, McBreen060505} and has been predicted based on theoretical grounds even before the detection of these two events \citep{Woosley1993, Arnett1996, Fryer2006, Fryer2007, Nomoto2007, Tominaga2007}. On the other hand, so far there is no strong observational evidence that these explosion mechanisms are actually realized in nature, whereas so far all Type II GRBs at sufficiently low redshift \citep[$z\lesssim0.7$,][]{Zeh2004} that were not highly extinguished or had very bright host galaxies showed photometric (or even spectroscopic) SN signatures. We therefore still deem this to be a criterion for discerning between the two types, and note again that it is just one of the multiple selection criteria.

\begin{figure}[!t]
\epsfig{file=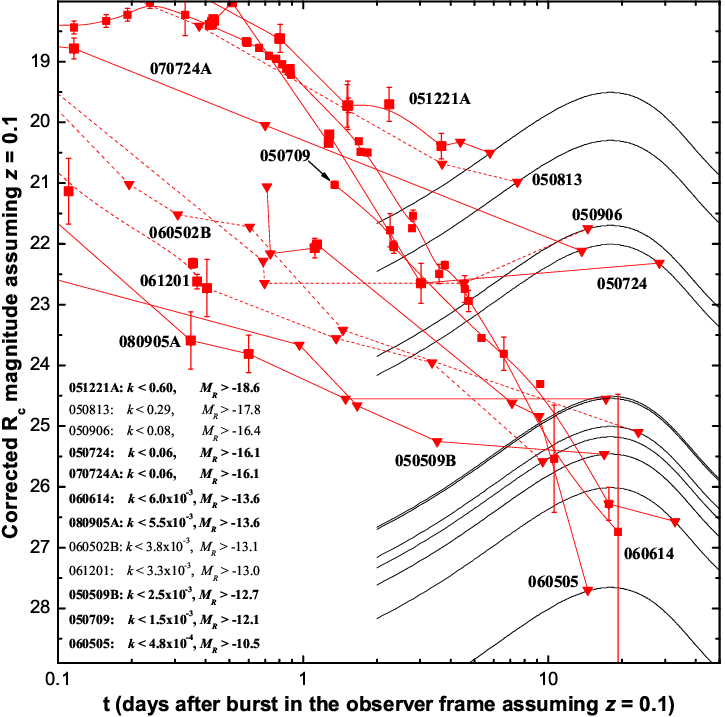,width=0.47\textwidth}
\caption[]{Deep late detections or upper limits of Type I GRB afterglows, all shifted to $z=0.1$, and compared with the $R$-band light curve of SN 1998bw at $z=0.1$. Here, we conservatively assume that the late detections derive from SN light only and there is no more afterglow contribution. For GRBs 051221A and 050813, the limits on an accompanying SN are not very strong, but all other Type I GRBs in this figure, including the temporally long events GRB 060505 and GRB 060614, give extremely stringent limits on any accompanying SN emission. The faintest upper limits are less luminous than any confirmed SNe at peak. In four cases (GRBs 050813, 050906, 060502B and 061201), the redshifts are insecure and thus the results are less significant, we have marked these cases with dashed lines and the results are not bolded.}
\label{SNfig}
\end{figure}

The results from our analysis of SN limits, including GRB 060505 and GRB 060614, are shown in Figure \ref{SNfig} and given in Table \ref{tabTypeI}. The limits for GRB 051221A and GRB 050813 are not very strict, as both GRBs lie at a redshift ($z=0.5-0.7$, note that the redshift of GRB 050813 is not secure) where it also becomes challenging to detect the SN signature in Type II GRB afterglows \citep{Zeh2004}. Furthermore, in both cases, observations were not extended to a time when a hypothetical accompanying SN would have probably peaked (assuming a similar rise time as SN 1998bw). The limits for GRBs 060614, 080905A, 050509B, 050709, and 060505 are much stricter, and fainter than any Type II SN known \citep[not to mention broad-lined Type Ic SNe;][and references therein]{Ferrero060218}\footnote{We point out that it has been claimed by \cite{PastorelloM85} that the ``luminous red nova'' M85 OT2006-01 \citep{KulkarniM85, RauM85, OfekM85} may actually be an extremely subluminous Type IIp SN with $M_R=-12.1$ during the plateau. Such an event would indeed not be detectable in almost all late Type I GRB afterglow light curves, except for GRB 060505, but here, see \kref{Hybrid}.}. The limits for GRBs 050724 and 070724A are intermediate between the two extremes, fainter than broad-lined Type Ic SNe, but still comparable to fainter Type II SNe. Three further GRBs with insecure redshift are found among the intermediately deep (GRB 050906) and very deep (GRB 060502B, GRB 061201) limits, though these results must be seen with caution. Our limits are in accordance with those found by other authors for GRBs 050509B \citep{Hjorth050509B, BersierGCN050509B}, 050709 \citep{Hjorth050709, Fox050709}, 050724 \citep{Malesani050724}, 050906 \citep{Levan050906}, 050813 \citep{Ferrero050813}, 051221A \citep{Soderberg051221A}, 060505 \citep{Ofek060505, FynboNature}, 060614 \citep{GalyamNature, FynboNature, DellavalleNature}, and 080905A \citep{Rowlinson080905A}. The limits for GRBs 060502B and 061201 stated here are derived for the first time in this paper, and our limits for GRB 070724A are much shallower than what \cite{Kocevski070724A} derive, as we correct for the high extinction along the line of sight \citep{Berger070724A}. Additionally, \cite{D'AvanzoThree} report a limit of $M_B=-15.1$ for any classical SN light following GRB 071227. Of the cases with secure redshifts and significant limits (therefore excluding GRB 051221A), three are short GRBs according to the old temporally-based classification scheme (GRB 050509B, GRB 070724A and GRB 080905A, note that GRB 070724A is more of a ``short/soft GRB'' than a ``short/hard GRB'', see Appendix \kref{AppA}), and three have IPC+ESEC-shaped light curves (GRB 050709, GRB 050724, and GRB 071227), which are generally accepted as a special subclass of non-collapsar events under the old classification also. All of these GRBs are also classified as Type I GRBs in our work, even independently of the deep upper limits on any SN contribution.
The missing bright late-time SN signal of Type I GRBs is thus a substantial phenomenological difference compared to the late-time evolution of Type II GRBs \citep[see also][]{Hjorth050509B, Fox050709}. On the other hand, even the very strict limit, $M_R\gtrsim-10.5$, on a SN accompanying GRB 060505 \citep[][this work]{Ofek060505}, which yields M($^{56}$Ni) $\lesssim  1\,\times\,10^{-4}$ $M_\odot$, cannot exclude the theoretical model of a collapsar with a very low jet-energy deposition \citep{Nomoto2007, Tominaga2007}. Furthermore, the less-constraining upper limits cannot exclude SNe similar to the faintest local core-collapse events \citep[see][but note that these are not the stripped-envelope SNe one would expect to be associated with the GRB phenomenon]{Richardson2002, Pastorello2004}. Still, there must exist a broad gap in peak luminosity between these faint SNe (if they exist at all, as we have pointed out) and the traditional SNe associated with Type II GRBs, which, independently of other criteria, would be a hint that these events derive from a different progenitor than Type II GRBs. Therefore, while it cannot be conclusively stated that the lack of SN emission down to very deep limits is ``smoking-gun'' evidence for a non-collapsar event, it gives additional support to such a classification in combination with other criteria, which has been the approach of the \cite{Zhang080913} classification scheme that we have employed.

It must be stressed that only in two cases (both with secure redshift) detections of the optical transient at the time of the suspected SN maximum at $t>10$ days have been reported in the literature (for GRB 050709 and GRB 060614; even though, after host subtraction, with a large error bar for the latter), but no late-time follow-up observations weeks after the suspected SN peak have been published so far. This leaves open the question if this positive detection was the late afterglow light or in fact an underlying faint SN component, even though the error bar is small enough for GRB 050709 only to tackle this question seriously. In all other cases only upper limits are available at the suspected SN maximum around $(1+z)\times 15...20$ days after the corresponding burst, if at all.

Clearly, the upper limits we can set will be misleading if the light curve evolution of any kind of SN following a Type I GRB differs substantially from the one of GRB-SNe of Type II GRBs, i.e., with respect to peak time and stretch factor. This brings us to the mini-SN model.

\subsection{Constraints on the Mini-SN/Macronova Model}
\label{Minicon}

\subsubsection{Power-law Decay}
We first consider the power-law decay model, which was considered by LP98ab as the most likely case. LP98ab demonstrated that, depending on the chosen model parameters, a mini-SN can peak at much earlier times than a normal core-collapse SN. Therefore, given the currently available data base (Figure \ref{SNfig}), one has to distinguish between those Type I GRBs with and those without detected early afterglow light. For example, a relatively bright mini-SN peaking at, say, $R_C$=22 about 0.5 days after the burst could have escaped detection in the early light curve of the afterglow of GRB 051221A. Faint mini-SNe could also be underlying the afterglow light curves of GRB 050709, GRB 060614 and GRB 080905A without being recognized because of lack of spectral information.

\begin{figure}[!t]
\epsfig{file=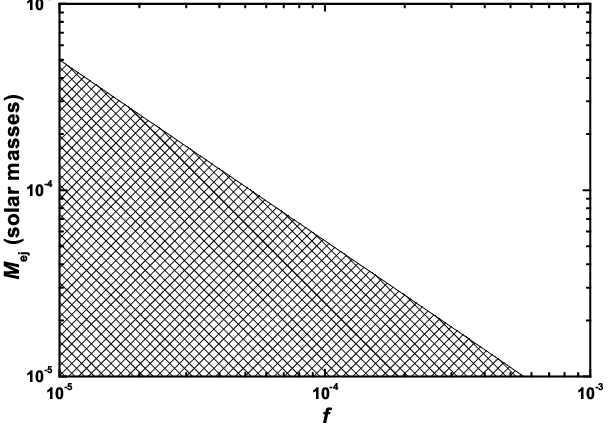,width=0.47\textwidth}
\caption[]
{Constraints on the parameter space $(f, M_{\rm ej})$ of ejecta of the GRB 050509B assuming a heating source that decays according to a power-law. Shown here is the result obtained based on the observed upper limit of $R>23.7$ at 0.96 days after the burst (see Figure \ref{SNfig}). The shaded region is the one allowed.}
\label{LiPa}
\end{figure}

Much tighter constraints can be set for those Type I GRBs with deep upper limits. Figure \ref{Bigfig2DetSec} shows that up to about 2 weeks after the event, the observed upper limits on any optical afterglow following GRB 050509B are the deepest limits for any Type I GRB with a secure redshift obtained so far (the detections of GRB 080905A at around 0.5 d are deeper, but the early limits are less constraining). In addition, the very likely association of GRB 050509B with a giant elliptical galaxy \citep{Gehrels050509B} and the offset from this host makes it unlikely that the faintness of the afterglow/mini-SN was due to internal extinction in the host galaxy. For this burst, the strongest constraint comes from the observed upper limit around 0.96 days after the event ($R>23.7$; note that this magnitude limit includes a transformation to a redshift of $z$=0.1; Figure \ref{SNfig}). Figure~\ref{LiPa} shows the allowed parameter space ($f, M$) (see \kref{Mini} for the definitions of these quantities) for any mini-SN following GRB 050509B, assuming an expansion velocity of $\beta=v/c=1/3$ (following LP98ab) and a matter opacity identical to Thomson scattering. If, within this model, log $f$ lies somewhere between $-$3 and $-$5 then for this event the ejected mass must have been less than 0.001 $M_\odot$.

\subsubsection{Exponential-law Decay}
K05 discussed in detail the model of an explosion where the decay of free neutrons is the internal heating source. Moreover, by adding the contribution of the thermal pressure to the equation describing the expansion of the ejected envelope, K05 made the approach of LP98ab more general. Unfortunately, in such a case no analytical solution is known.

Following K05 we assumed that the energy source decays according to an exponential law with a half-life time of 10.4 min but we fixed $v/c=0.3$. The evolution of the neutron-rich ejecta is then photon-pressure dominated so that the analytic solutions given in LP98b (their Equation 20) can be used.\footnote{Note that this analytic solution is not included in the published Letter, LP98a, but only included in the on-line available preprint version, LP98b.} For the parameter range $(f, M_{\rm ej})$ investigated here the ejecta  is usually very hot ($>10^4$ K) at the time of the peak of the bolometric luminosity. The luminosity in the optical bands then peaks at a later time (see also K05, their Figures 5 and 6), usually around 1 day. Unfortunately, for small ejected masses the envelope becomes rapidly optically thin so that our approach to calculate the optical luminosity via the assumption of black body radiation (see \kref{AppC} of the Appendix concerning the method) becomes less secure in such cases.

\begin{figure}[!t]
\epsfig{file=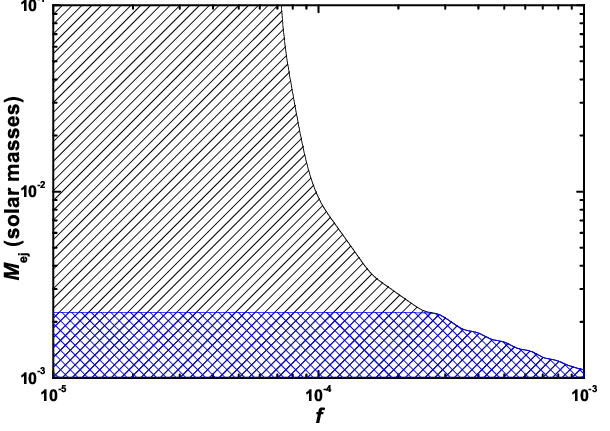,width=0.47\textwidth}
\caption[]{Same as Figure \ref{LiPa} but assuming that the ejecta is heated by neutron decay. The shaded region is the one allowed. The blue, crosshatched region stands for cases where the ejecta is optically thin, so that the results obtained are less secure.}
\label{macronova}
\end{figure}

Figure~\ref{macronova} shows the allowed parameter range $(f, M_{\rm ej})$ again based on the observed upper limit of $R>$23.7 at 0.96 days for any optical transient that followed GRB 050509B. For $f=3\,\times\,10^{-4}$, as it follows from the calculations given in K05 (their equation 43), the ejected mass cannot have been larger than about 0.002 $M_\odot$. This constraint on $M_{\rm ej}$ is still in the lower range of the results obtained by numerical studies of
neutron star mergers \citep{OJM2007}.

Recently, \cite{Kocevski070724A} also published an analysis on the mini-SN model, using deep upper limits on the afterglow of GRB 070724A which set limits on different parameter combinations than in our case. We caution though that the discovery of the very reddened afterglow of this event \citep{Berger070724A}, for which we find $\AV\approx0.9$ mag (Appendix \kref{AppA}), implies that the upper limits presented by \cite{Kocevski070724A} are much less constraining than presented by those authors. Also, the peculiar late-rising optical transient associated with GRB 080503 was initially suspected to be a mini-SN, but this was later ruled out by contemporary X-ray observations \citep{Perley080503}.

\section{DISCUSSION}
\label{Disc}
\subsection{The Afterglow Luminosities of Type I GRB versus Type II GRB Afterglows}

Several years before the first detection of a Type I GRB afterglow, \cite{Panaitescu2001} predicted that the discovery and follow-up of Type I GRB afterglows would be a big observational challenge. Based on the observational fact that typical Type I GRBs show a fluence more than an order of magnitude smaller than typical Type II GRBs, they predicted that the afterglows should be 10 to 40 times fainter, with radio afterglows hardly detectable and X-ray afterglows giving the best chance for detection. Furthermore, a low density external medium, as might be expected from merger progenitor models \citep[][but see \citealt{Nysewander2009}]{Nakar2007}, would further complicate the chances for follow-up, as would less collimated jets. At the beginning of the \emph{Swift} era, \cite{Fan2005} presented further calculations prediciting the magnitudes of Type I GRB afterglows, also coming to the conclusion that they would be significantly fainter than Type II GRB afterglows if merging compact objects were the progenitor (they also discussed two models which produced short-duration GRBs from the collapse of massive stars, these can be ruled out in most cases nowadays). Basically, their predictions have been observationally confirmed. We have shown, however, that the factor is around 210 (130 to 340), not only 10 to 40. One reason for this discrepancy is that many \emph{Swift}-detected Type I GRBs have up to orders of magnitude less isotropic energy release than the $5\times10^{51}$ erg \citet[][\citealt{Fan2005} also assumed a similar value]{Panaitescu2001} used in their modeling (of the GRBs with secure redshifts, only two, namely GRB 070714B and GRB 090510, exceed this energy, with GRB 051221A coming close, Table \ref{tabTypeISample}). The additional detrimental effects of low density external media \citep[e.g.,][]{Panaitescu2006, Nakar2007} and large jet opening angles \citep[e.g.,][]{Grupe050724} have also been shown to play crucial roles\footnote{We caution that the jet opening angle only plays an important role in a comparative sense if a standard jetted energy reservoir is assumed \citep{Frail2001, BFK2003}. With the discovery of both intrinsically subluminous \citep[e.g., XRF 060218,][]{Amati060614, Cobb060218, Liang2007a, Pian060218, Soderberg060218} and ``superluminous'' GRBs \citep[e.g., GRB 050820A, GRB 050904, GRB 070125, GRB 080319B, GRB 090902B and GRB 090926A,][]{Cenko050820A, Tagliaferri2005, Frail050904, Chandra070125, Updike070125, Cenko060418, McBreenLAT, CenkoLAT, Rau090926A}, the idea of a standard energy reservoir for Type II GRBs is untenable \citep[see also][]{Kocevski061028, Liang2008}. Furthermore, there is as yet only little evidence of a standard energy reservoir for Type I GRBs \citep[e.g.,][]{Soderberg051221A}, nor that the opening angles of Type I GRBs are significantly larger as an ensemble in comparison to those of Type II GRBs \citep{GaoDai}.}, these values were also overestimated in the assumptions of \cite{Fan2005}. Even very energetic Type I GRBs at redshifts comparable to typical Type II GRBs, such as GRB 060313 \citep{Roming060313}, and the aforementioned GRB 070714B  and GRB 090510, have optical afterglows that are comparable to faint Type II GRB afterglows only. The predictions of \cite{Panaitescu2001} concerning radio and X-ray afterglows have also proven to be correct, as only two Type I GRBs have been detected in the radio \citep{Berger050724, Soderberg051221A}, whereas most of those which \emph{Swift} was able to slew to immediately have X-ray afterglows \citep[e.g.,][]{NakarReview, Nysewander2009}.

To access the reason of the faintness of Type I GRB optical afterglows, we use the standard external shock model \citep{MeszarosRees1997, SPN1998}. For merger-like events, the circumburst medium is expected to have a constant density. With typical parameters, the optical band should satisfy $\nu_m < \nu_{opt} < \nu_c$, where $\nu_m$ and $\nu_c$ are the minimum injection synchrotron frequency and cooling frequency of relativistic electrons,
respectively. The optical afterglow flux density in this regime is \citep{Panaitescu2001}
\begin{equation}
F_\nu \propto \epsilon_B^{(p+1)/4} \epsilon_e^{p-1} E_{K,iso}^{(p+3)/4} n^{1/2} f_p D_L^{-2}
\label{Fnu}
\end{equation}
where $f_p \propto [(p-2)/(p-1)]^{(p-1)}$ \citep{Zhang2007}. Other notations follow the convention of the standard afterglow model: $E_{K,iso}$ is the isotropic kinetic energy of the blastwave, $n$ is the circumburst medium density, $\epsilon_e$ and $\epsilon_B$ are the fractions of the shock internal energy carried by electrons and magnetic fields, respectively, $p$ is the spectral index of the relativistic electrons, and $D_L$ is the luminosity distance of the burst. The fainter afterglows of Type I GRBs are due to the combination of a lower fluence and a lower energy density as expected for the merger scenarios \citep{Panaitescu2001, Nysewander2009}. The derivation of \cite{Panaitescu2001} was based on two assumptions: Type I GRBs have similar radiative efficiency as Type II GRBs, and $E_{K,iso}$ of Type I GRBs is on average 20 times smaller than that of Type II GRBs. With the recent observations of Type I GRBs, it is clear that the first assumption holds, i.e., for a sample of Type I GRBs studied, the radiative efficiency is not very different from that of Type II GRBs \citep{Zhang2007, NakarReview, Berger2007, Nysewander2009}.  However, the second assumption, which was based on the fact that Type I GRBs have a $\sim 20$ times smaller fluence than Type II GRBs and the implicit assumption that both populations have a similar mean redshift, is no longer justified.
Leaving out the $E_{K,iso}/D_L^{-2}$ factor in Equation (\ref{Fnu}) which accounts for the fluence factor discussed by \cite{Panaitescu2001}, there is an additional $\propto E_{K,iso}^{(p-1)/4}$ dependence. This accounts for another factor of $100^{0.3} \sim 4$ reduction of Type I GRB flux (assuming a typical value of $p \sim 2.2$) with respect to the estimate of \cite{Panaitescu2001}. This is in agreement with the results presented in this paper. In some cases, an even lower density $n$ (to be consistent with the intergalactic medium outside the host galaxy, as expected to happen for some Type I GRBs with large kick velocities) is needed to account for the faintness of the afterglows \citep{NakarReview}. We caution here, though, that \cite{Nysewander2009} have derived results which can be interpreted as that Type I GRBs and Type II GRBs occur in similarly dense environments \citep[but see][for critiscism of the methodology of \citealt{Nysewander2009}]{Norris2011}.

The larger spread of $F_\nu$ for Type I GRBs than Type II GRBs is less straightforwardly interpreted. Both types of GRBs should follow the same parameter dependences as shown in Equation (\ref{Fnu})\footnote{In principle, some Type II GRBs may have a stellar wind medium \citep{ChevalierLi1999}. Analyses of \emph{Swift} GRB X-ray afterglows however suggest that most bursts are consistent with a constant-density medium \citep[][though note that \citealt{StarlingPaperII} come to the opposite conclusion using a sample of pre-\emph{Swift} GRBs]{Zhang2007, Liang2007b, Schulze2010}.}. One has to argue that the scatter of the parameters is larger for Type I GRBs than Type II GRBs. One factor of $F_\nu$ scatter is due to that of $E_{\rm K,iso}$ (with a dependence of $\propto E_{\rm K,iso}^{(p+3)/4}$).  If it actually is a high-$z$ type I GRB (which we consider doubtful, see \kref{Hybrid}), GRB 060121 is an example that has a much larger $E_{\rm K,iso}$ than its low-$z$ brethren, though it may be highly collimated \citep{Levan060121, Ugarte060121}. Even if we discount it, GRB 070714B and GRB 090510, as has been pointed out above, are much more energetic than typical Type I GRBs. A second factor that causes the larger scatter of $F_\nu$ for Type I GRBs is the circumburst medium $n$ (with a dependence $\propto n^{1/2}$).  Since merger events can happen in all types of galaxies and either inside and outside the hosts, as suggested by the data, the ambient density could have a large scatter \citep[but see][]{Nysewander2009}. While mergers inside star-forming galaxies may have a medium density comparable to that of Type II GRBs, those events outside the hosts (due to large kicks received during the births of one or two neutron stars in the system) could have a tenuous medium, which tends to give rise to a ``naked'' burst \citep[e.g.,][]{Laparola051210, Perley080503}. Another possibility that leads to a low density circumburst medium and a large offset without the need for high kick velocities are mergers in globular clusters \citep{GPZM2006, Salvaterra2007, Guetta2009, Salvaterra2010, Berger2010, Lee2010, Church2011}, though at least some Type I GRBs show high X-ray column densities and thus cannot reside in globular clusters \citep{D'AvanzoThree}. A more speculative possibility is the scatter of shock parameters. While for Type II GRBs $\epsilon_B$ may be mainly determined by the post-shock instabilities that generate the in-situ fields \citep{MedvedevLoeb1999}, the existence of a pulsar wind bubble \citep[for reviews, see e.g., ][]{GaenslerPWB, SlanePWB, BucciantiniPWB} before the merger events would introduce a background magnetic field which would be compressed by the shock to power synchrotron emission \citep[for GRBs and pulsar wind nebulae, see][]{Konigl2002, Guetta2004}. This extra complication may introduce a larger scatter of $\epsilon_B$ and hence $F_\nu$ (with a dependence $\propto \epsilon_B^{(p+1)/4}$). But the existence of such a bubble can be ruled out for all but the youngest merging systems, though we caution that they may make up a significant fraction of the population \citep[e.g.,][but see \citealt{LeiblerBerger}]{Belczynski2006, Belczynski2007, TrojaSN}.

\subsection{SN Light in Type I GRB Afterglows}

Classical SN 1998bw light with its peak around $(15\; \tn{to}\; 20) \times (1+z)$ days after a burst is still the clearest signature of a Type II event. Type I events, on the other hand, should not have such a pronounced SN signal and in fact none has ever been found so far. The deepest limit obtained so far for a potential SN 1998bw component that followed a burst is for GRB 060505 with $M_R \gtrsim -10.5$.

As it has already been shown by LP98ab and K05, getting very deep upper limits between around some hrs and 1 day after a short burst might provide the strongest constraints on extra light related to the ejecta from merging compact objects. For example, for certain models discussed here the current data set, given by the upper limits on an optical transient following GRB 050509B, already excludes a mass ejection of more than about 0.001 solar masses. Unfortunately, since the peak luminosity is proportional to $f\, M_{\rm ej}^{1/2}$  (LP98a, their Equation 13) getting deeper flux limits basically  means setting tighter constraints on $f$ and not so much on $M_{\rm ej}$.

Given the very small signal strength we potentially expect from a mini-SN or macronova, observing in white light, i.e., unfiltered, is perhaps the very best strategy at first. Since short bursts related to elliptical host galaxies are the best candidates for having faint afterglows\footnote{\cite{Niino2008} and \cite{Zemp2009} have simulated Type I GRBs in galaxy clusters. They find that under the assumption that the dark matter subhalos of the cluster galaxies have been stripped, and Type I GRB progenitors exhibit expected kick velocities, a large number of intracluster Type I GRBs would be expected. While the typical ICM density is lower than the ISM density of a galaxy, it is high enough to allow for reasonably bright afterglows, and thus localization and clear association with the cluster. Type I GRBs from field galaxies, on the other hand, if they are ejected and thus explode surrounded by the rarified IGM, are not expected to produce afterglows at all, allowing no association with any galaxies (BAT detection only). Thus far, of course, observations show that associations of Type I GRBs with galaxy clusters seem to be rare \citep{Gehrels050509B, BergerClusterSGRB}, and most hosts are typical field galaxies \citep{Berger2009}.}, any additional signal might be best detectable or constrained in these cases.

Additional strong constraints on decay-driven light in Type I GRB optical transients would come from spectroscopy or broadband optical/NIR photometry. This would allow the detection of a component in the transient light that deviates from the expected afterglow synchrotron spectrum (which may be additionally affected by dust), or allow the placement of upper limits thereon. Acquiring such spectroscopy will be challenging, though. Thus far, only a few Type I GRBs have had their afterglows detected within the first minutes after the GRB (GRBs 060313, 060614, 061201, 070429B, 070714B and 090510), and in all cases, the early evolution was flat and the afterglow was around or fainter than 20th mag upon discovery. Only GRB 060614, for which the identification as a Type I GRB remains subject to debate, had an afterglow which was bright enough at observation time to allow 8m-class telescopes to obtain high S/N spectroscopy. This case illustrates a second quandary concerning the spectroscopy of Type I GRB afterglows, and the identification of absorption lines and thus unambiguous redshifts. The first spectrum of the afterglow of GRB 060614 contained no lines in emission or absorption \citep{FugazzaGCN060614}, and a redshift could only be determined after the afterglow had faded enough to allow host-galaxy emission lines to be detected \citep{PriceGCN060614}. Even so, the association with the galaxy is somewhat controversial \citep{Schaefer060614, Cobb060614} due to the quite large offset from its host galaxy in terms of half-light radius or brightest pixel distribution \citep{GalyamNature}, but this is based on \emph{a posteriori} statistics; \cite{GalyamNature} use HST imaging to determine a very low probability ($6\times10^{-6}$) of a chance association. Indeed, it is this offset, and the associated low-density medium surrounding the progenitor, which is probably the source of the low column densities of any absorption lines. This may turn out to be a serious problem for the determination of absorption line redshifts, as the typical observed magnitudes of Type I GRB afterglows preclude the use of echelle spectroscopy, which is a better tool to detect low-column-density absorption lines. Thus, we conjecture that the first successful (i.e., high continuum S/N) spectroscopy of a Type I GRB afterglow will probably not yield usable absorption lines that allow a redshift determination. Indeed, \cite{Stratta061201} reported that spectroscopy of the afterglow of GRB 061201 revealed neither absorption nor emission lines, but pointing constraints limited the exposure time to one hour, and the S/N is low. Similar failures have been reported for GRB 070707 \citep{Piranomonte070707} and GRB 060313 (Hjorth et al., in preparation). The only possible counterexample (for which the classification is still unclear, though) is the recent event GRB 100816A, for which very-low-column-density absorption lines were discovered at $z=0.8049$ \citep{TanvirGCN100816A, GorosabelGCN100816A}. No publications beyond the GCN Circulars have been published to this date, though.

\subsection{Energetics and Correlations}
\label{EaC}
Our unique sample of Type I and Type II GRB afterglow luminosities allows us to look for correlations between different parameters. By now, there is significant evidence \citep{Amati060614, Amati080916C, Piranomonte070707, OhnoWAM, KrimmBATWAM, Ghirlanda2009, Zhang080913} that Type I GRBs do not obey the relationship between the peak energy of the gamma-ray spectrum and the isotropic energy release \citep[``Amati relation'',][though possibly they lie on a parallel relation at an offset to that of the Type II GRBs]{AmatiRelation}, while it seems they do obey \citep{Ghirlanda2009, GhirlandaGBM} the relation between the peak energy and the isotropic peak luminosity \citep[``Yonetoku relation'',][]{YonetokuRelation}. To achieve a comparison with the energies of the GRBs, we compile the fluences and Band function (or cut-off power law) parameters for our Type I GRB sample (Table \ref{tabTypeISample}) and add these parameters for the Type II GRB sample (Paper I). In total, our sample encompasses 39 Type I GRB events (or 38, as GRB 060121 is included twice at different redshifts) and 98 Type II GRB events. Using the given spectral parameters and the redshifts, we derive $k$-corrections for the rest-frame bolometric bandpass of 1 to 10000 keV following the method of \cite{BloomBol}. Using the bolometric correction, the fluences and the luminosity distances, we then derive the bolometric isotropic energy $E_{iso,bol}$ for all GRBs.

\subsubsection{The Bolometric Isotropic Energy versus the Optical Luminosity}
\label{Bollum}

\begin{figure}[!t]
\epsfig{file=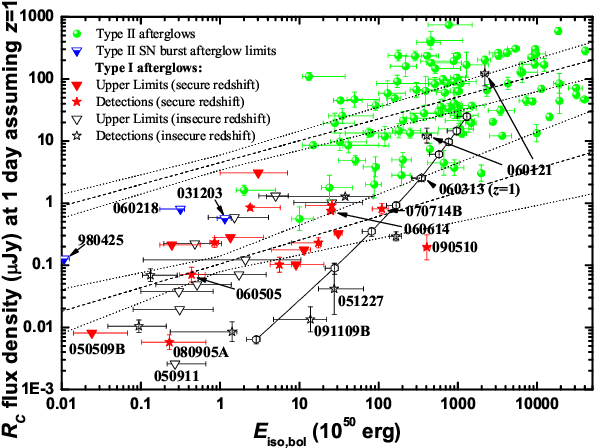,width=0.47\textwidth}
\caption[]{$R_C$ flux densities of the Type I and Type II GRB afterglows measured in the observer frame at 1 day after shifting the afterglows to $z=1$ (Table \ref{tabTypeI}, Paper I, K06) plotted against the isotropic energy of the GRBs (Tables \ref{tabTypeISample}, \ref{tabTypeIISample}, and Table 2 in Paper I). We differentiate between Type II GRB afterglows (green circles, Paper I and this work), Type II SN-GRB upper limits (blue half-filled triangles, Paper I, this work), Type I GRB afterglow detections (stars) and upper limits (triangles). Type I GRBs with redshifts we consider secure have filled red symbols, those with insecure redshifts have black open symbols. There is a positive correlation visible. Bursts with higher isotropic energy tend to have more luminous afterglows at a fixed time. We plot two Monte Carlo fits, the upper one to the Type II sample (Paper I) and the lower one to the Type I sample with detections and secure redshifts, as well as their $3\sigma$ confidence intervals. Both fits have very similar slopes but a different normalization, indicating different typical circumburst densities. The SN-GRBs have upper limits on their afterglow luminosity which are intermediate between the two fits. We also illustrate the effect of different redshifts (from $z=0.1$, bottom, to $z=2.0$, top; in steps of 0.2 or 0.3) for GRB 060313. See text for more details.}
\label{magEiso}
\end{figure}

The plot of bolometric isotropic energy release (Table \ref{tabTypeISample} as well as Table \ref{tabTypeIISample} for the additional Type II GRBs presented in this paper) versus the flux density of the afterglow at one day after the GRB assuming $z=1$ (converted from Table \ref{tabTypeI}) is shown in Figure \ref{magEiso}. This is an expansion of Figure 7 in Paper I. We differentiate between five data sets. All Type II GRB afterglows have detections and a secure redshift (while a few are photometric, their errors are small). In the case of Type I GRB afterglows, we differentiate between detected afterglows and upper limits, and between secure and insecure redshifts. Taking the complete data set, there is a positive correlation visible. GRBs with larger isotropic energy output tend to have brighter optical afterglows at a fixed late time after the GRB, when possible additional early emission processes like reverse-shock flashes contribute only negligibly. This correlation is very similar to the one found by \cite{Berger2007}, who compared the isotropic energies (without performing a $k$-correction) with the X-ray luminosities at a fixed datum, and it confirms the weak correlation already seen in the Type II GRB afterglows alone (Paper I) (significant at the $4.1\sigma$ level). We find a tighter correlation in this case. Using only the Type I GRB afterglow detections with secure redshifts, we find, using Kendall's rank correlation coefficient $\tau$, that $\tau=0.42$. For Spearman's rank correlation coefficient $\rho$, we find $\rho=0.64$. Due to the much lower number of data points, the significance is still very low, though, only $1.7\sigma$. We also derive maximally tight and maximally scattered data sets (see Paper I for more details), and find $\tau=0.47$, $\rho=0.76$, and $\tau=0.29$, $\rho=0.56$ for the maximally tight and maximally scattered data sets, respectively. The significances are not overly different this time, $1.9\sigma$, and $1.2\sigma$, respectively.

Two further teams have presented similar results to ours while this paper was in revision, \cite{Gehrels2008} and \cite{Nysewander2009}. Both added X-ray data to their studies. The sample of \cite{Gehrels2008} is \emph{Swift}-era only and smaller than our total sample, whereas \cite{Nysewander2009} use a Type I GRB sample very similar to ours, but a much larger Type II GRB sample which also includes upper limits. In comparison to our study, they neither perform extinction correction of the afterglows in the source frame (which would not be possible in many cases anyway), nor do they derive the bolometric isotropic energy release, especially the latter may influence their results.

At first glance, the Type I and Type II samples form a homogeneous sample, with the brightest and most powerful Type I GRBs (e.g., GRB 060614, GRB 070714B, GRB 060313 if at $z=1$; note that the Type II candidate GRB 060121 lies ``deep'' within the Type II GRB cloud) overlapping with the faintest Type II GRBs (e.g., XRF 060512, XRF 050416A, GRB 070419A, see Paper I). One exception is the most energetic (of those with secure redshifts) of all Type I GRBs, GRB 090510, which lies a whole order of magnitude under the faintest Type II GRBs of comparable energy. The Type II GRB afterglows have a very large scatter, e.g., the spread around $E_{iso,bol}\approx10^{53}$ erg is almost a factor of 200 (over 5 mag), and in most cases, the errors on the flux density of the optical afterglows are very small. Without several low-luminosity events such as those mentioned before, the significance is even lower (Paper I). This large scatter is probably due to several underlying causes, with the diverse jet opening angles probably having the strongest impact. A large spread of circumburst densities may also play a role \citep[see, e.g.,][]{Caito060614}. Similar to Paper I, we use a Monte Carlo method (30,000 runs each) to fit the Type I GRB afterglows with detections and secure redshifts while accounting for the two-dimensional asymmetric error bars. Analog to the fit presented in Paper I, we find the following correlation:

\begin{equation}
\frac{F_{\rm opt}\;({\rm at}\; t=1\; {\rm day})}{1 \mu\rm Jy} = \qquad\qquad\qquad\qquad\qquad\qquad\nonumber\\
\end{equation}
\begin{equation}
10^{(-0.978\pm0.041)}\times \left(\frac{E_{iso,bol}}{10^{50}\;erg}\right)^{(0.383\pm0.039)}.
\end{equation}

Adding the additional Type II GRBs presented in this paper to the sample of Paper I, we update their Equation 1 to:

\begin{equation}
\frac{F_{\rm opt}\;({\rm at}\; t=1\; {\rm day})}{1 \mu\rm Jy} = \qquad\qquad\qquad\qquad\qquad\qquad\nonumber\\
\end{equation}
\begin{equation}
10^{(0.653\pm0.040)}\times \left(\frac{E_{iso,bol}}{10^{50}\;erg}\right)^{(0.352\pm0.012)}.
\end{equation}

This shows that while the slope is similar (we find slopes of 0.35 and 0.38 for Type II and Type I afterglows, respectively, which is identical within error bars), the normalization is different. At $10^{50}$ erg, the difference in flux density is a factor $43^{+9}_{-7}$; and $40^{+14}_{-10}$ at $10^{51}$ erg, where the two data clouds overlap.

As discussed before, assuming the radiative efficiencies and blastwave physics to be similar for both central engine types (also, the jet opening angle distribution needs to be similar, it is as yet unclear if this is the case), this is an indication that the typical circumburst density around Type I GRB progenitors is lower than for collapsar-induced GRBs. As the normalization difference is $\propto n^{1/2}$, this implies that the typical ambient density around Type I GRB progenitors is roughly a factor of $\approx1700$ less, albeit with large error margins (in the range of 900 to 3000). This result is markedly in contrast to that of \cite{Nysewander2009}, who perform a very similar fit and find that the normalization for the Type I and Type II GRBs is extremely similar, the afterglow luminosities scale almost exclusively with the prompt energy release. If we assume all afterglows (Type II and Type I detection with secure redshift) to be one population, we derive $\tau=0.41$ (significance $6.0\sigma$), $\rho=0.58$, significantly higher than for the Type II GRBs alone. As pointed out above, \cite{Nysewander2009} do not use bolometric energy releases. Since Type I GRBs typically have harder spectra and higher peak energies \citep{BaratLestrade, Ghirlanda2009, Nava2010, GhirlandaGBM, GoldsteinBATSE, Guiriec090510}, their bolometric corrections would be higher, moving them further to the right in the plot in comparison to their position as derived by \cite{Nysewander2009}. On the other hand, a correction for line-of-sight extinction, which \cite{Nysewander2009} also do not perform, moves data points up in the plot, partly canceling the aforementioned effect (but typically, this correction will be more significant for the Type II GRB afterglow sample, moving it away from the Type I GRB afterglow sample). Furthermore, \cite{Nysewander2009} show that the optical-to-X-ray flux ratios also point to a similar circumburst density for both types of GRB. Though again, one must be cautious, as these will be influenced by a correction for extinction. Any extinction correction will increase the optical-to-X-ray flux ratio, and this effect may be stronger for Type II GRBs. On the other hand, most Type II GRBs from the sample of Paper I exhibit only low line-of-sight extinction, so the effect cannot be too strong. Finally, a spread of efficiencies is also possible \citep[for a detailed analysis, see][]{Zhang2007}, which may also induce the large scatter in the Type II sample. Note that since many Type I GRBs only have upper limits on the optical luminosity, the reduced scatter that seems to be visible in Figure \ref{magEiso} is probably not real (indeed, GRB 080905A already represents a strong outlier).

We also plot the effect of an unknown redshift. Here, we use GRB 060313 as an example, since it has a well-determined prompt emission spectrum and a well-observed afterglow too. We determine the isotropic bolometric energy release assuming the GRB actually lies at $z=0.1, 0.3, 0.5, 0.7, 1.0, 1.3, 1.5, 1.7,$ and 2.0, as well as the magnitude shift $dRc$ for these redshifts \citep[ignoring the fact that UVOT detections in all bands imply $z\lesssim1.3$,][]{Roming060313}, and then use the shifted light curve to determine the flux density at 1 day assuming $z=1$. The results are shown as data points connected by a spline. They rise more rapidly than the slope of the correlations, implying that an unknown redshift will have a significant effect on the scatter and on the fit results if one were to add these additional GRBs.

Finally, we also undertake a rough study into the optical luminosity of three local-universe Type II ``SN-GRBs'': GRB 9080425, GRB 031203 and XRF 060218 (see Paper I and references therein for more information on this special subclass). These events were subluminous energetically (Paper I and references therein), were accompanied by strong broad-lined Type Ic SN emission \citep[e.g.,][]{Galama1998, Malesani2004, Pian060218} but showed no sign of any ``normal'' forward-shock afterglow \citep[there may be a very faint component contributing to the early emission of GRB 031203,][]{Malesani2004}. Still, radio observations revealed that at least a small amount of their ejecta must have achieved relativistic speeds in contrast to typical (or even non-GRB-associated) Type Ic SNe \citep{Kulkarni980425, Soderberg031203, Soderberg2006, Soderberg060218}. To also study their optical luminosity compared to their isotropic energy release, we take the energetics given in Paper I. The optical data need special care.

GRB 980425 \citep[data taken from][]{Galama1998} is detected in $V$ and $R_C$ only at early times, and if we transform the light curves temporally to $z=1$, the earliest detections are at 1.4 days. The early data shows a flat behavior, though, so we will not create too large an uncertainty by back-extrapolating it with $\alpha=0$. We find a very blue spectral slope, $\beta\approx-0.3$, which is a strong indication that this is not classical forward-shock afterglow radiation (which typically has $\beta\approx0.5-1.1$). Hereby, we have assumed no host galaxy extinction. For GRB 031203, we take $I_C$ data from \cite{Cobb031203} and $K$ data from \cite{Malesani2004}. We follow \cite{Mazzali031203} and assume $E_{(B-V)}=1.07$ for the combination of Galactic and host galaxy extinction. We find a flat spectral slope, $\beta\approx-0.05$ (note that the fact that this is the reddest value of the three may indicate an afterglow contribution as \citealt{Malesani2004} proposed). The most secure optical data comes for XRF 060218, at the studied time, the event is at the peak of the shock-breakout \citep{Campana060218}, the temporal evolution is flat and UVOT detects it with high S/N in all six color filters, giving us a $UVW2\;UVM2\;UVW1\;UBV$ SED using the data from \cite{BrownUVOT}. We find an extremely blue spectral slope, $\beta\approx-1.2$, after correcting for the total $E_{(B-V)}=0.169$ \citep{Ferrero060218}.

As we find negative spectral slopes in all cases, this implies that the contribution by a forward-shock afterglow is negligible. All we can do is declare the magnitude of the optical transient at one day after the GRB in the $z=1$ frame (after also using the spectral slope to transform it into the $R_C$ band in the cases of GRB 031203 and XRF 060218) to be an upper limit on any classical afterglow contribution, and then use our shifting method (\kref{Shift}) to transform these upper limits to $z=1$. As all three events are very nearby, $z\lesssim0.1$, the $dRc$ values are large and positive. We find $dRc=+10.3$, $dRc=+5.0$, and $dRc=+6.7$ for GRB 980425, GRB 031203 and XRF 060218, respectively, with the errors being $\approx0.3...0.5$ mag. The three data points are included in Figure \ref{magEiso}. All three events emitted less energy isotropically than any Type II GRBs of Paper I, and are thus comparable to the Type I GRBs (GRB 980425 is even slightly less energetic than GRB 050509B). The upper limits on the afterglow contribution in all three cases are intermediate between the extrapolated trend of the Type II GRB afterglows (but still in agreement with the Type II GRB afterglow luminosity spread, and only GRB 980425 is definitely fainter than any detected Type II GRB) and the actual Type I GRB afterglow measurements. Since all three events are indubitably linked to massive star formation, a circumburst medium density typical of the high values of other Type II GRBs would be expected, so this is probably not the reason for the optical subluminosity. It may be due to the very small amount of energy which is actually contained in the relativistic part of the outflow. Of course, we caution that the true afterglow luminosities may lie way below our upper limits, within the Type I GRB cloud or even below.

\subsubsection{The Optical Luminosity versus the Host Galaxy Offset}
\label{OFFSET}

\begin{figure}[!t]
\epsfig{file=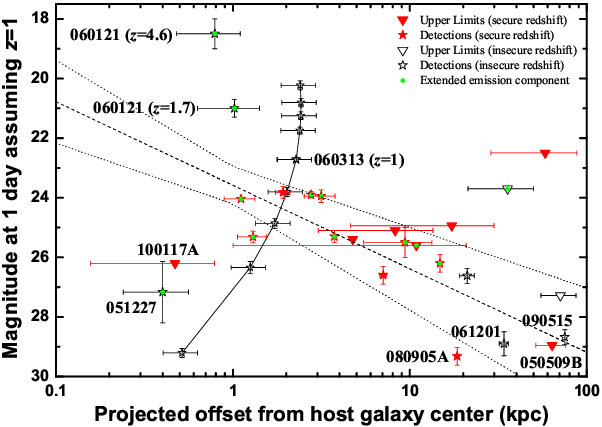,width=0.47\textwidth}
\caption[]{Magnitude of the Type I GRB afterglows measured in the observer frame at one day after shifting the afterglows to $z=1$ (Table \ref{tabTypeI}) plotted against the offset to the (assumed) host galaxy of the GRB (Table \ref{tabTypeISample}). The labeling is identical to Figure \ref{magEiso}. For the Type I GRB afterglow detections with secure redshifts, we find a correlation between the two quantities, which could be expected, since larger offsets typically imply lower circumburst densities and thus lower afterglow luminosities. The upper limits with secure redshifts are also in agreement with the correlation, with the exception of GRB 100117A. Further strong outliers (with uncertain redshifts) are GRB 051227 and the high redshift solution of GRB 060121. We illustrate the effect of different redshifts (from $z=0.1$, bottom, to $z=2.0$, top; in steps of 0.2 or 0.3) for GRB 060313. Clearly, an uncertain redshift has a strong effect on the scatter of the correlation. GRBs with green dots are those with an extended emission component. While most of these are found at small offsets, as claimed by \cite{Troja2007}, there are several at much larger offsets, and, conversely, several GRBs without an ESEC at small offsets.}
\label{magOffset}
\end{figure}

In the case of Type II GRBs, it has been shown that they occur almost exclusively at small offsets from their host galaxies \citep{Bloom2002}, and that their locations usually mark the brightest pixels in the host-light distribution, pointing to their origin in star-forming regions \citep{FruchterNature}. Right from the first Type I GRB localization, it was clear that this paradigm would not hold for this class of GRBs, as GRB 050509B was localized to the outskirts of its host galaxy \citep{Gehrels050509B}. While some Type I GRBs lie at small offsets, within their host light, which may point to low kick velocities or fast merger channels\footnote{\cite{Lee2010}, who study the dynamic creation of compact object binaries through two-body and three-body interactions in deep-core-collapse globular clusters, also suggest that Type I GRBs with very faint (or undetected) underlying host galaxies \citep[for examples with optical localizations, see][]{Berger2010} may have formed in nuclear-core-collapse dwarf galaxies.} \citep{Graham070714B, Piranomonte070707, D'AvanzoThree}, typically, the offsets have been found to be much larger than for Type II GRBs, indeed in agreement with predictions from the NS-NS merger models \citep{FongHST, Berger2010}. Furthermore, they trace their host light uniformly, indicating no preferred explosion environments. On the other hand, \cite{FongHST} caution that Type I GRB host galaxies are also larger, so the relative offsets of Type I and Type II GRBs are very similar, actually. \cite{Salvaterra2010} study the detectability of Type I GRB afterglows in different scenarios, from primordial binaries with high kick velocities to dynamically formed binaries in different types of globular clusters, including intra-cluster globular clusters (ICGCs). They find the afterglows in the latter cases should be detectable, as the gas density within such ICGCs is still appreciably higher than the inter-cluster medium (ICM). \cite{Berger2010} present several recent examples of Type I GRBs with optical afterglows which do not have any underlying host galaxies down to very deep limits (though note that almost none of the observations would have detected the host galaxy of GRB 070707, \citealt{Piranomonte070707}). They obtained spectroscopy of multiple galaxies in the surroundings and find that often, very faint galaxies (without redshifts) lie close to these GRBs, but these are statistically less likely to be the host galaxies than more distant, bright galaxies. Therefore, it is likely these GRBs exploded in globular clusters in the outer halos of nearby galaxies.

Figure \ref{magOffset} shows the afterglow magnitude of Type I GRBs (the same data as in Figure \ref{magEiso}) plotted against the offset from their host galaxy. Once again, we differentiate between detections and upper limits and secure and insecure redshifts. If we concentrate on the secure redshifts only, a clear correlation emerges, with larger offsets implying fainter magnitudes. Another Monte Carlo analysis, using the detections with secure redshifts only (additionally, note that, with the exception of GRB 100117A, the upper limits with secure redshifts all agree with the fit), finds in 30,000 runs:

\begin{equation}
m_{\rm R_C}\;({\rm at}\; t=1\; {\rm day})\; {\rm mag} = \qquad\qquad\qquad\qquad\qquad\qquad\nonumber\\
\end{equation}
\begin{equation}
(23.59\pm0.62)\; {\rm mag} + \left(\frac{\rm Offset}{kpc}\right)\times{(2.80\pm0.77)}.
\end{equation}

The high significance of the correlation is shown by non-parametric rank correlation tests, we find $\tau=0.64$, and $\rho=0.84$. Due to the low number of data points, the significance is still not very high ($2.7\sigma$). The correlation once again indicates the probable effect of the density of the circumburst medium on the kinetic-energy conversion efficiency and thus the afterglow magnitude. Two events are outliers of the correlation: GRB 080905A is once again (see \kref{Bollum}) an outlier, though only slightly. For the upper limits with secure redshifts, GRB 100117A is a strong outlier, a very faint afterglow at small offset from the core of its host galaxy. Possibly, this is related to the host being an elliptical galaxy \citep{Fong100117A}, though the one other unambiguous association between a Type I GRB and an elliptical galaxy, GRB 050724 \citep{Barthelmy050724, Berger050724, Gorosabel050724}, lies on the correlation. Here, though, the afterglow is measured during a strong X-ray/optical flare, and this emission may not be of forward-shock origin \citep{Malesani050724}, so a dependence on the external medium density need not be given. Anyway, no significant claims can be made with a sample of two (or three, if 050509B is counted) events.

For the cases with insecure redshift, the scatter is much larger, with GRB 051227 (faint afterglow centered on the host) and the high-redshift solution of GRB 060121 (extremely bright afterglow with a moderate host offset) being the strongest outliers. As we are only able to measure the offset in projection, this can have a strong effect, e.g., GRB 051227 may have occurred at a much larger offset but right in front of its host galaxy.

Again, we use GRB 060313 to analyze the effect of an unknown redshift. The derived track is roughly perpendicular to the correlation, implying a strong dependency on redshift. The track of GRB 060313 crosses the correlation at roughly $z\approx0.6$. Interestingly enough, the $z=1.7$ solution of GRB 060121 is quite close in both afterglow magnitude and host offset to GRB 060313 at a similar redshift, at first glance implying a similar track for GRB 060121 and a naive ``prediction'' of a redshift around $z\approx0.6$. Independent of the validity of the correlation as a rough redshift indicator, the track for GRB 060121 would be different, though, as the red afterglow would imply a strong extinction correction at such a low redshift \citep{Ugarte060121}, which would correct the afterglow magnitude up again (no extinction correction has been assumed for GRB 060313).

\cite{Troja2007} claim that ``all SHBs with extended-duration soft emission components lie very close to their hosts.'' and posit that this is an indication of two different progenitor classes of Type I GRBs, with the low-offset GRBs being NS-BH mergers \citep[as stated before, simulations show that the fraction of NS-BH mergers should be low,][whereas the result of \citealt{Troja2007} would imply a large fraction]{O'Shaughnessy2008, Belczynski2008}. \cite{FongHST} use their sample of afterglow offsets derived from HST observations to place doubt on this claim, finding no strong dichotomy \citep[see also][]{Cui2010}. Using our large sample of offsets (of which many have been taken from \citealt{FongHST}), we similarly find only marginal evidence for the claim of \cite{Troja2007}. In Figure \ref{magOffset} we have indicated the GRBs with extended faint emission with green dots \citep[note that in some cases, it is unclear if extended emission exists,][see \kref{AppA} of the Appendix as well as the footnotes of Table \ref{tabTypeISample}]{Norris2009}. Of those with secure redshifts, the largest offsets are for GRB 071227 \citep[though this event does still lie in the light of its host galaxy, an edge-on spiral galaxy,][]{D'AvanzoThree, FongHST} and GRB 061210 (though this offset has a large error bar due to it being an XRT-only detection). GRB 090510 lies at a marginally smaller offset, with a detection and a more precise position. GRB 051210 has an even larger offset, though its redshift is not secure. \cite{Perley080503} also point out that GRB 080503, the epitome of Type I GRBs with extended emission, lies at a large offset to any possible host galaxy detected in deep HST imaging. Note that GRB 081211B, which is not part of our sample, might be a very similar case \citep{PerleyGCN081211B}. Conversely, GRB 051221A has a relatively small offset and no extended emission. GRB 100117A is an even stronger case, note that while this plot gives only an upper limit, its afterglow is detected at an earlier time and well-localized.

Note that there are also other approaches which try to discern two different classes of Type I GRBs. \cite{Rhoads2008} finds a possible anti-correlation between the prompt isotropic energy release $E_{iso}$ and the mass of the host galaxy, interpreting this as the effect of two populations, with brighter GRBs resulting from younger populations. \cite{Church2011} recast this link to also be a correlation between energy release and offset, the latter being influenced by the host galaxy mass. \cite{Sakamoto2009} find two classes from their X-ray afterglow properties, one with short-lived, faint X-ray afterglows, the other with long-lived, bright X-ray afterglows which are similar to those of Type II GRBs. \cite{Norris2009}, analyzing the extended emission using a Bayesian Blocks method, find that 3/4 of all Type I GRBs have no ESEC, and that there is a real cutoff mechanism below a certain ratio of peak intensity to ESEC intensity. \cite{Norris2011} extend this research, confirming the results of \cite{Sakamoto2009} and finding further evidence for two classes of Type I progenitors. Further research into these possible dichotomies is beyond the scope of this paper. Finally, we point out that \cite{Kocevski070724A} looked for a correlation between afterglow luminosity and the star-formation rate of the GRB host galaxies, and found none.

\subsubsection{The Bolometric Isotropic Energy versus the Duration}

\begin{figure}[!t]
\epsfig{file=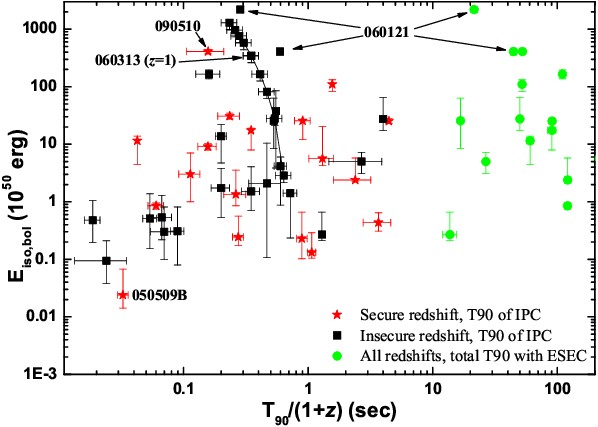,width=0.47\textwidth}
\caption[]{Bolometric isotropic energy of Type I GRBs plotted against the $\TNT$ of the Type I GRBs (Table \ref{tabTypeISample}). We differentiate between the duration of the Initial Pulse Complex (IPC) and the total duration in those cases where an Extended Soft Emission Complex (ESEC) exists (green circles). For the IPC $\TNT$, we further differentiate between those GRBs with a redshift we consider secure (red stars) and with an insecure redshift (black squares). While there is a weak correlation visible, where longer GRBs have higher isotropic energies, it is not statistically significant. Once more, we illustrate the effect of different redshifts (from $z=0.1$, bottom, to $z=2.0$, top; in steps of 0.2 or 0.3) for GRB 060313. Again, an uncertain redshift has a strong effect on the scatter of the correlation.}
\label{T90Eiso}
\end{figure}

\cite{Berger2007} researched a possible correlation between $\TNT$ and $E_{iso}$, and found tentative evidence for a correlation between the two parameters. With our larger sample, we repeat this analysis. We correct the $\TNT$ times for the redshift\footnote{We caution that, in lieu of a complicated analysis of the prompt emission, we simply derive $\TNT/(1+z)$. A more correct approach would need to involve the modeling of detector thresholds and a temporally resolved spectral analysis of the prompt emission to determine which parts would still be detectable at different redshifts. This is especially important for the ESEC component, which typically has both a very low peak flux as well as soft emission, and thus rapidly becomes undetectable with rising redshift. See \cite{Zhang080913} for a detailed analysis of two GRBs, GRB 080913 and GRB 090423, and more detail on why such an analysis especially with BAT data is complicated and often unfeasible. Considering we find no significant evidence for a correlation, a more detailed analysis is beyond the scope of this work.}, and, in contrast to \cite{Berger2007}, our isotropic energies are bolometric. In the case of GRBs which have an extended soft emission component (ESEC), we separate this total $\TNT$ from the duration of the initial pulse complex (IPC) only, which is shorter than 5 s in all cases. Figure \ref{T90Eiso} shows the plotted data. Disregarding the $\TNT$ values which include an ESEC, a weak correlation seems to be visible, both in the sample with secure redshifts only and in the complete sample, but the scatter is very large, and we caution that biases may be involved. Rank correlation tests also show that no significant correlation exists, we find $\tau=0.06$ (significance $0.33\sigma$), $\rho=0.26$ for the cases with secure redshifts, and $\tau=0.18$ (significance $1.6\sigma$), $\rho=0.32$ for the whole sample. GRB 090510 is a strong outlier, indicating an extremely high peak luminosity.

We once more use GRB 060313 to derive a redshift track. Again, this GRB is very suited for this analysis, as it was exceedingly bright and had the highest (lower limit) ratio of IPC to ESEC emission \citep{Roming060313}, therefore our naive $\TNT$ transformation with redshift is expected to be adequate. Similar to the effect of an unknown redshift on host galaxy offset (Figure \ref{magOffset}), the track is roughly perpendicular to the weak trend seen in Figure \ref{T90Eiso} and thus redshift uncertainty may strongly contribute to scatter. In this case, GRB 060313 agrees with the values of other GRBs only for low redshifts $z\lesssim0.5$. If a redshift $z\approx1$ is confirmed spectroscopically, it will be a strong outlier in this plot, indicating the lack of a true correlation, similar to GRB 090510, which it resembles.


\subsubsection{The Optical Luminosity as a Function of Redshift}

\begin{figure}[!t]
\epsfig{file=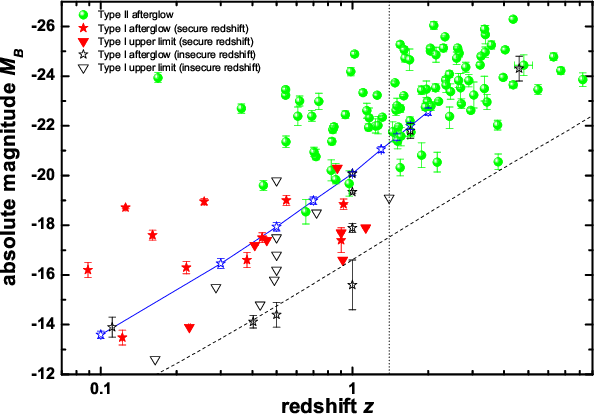,width=0.47\textwidth}
\caption[]{Absolute magnitude $\MB$ of the Type I (Table \ref{tabTypeI}) and Type II GRBs (Table 4 of Paper I) at one day assuming $z=1$ versus their redshift $z$. A ``zone of avoidance'' for faint afterglows at high redshifts is visible, indicating a bias, both due to the detectors (telescope) and selection criteria. This is supported by plotting (dashed line) a line of constant observer frame luminosity, which parallels the detection edge. Deep, dedicated observations with 8m class telescopes are able to find even fainter afterglows, though, such as those of GRB 090515, GRB 091109B and GRB 051227 (from left to right, just beneath the dashed line). We plot the redshift track of GRB 060313, in this case, an uncertain redshift has almost no influence on the position compared to the detection edge. The vertical dotted line lies at $z=1.4$ and denotes the separation between low-$z$ and high-$z$ GRBs (``type A'' and ``type B'', respectively, see K06). Clearly, with three exceptions (GRB 030329 at $z=0.17$, GRB 071010A at $z=0.99$, and GRB 991216 at $z=1.0$), the nearby afterglows are fainter than the more distant ones. The very faint afterglow at $z=3.8$ is GRB 050502A, which decayed rapidly \citep[][Paper I]{Yost050502A}.}
\label{MBz}
\end{figure}

In Figure \ref{MBz}, we plot the absolute magnitude $\MB$ of all Type I and Type II GRBs in our sample over the redshift of the GRBs. There is clearly a ``zone of avoidance'' in the lower right corner. If we plot the constant observer frame luminosity $C-5log(D_L)/log(10)$ (with the normalization constant $C=2.5$ in this case), shown as a dashed line, it becomes clear that this effect is due to the optical detector threshold, in this case the limiting magnitude that the telescopes used for observations can reach. This is similar to the detector threshold bias in high-energy observations \citep{ButlerBAT}. Another point, different from detector thresholds, is how much effort is (can be) actually invested into obtaining deep observations. GRB 050509B is a good example, being the first well-localized Type I GRB, it triggered an unprecedented observing campaign, yielding very deep early limits (Figure \ref{Bigfig1ULSec}). Sometimes even luck plays a role, for example, the extremely deep detection and upper limits of GRB 080503 at early times were mostly due to exceptional seeing during the observations \citep{Perley080503}, the same is true for GRB 090515 \citep{Rowlinson090515}. Another strong bias in the case of Type II GRB afterglows comes from the sample selection criteria, especially the need for a spectroscopic redshift, which favors afterglows that are bright in the observer frame (K06, Paper I). For Type I GRBs, this bias is reduced, as all redshifts have been derived from host galaxy spectroscopy, but here, the need for (at least) an X-ray afterglow detection to determine the host identification with sufficient significance yields a similar effect. Several outliers under this threshold are visible, GRB 090515 at $z=0.403$, GRB 091109B at $z=0.5$ (assumed) and GRB 051227 at $z=1$ (assumed). These afterglows were only discovered due to very deep and quite rapid observations with 8m-class telescopes \citep[][also note the ultra-deep limit on GRB 050911 we derive from VLT observations in this work]{Rowlinson090515, LevanGCN091109B, MalesaniGCN091109B, D'AvanzoThree, BergerHighzSGRB}. In these cases, the redshift assumption is almost irrelevant, as changing the redshift will move the data point more or less parallel to the threshold line. We illustrate this again with GRB 060313. Interestingly, the absolute magnitudes of two Type I GRB afterglows with uncertain redshifts, both of them bright high-fluence events, lie exactly on this line: GRB 061201 (which is quite similar to GRB 060313, App. \ref{AppA}) at $z=0.111$, and the $z=1.7$ solution of GRB 060121. GRB 070707 at $z=1$ also lies not far beneath it.


\subsection{``Hybrid Indicator'' GRBs in the Light of their Optical Afterglow Luminosities}
\label{Hybrid}

In this chapter, we will discuss three events that are in our sample which are contested. They have ``hybrid indicators'', with some of the population indicators, by themselves, pointing to a Type I (merger population) origin and some pointing to a Type II (collapsar population) origin. In the literature on these events, they have been described to be both Type I and Type II GRBs, depending on the work chosen. In \kref{dataI}, two of these events have been classified as Type I/Type I candidate GRBs, while the other has been classified as a Type II candidate. We can now add the optical afterglow luminosity at one day after the GRB assuming $z=1$ as a further criterion to help in classification, as we have shown that Type I GRB afterglows are typically a factor $100-300$ fainter than Type II GRB afterglows.

\subsubsection{GRB 060614}

GRB 060614 is the much-discussed example of a temporally very long GRB ($\TNT=102\pm5$ s) that nonetheless seems to belong to the Type I GRB population, having negligible spectral lag while being subluminous at the same time \citep{GehrelsNature, Mangano060614, Zhang080913}, a host galaxy with a small specific star-formation rate \citep{GalyamNature, DellavalleNature} which, while having a low mass for a typical Type I GRB host galaxy, also has a mean stellar population age much larger than those of Type II GRB hosts \citep{LeiblerBerger}; a large offset in terms of half-light radius and brightest pixel distribution \citep{GalyamNature} and a missing SN component down to $M_R\gtrsim-13.6$ \citep[][this work]{FynboNature, GalyamNature, DellavalleNature}. The medium the GRB jet propagates into is of constant density, but this finding carries no weight as most Type II GRBs also show constant density surroundings \citep[][and references therein]{Schulze2010}. The energetics alone are not a decisive factor, since the GRB lies in the transition zone between Type II and Type I GRBs (Figure \ref{magEiso}). The prompt-emission light curve has been shown to be an extreme IPC+ESEC form, similar to other Type I GRBs but at higher luminosity \citep{Zhang060614}. One difference to other Type I GRBs is that it does obey the Amati relation \citep{Amati060614}, although only in terms of the integrated spectrum, the IPC alone is strongly offset as is the case for other Type I GRBs \citep{AmatiCONF}, though the location versus the Amati relation alone is also not a strong indicator (\kref{GRB060505}). \cite{Lü2010} find the IPC lies in the Type I GRB region, though at the temporally long edge. Furthermore, it clearly does not follow other luminosity indicators \citep{Schaefer060614}. \cite{Perley080503} also state that the strong resemblance of light curve shapes between GRB 080503 and GRB 060614 is a further indicator that this is a Type I event.

We find that in accordance with the relatively high isotropic energy release, the afterglow luminosity at late times is also quite high -- for a Type I GRB. Even so, it does not become more luminous than the faintest Type II GRB afterglows. We thus do not contradict earlier interpretations, and while no absolute consensus can be reached, we consider that there are more indications that this GRB did not result from massive stellar death than evidence supporting such a SN-less demise. Still, the extreme light curve shows the need to develop merger models that are able to accommodate such long periods of sustained bright emission. Mergers involving white dwarfs may be a solution through the creation of long-lasting tidal tails \citep{King060614, Lee2010}.

\subsubsection{GRB 060505}
\label{GRB060505}
With the discovery of significant spectral lag for this event \citep{McBreen060505}, this event has become of crucial importance. In light of all data on this GRB, one of two indicators of progenitor affiliation that have been considered to be ``golden'' \citep{DonaghyHETE}, namely the lack of a SN signature to deep levels and negligible lag combined with low luminosity (essentially, being significantly in disagreement with the lag-luminosity correlation) must be incorrect in at least some cases. Either not all merger-population GRBs show negligible spectral lag, or not all collapsar-population GRBs show a SN component. \cite{McBreen060505} argue that GRB 060505 is ``likely a member of the long-duration class'' but base this argument on the non-negligible spectral lag alone (linked with the relatively long duration), while \cite{Zhang080913} show that the GRB does not lie on the lag-luminosity correlation, although its position does not agree with that of Type I GRBs in general (obviously, due to the non-negligible lag) and shares a space with several SN-associated low redshift events while at the same time being the ``most SN-less'' GRB ever. Therefore, spectral lag alone does not seem to be a significant distinguishing criterion anymore. The $\varepsilon$-criterion of \cite{Lü2010} also places this GRB into the ``cloud'' of Type I GRBs, though on the right edge due to the high $\TNT$, beneath GRBs 051227 and 060614.

In comparison to the extreme length of GRB 060614, the $\TNT=4.8$ s of GRB 060505 \citep{McBreen060505} is still marginally in agreement with a long tail of the Type I GRB distribution \citep{DonaghyHETE}, and precursors have also been discovered in some unambiguous Type I GRBs \citep{TrojaPrecursor}. The fact that the host environment, a low-metallicity super starcluster in a spiral galaxy, strongly resembles the typical blue starburst host galaxies of Type II GRBs \citep{Thoene060505, LeiblerBerger} is also not a definitive argument against this being a Type I event \citep{Ofek060505}, as by now the majority of Type I GRB host galaxies have been found to be actively star-forming \citep{BergerHighzSGRB}\footnote{We caution that the redshifts derived from host galaxy observations are strongly biased toward star-forming galaxies, as their emission lines are detectable at much higher significance than absorption lines in non-star-forming hosts \citep[see the case of GRB 051210 in][]{BergerHighzSGRB}. Furthermore, there are indications that offsets are larger in the case of massive elliptical hosts \citep[such as for GRB 050509B and possibly GRB 060502B,][]{Gehrels050509B, Bloom060502B}, making the association with these galaxies less secure \citep{Troja2007}.}. Also, the negligible offset from the star-forming region is not a conclusive argument for a Type II event, as \cite{Belczynski2006} show that compact-object mergers can occur within just a few million years after a starburst via a common-envelope phase channel \citep[see also][and \citealt{TrojaSN, Dokuchaev2011} for possible channels where a Type I GRB can occur within just days to a few years of the second SN in a binary]{Belczynski2010}. On the other hand, the fact that the GRB does not obey the Amati relation \citep{Amati060614, KrimmBATWAM} is also not a strong indication of this being a Type I event, as several clear collapsar events (GRB 980425, GRB 031203) are also not in accordance with the Amati relation. The one argument for this being a ``likely'' Type II GRB is the significant spectral lag, and the one argument for this being a Type I GRB is the deepest non-detection of a SN in a GRB-afterglow light curve ever.

From a theoretical standpoint, there is no compelling reason for Type I GRBs to have negligible spectral lag. \cite{Salmonson2000} and \cite{IokaNakamura2001} interpret the lag-luminosity correlation \citep{Norris2000} as a kinematic effect, dependent on the viewing angle from which we see the jet and on the Lorentz factor. One may now speculate that the jets of Type I GRBs have higher Lorentz factors, and thus smaller lags, as they propagate into a cleaner environment, since they do not have to penetrate a heavy stellar envelope and are thus less affected by baryon loading\footnote{\cite{Zhang080913} discuss several caveats to this assumption in their \S4.6. They find that the lag may be similar to the pulse width, and therefore carry no information about the progenitor, unless the stellar envelope of the progenitor influences the variability timescale and thus the pulse width. Also, the direct link between the Lorentz factor and the lag is based on assumptions that are not physically justified.}. A test of this hypothesis awaits the measurement of the Lorentz factors of Type I GRB jets, something that is non-trivial even for the much brighter afterglows of Type II GRBs \citep[e.g.,][]{Molinari060418}. Recent estimations of an extremely high Lorentz factor for GRB 090510 \citep[$\Gamma\gtrsim1000$,][but see \citealt{Hascoet2011}]{Abdo090510SUB, Ghirlanda090510, Ackermann090510} may point toward a verification of this hypothesis, but we caution GRB 090510 was anything but a typical Type I GRB \citep[and even its classification as a Type I event has been called into question,][]{Panaitescu090510}. Concerning a missing SN component, we have already pointed out that several authors have proposed the ``fallback black hole'' scenario which results in a GRB without a bright accompanying SN \citep{Fryer2006, Fryer2007, Moriya2010}. \cite{Nomoto2007} and \cite{Tominaga2007} show that GRB-producing relativistic jets can be launched with negligible Ni$^{56}$ production, leading to the absence of SN emission. But it seems that such events must be either rare or usually very subluminous, thus evading detection. Recently, \cite{Valenti2008ha} reported observations of SN 2008ha, an extremely underluminous SN which they classified as a Type Ic SN. One of their interpretations of their measurements was that this was an almost-failed SN in which a very massive star (i.e., a possible Type II GRB progenitor) collapsed to a fallback black hole, which would have been evidence that such events do exist (note that SN 2008ha was not connected to a GRB, and its luminosity of $M_R\approx-14.3$ at peak does exceed the limits found for GRB 060614 and GRB 060505). \cite{Moriya2010} have also shown with numerical modelling that fallback black holes can indeed produce very subluminous SNe such as SN 2008ha (though from less massive stars, and it is unclear if such a collapse can launch a jet). But \cite{Foley2008ha}, presenting more data, favored the partial deflagration of a white dwarf, a model which was strongly favored by additional evidence presented in \cite{Foley2008haL}. Therefore, so far there is no robust observational evidence for the existence of such failed SNe.

\cite{Xu060505060614} study both GRB 060614 and 060505 with broadband modelling, and come to the conclusion: ``Hence, from the properties of the afterglows there is nothing to suggest that these bursts should have another progenitor than other L[ong ]GRBs.'' We consider this misleading, as one would not expect the afterglow properties \cite{Xu060505060614} study, such as decay slopes and the optical-to-X-ray luminosity ratio, to be different in Type I and Type II GRBs \citep{NakarReview, Nakar2007, Nysewander2009}. Here we show that the afterglow luminosity, on the other hand, differs strongly from that of all Type II GRBs presented in Paper I.

Despite its lag and rather long duration, the flowchart of \cite{Zhang080913} classifies this as a Type I GRB candidate, and now we find that it shows an intrinsically extremely faint afterglow that is as much an outlier in comparison to the Type II GRB afterglows as GRB 060121 is an outlier compared to the Type I GRB afterglows (\kref{GRB060121}). If this truly is a Type II event, we are left with a uniquely subluminous GRB, one that is faint in the prompt emission, in the afterglow \emph{and} in the SN emission, the latter implying that only a small amount of energy is deposited in the subrelativistic ejecta too, in strong contrast to the other subluminous local universe events. Therefore, if the progenitor is of similar mass as a typical Type II GRB collapsar, most of the kinetic and rest-mass energy of the collapsing core must fall rapidly, without significant emission, through the event horizon of the central engine. The alternative possibility is that this is a merger-event, probably from a rapid channel \citep{Belczynski2006, Belczynski2010}, that for some reason does not show a typical sub-second spike of emission, but a more extended light curve with significant lag, and is otherwise typically subluminous in terms of prompt and afterglow emission.

\subsubsection{GRB 060121}
\label{GRB060121}
\cite{DonaghyHETE} present a detailed analysis of the prompt emission properties of this GRB. They find $\TNT=1.60\pm0.07$ s in the energy range $85-400$ keV, and $\TNT=1.97\pm0.06$ s in the energy range $30-400$ keV, at the ``borderline'' of the classic BATSE short GRB definition. Furthermore, the spectral lag is negligible, and the prompt light curve shows the IPC + ESEC shape. The fluence is among the highest in the Type I sample, but much smaller than bright Type II GRBs (Paper I). The observed afterglow is extremely faint and very red. The host galaxy offset is larger than for a typical Type II GRB \citep{FongHST}. Initially, therefore, this event was discussed as a short/hard GRB in the literature, albeit an extremely energetic example \citep{Levan060121, Ugarte060121, BergerHighzSGRB}. \cite{Zhang080913}, though, find that this event obeys the Amati relation, in contrast to other Type I GRBs, and that while the spectral lag is negligible, the event obeys the lag-luminosity relation due to its extremely high luminosity. Following their flowchart, it is found to be a Type II candidate, as we have shown (\kref{dataI}). Also, we find that applying the $\varepsilon$-criterion of \cite{Lü2010}, this GRB lies in the same region as the ultra-high-$z$ GRBs 080913 and 090423 (assuming $z=4.6$), and clearly above the Type I GRB region, in agreement with the result of the flowchart.

What makes this event extraordinary if it were a Type I GRB is the implied very high redshift \citep{Ugarte060121, Levan060121, BergerHighzSGRB}. If the GRB really lies at $z\sim4$, then the isotropic energy release is comparable to the more powerful Type II GRBs (Paper I), and the afterglow luminosity is typical for a Type II GRB too. Even if one assumes $z=1.7$ \citep{Ugarte060121}, the event is an outlier in comparison to the other Type I GRBs, and the additional problem of the high line-of-sight extinction that is needed \citep{Ugarte060121} emerges, which would be rather peculiar at the large offset (but see GRB 070809). A yet lower redshift (e.g., $z=0.5$ which we assume for some other GRBs in the sample) eases the energy problem, but the extinction has to be increased even more, and the inferred low luminosity of the host galaxy becomes an additional factor to consider. In any case, the afterglow light curve points to extreme collimation (\citealt{Ugarte060121} find $\theta_0=0.\!^\circ6$ for the $z=4.6$ case from broadband modelling), which is hardly achievable within the context of compact object mergers \citep[][but see \citealt{Rezzolla2011}]{Aloy2005}.

If this is a Type I GRB, it indicates that in rare cases the isotropic energy release is comparable to Type II GRBs (note that at least the $z=1.7$ solution is similar in terms of energetics to the bona fide Type I GRB 090510, though the latter is a truly extreme event, but here, see \citealt{Panaitescu090510}), and the afterglow luminosity is not an indicator of the progenitor population. If this is a Type II GRB, which seems more likely all in all, then the problem emerges of how to explain the extremely short prompt emission, $\sim0.3$ s at high energies\footnote{\cite{Zhang080913} emphasize that the duration of the prompt emission reflects the duration of the relativistic jet, which can be shorter than the duration of the jet as well as the duration of the central engine activity.} in the rest frame assuming $z=4.6$, in the framework of the collapsar model. \cite{Zhang2003} show that, under special conditions, the jet breakout from the massive star can produce a bright short emission spike, which is then followed by the lower-luminosity long GRB \citep[see also][]{Lazzati2009}. This is exactly the IPC+ESEC light curve\footnote{Also note that the highest-redshift GRBs 080913, 090423 and 090429B showed X-ray flaring activity for several 100 s (in the observer frame) after the temporally short prompt spikes \citep{Greiner080913, Tanvir090423, Cucchiara090429B}, such activity might be detected as a low-level extended emission component in soft $\gamma$-rays.} seen for GRB 060121 (but also for events like GRB 050724 which are clearly not associated with massive stars). But these authors also note that the initial bright spike should dominate only in flux, not in fluence, as is the case for GRB 060121, where the extended emission does not contribute to $\TNT$ at all. A host galaxy redshift might help to solve the affiliation of this enigmatic event, but the extremely faint host \citep[$R_C\approx26.5$,][]{Levan060121, BergerHighzSGRB} may prevent such a measurement before the next generation of large optical telescopes. In any case, independent of which population it actually belongs to, this event probes the envelope of known progenitor models.

\section{SUMMARY \& CONCLUSIONS}
\label{End}

We have compiled a complete set of optical/NIR photometry of \emph{Swift}-era Type I GRB afterglows, both detections and upper limits, creating a total sample of 38 GRBs, considering events up to the end of 2010 January. Using the methods of \cite{ZKK} and K06, and assuming reasonable values in cases where parameters like the redshift and the spectral slope are unknown, we analyzed the light curves and derived the luminosity distribution of the optical afterglows of Type I GRBs. Furthermore, we collected data on the energetics, and other prompt emission parameters, of the GRBs, as well as host galaxy offsets. With this sample, we are able, for the first time, to compare the parameter spaces of Type I GRBs to those of Type II GRBs from the sample of Paper I (both pre-\emph{Swift} and \emph{Swift} GRBs), both in terms of optical luminosity as well as in terms of energetics. To summarize, we come to the following results.

\begin{itemize}
\item Observationally, the optical afterglows of Type I GRBs are typically fainter than those of Type II GRBs. Many Type I GRBs do not have any optical detections at all, and often these non-detections reach upper limits much deeper than the magnitudes of our (biased) Type II GRB afterglow sample at similar times. Type II GRBs not detected to similar depths are usually dark GRBs, which are very likely strongly extinguished or lie at very high redshifts.
\item The luminosity distribution of Type I GRB afterglows shows a similar scatter to that of Type II GRBs. The fact that many Type I GRBs have upper limits on their optical afterglows implies that the luminosity distribution is probably even broader than what we find. With few exceptions, the results for assumed redshifts agree with those for GRBs with redshifts we consider secure, implying that our assumptions were not too far off the mark. We find that, comparing the most secure samples, the afterglows of Type I GRBs are, in the mean, almost 6 mag fainter than those of Type II GRBs. This is further support that Type I and Type II GRBs have different progenitors, exploding in different environments.
\item We find no evidence for the existence of radioactive-decay-driven SN emission in the light curves of Type I GRB afterglows, confirming earlier studies, and adding several more examples of strong upper limits.
\item We research the parameter space of Li \& Paczy\'nski mini-SNe, driven both by the decay of radioactive elements as well as neutrons, and strongly rule out the brightest emission models. Again, no evidence is found for such additional emission.
\item Using our knowledge of the energetics and typical afterglow luminosities of Type I GRBs, we explore theoretically why their afterglows are even fainter than predicted by \cite{Panaitescu2001}. The main reasons seem to be that \cite{Panaitescu2001} overestimated the energy release, also, generally the circumburst density is lower than expected.
\item In Paper I, we found a tentative correlation, with a large degree of scatter, between the isotropic bolometric energy release and the afterglow luminosity at a fixed late time, which was also confirmed by other authors \citep{Gehrels2008, Nysewander2009}. The Type I GRB results extend this correlation to smaller energies and lower luminosities. We find a different normalization of the fit, which can be explained by a strong difference in the density of the external medium into which the jets propagate.
\item Another correlation that confirms expectations is found between the host-galaxy offset and the afterglow luminosity. If confirmed by more data, it may be used as a rough redshift indicator, though we caution projection effects can play an important role. We find only marginal evidence for the claim of \cite{Troja2007} that all GRBs with extended emission have small host galaxy offsets, in agreement with other recent studies.
\item A correlation between the duration and the isotropic energy release is not detected in a significant way.
\item We discuss three anomalous GRBs, which have been assumed to be Type I GRBs in the parts of the literature, and Type II GRBs in others, in the light of the results on their optical luminosities.
\begin{enumerate}
\item GRB 060614, notwithstanding its extreme duration, is in good agreement with the upper end of the Type I GRB distribution in terms of energetics and afterglow luminosity, and thus seems to represent an extreme case of an Extended Soft Emission Component. We caution though that in this case its optical luminosity is not useful as an additional distinguishing criterion (as it is at the same time still in agreement with the faint end of the Type II GRB distribution), though most of the observational evidence points toward it being a Type I GRB.
\item GRB 060505 remains a puzzling object. The measurement of a significant spectral lag by \cite{McBreen060505} would place it with the Type II GRBs (in agreement with the environment), whereas the total lack of a SN, the very low isotropic energy release and the highly underluminous optical afterglow luminosity we derive are more akin to Type I GRBs, in agreement with the results of the classification schemes we use. We pose the question if the existence of significant spectral lag truly is a surefire indication that a GRB is a Type II event, especially since even with this lag, GRB 060505 is not in agreement with the lag-luminosity relation.
\item GRB 060121 is found to resemble Type II GRBs much more than Type I GRBs, but would then have an extremely short prompt emission spike. It therefore is another example of the breakdown of the old short/hard vs. long/soft classification, just that it likely is a Type II GRB ``masquerading'' as a Type I GRB, and not the other way around as in the two other cases.
\end{enumerate}
\end{itemize}

One of the main results of Paper I was that the afterglows of pre-\emph{Swift} and \emph{Swift} Type II GRBs are, all in all, very similar to each other. Here, we clearly find that this is not the case for Type II GRB and Type I GRB afterglows, the latter are much less luminous. The number of detected afterglows as well as the density of the follow-up observations is still too small to derive if the light-curve evolution is also significantly different. More than a decade has passed since the afterglow era began, but for a long time, it was the Type II GRB afterglow era only. The Type I GRB afterglow era is only six years old, and this study can only be a first exploration into a research area that is just now becoming accessible to us, thanks to an ever-evolving follow-up technology which allows us to be there faster, more precisely and deeper all the time. And only a much larger sample will tell where exactly peculiar ``hybrid indicator'' GRBs are to be found in the ever-expanding zoo that is the GRB parameter space.

\acknowledgments
D.A.K. thanks C. Guidorzi for helpful comments and A. Zeh for the fitting scripts. S.K. thanks Shri Kulkarni for helpful discussions. D.A.K., S.K. and P.F. acknowledge financial support by DFG grant Kl 766/13-2 as well as DFG grant Kl 766/16-1. B.Z. acknowledges NASA NNG 05GC22G and NNG06GH62G for support. A.R. acknowledges support from the BLANCEFLOR Boncompagni-Ludovisi, n\'ee Bildt foundation. The research activity of J.G. is supported by Spanish research programs ESP2005-07714-C03-03 and AYA2004-01515. S.S. acknowledges support by a Grant of Excellence from the Icelandic Research Fund. Furthermore, we wish to thank Scott Barthelmy, NASA for the upkeep of the GCN Circulars, Jochen Greiner, Garching, for the ``GRB Big List'', Robert Quimby et al. for GRBlog, and Daniel A. Perley et al. for GRBOX.

\newpage\clearpage
\appendix
\label{App}

\section{Observations}
\label{Observations}
\begin{itemize}
\item{{\bf GRB 050911}: We obtained two epochs of imaging of this GRB (which has no X-ray afterglow and thus only a BAT localization) with the FORS1 instrument on the Very Large Telescope (VLT) on Cerro Paranal, Chile. Three images of 180 s duration each were obtained in $V$ and $R_C$ for each epoch at low airmass and under excellent seeing conditions (0\farcs6 and 0\farcs8 for the first and second epoch, respectively). Calibration was done against Landolt standard stars. Image subtraction using the ISIS package\footnote{http://www2.iap.fr/users/alard/package.html} reveals no variable sources. We created faint artificial sources after calibrating the images and tested to which magnitude the subtraction recovered them. Note that since the two epochs are near together in time, a slowly varying afterglow would not be discovered this way. On the other hand, the X-ray afterglow was extremely faint, therefore it would be expected for an optical afterglow to lie under our limits. These are the deepest limits obtained for this GRB, and, if the GRB is indeed associated with the cluster at redshift $z=0.1646$ \citep{BergerClusterSGRB}, these are among the deepest limits ever obtained for any GRB afterglow.}
\item{{\bf GRB 051105A}: We obtained two epochs of imaging in the $R_C$ filter with the DOLoRes instrument on the Telescopio Nazionale Galileo (TNG) on La Palma, Canary Islands, Spain, at 0.56 and 2.55 days after the GRB, with $9\times300$ s and $9\times180$ s integration time, respectively. Conditions (especially seeing) were much worse in the second epoch, this limits the magnitude we can reach by performing image subtraction. We also obtained two epochs of imaging in the $R_C$ filter with the 1.34m Schmidt telescope of the Th\"uringer Landessternwarte Tautenburg (TLS) 0.48 and 1.46 days after the GRB, with exposure times of $21\times300$ s and $15\times300$ s, respectively. Both stacks reach a comparable limit. Photometric calibration has been derived assuming mag $R_C=17.0$ for the star U1200\_08571196 at coordinates RA, Dec. = 17:41:14.05, +34:55:35.1.}
\item{{\bf GRB 051210}: We observed the field of GRB 051210 with the VLT and the FORS1 instrument in the $R_C$ filter, 1.81 days after the GRB, for an integration time of $20\times180$ s. The host galaxy candidate is clearly detected. The X-ray afterglow position is clearly offset from the galaxy, and we derive a deep upper limit on this position, the deepest for this GRB. We also observed the host galaxy candidate of GRB 051210 on 2005 December 29 UT starting at 01:05:52 for a single exposure of 3600 s with the FORS2 spectrograph on the VLT, using the 300I grism with a slit width of 1\farcs0. In the reduced two-dimensional spectrum, no trace of the galaxy was detected.}
\item{{\bf GRB 060121}: The field of GRB 060121 was observed with the TNG in the $R_C$ filter at two epochs with the DOLoRes and OIG instruments. The afterglow fades clearly during the $\sim44$ hr span of the observations. The magnitude has been calibrated against the USNO B1.0 -- namely the R1 magnitude -- and the SDSS catalogs. Since we had no observations in a second filter, the color term has been neglected. This led to the following zero-points: 1st night: DOLoRes = $26.40\pm0.15$, OIG = $25.23\pm0.20$;  2nd night: DOLoRes = $26.20\pm0.20$. The difference in the value for DOLoRes in the two nights indicate that at least one of the night was not photometric. However, the magnitudes have been obtained through direct comparison with field objects. The calibration with the SDSS has been performed by using 13 selected field objects. Two of them are quite red and define, in the calibration relation, a color term slope of 0.35 which is quite strong. This leads to a zero-point $r_{SDSS}-r_{instr} = 26.4$ and $r_{AG} \sim 23.05$ for the first point in the SDSS system. By applying a correction term $R_{Johnson}-r_{SDSS}$ estimated through the field calibrating objects, we obtain $R_{AG} \sim 22.9$ in the traditional Johnson/Cousins system, thus confirming the values reported in the table. These observations mark the discovery of the afterglow \citep{MalesaniGCN060121}, though even earlier detections were later reported \citep{Ugarte060121}.}
\item{{\bf GRB 060313}: We observed the field of GRB 060313 with the ANDICAM (A Novel Dual Imaging CAMera) mounted on the 1.3m Small and Moderate Aperture Research Telescope System (SMARTS) telescope at the Cerro Tololo Interamerican Observatory (CTIO) in Chile. The final afterglow position lay outside the smaller NIR FoV, so no $YJHK$ observations could be derived. The afterglow was undetected in $BV$, clearly detected in $R_C$ and faintly detected in $I_C$. Calibration was achieved against Landolt standards.}
\item{{\bf GRB 060502B}: We obtained a single 900 s $R_C$-band image with the TLS telescope at 0.115 days after the trigger, at high airmass, in inclement weather conditions, twilight and under the influence of moonlight. These combined factors allow us to derive a shallow upper limit only. The galaxy \cite{Bloom060502B} suggest as a host is clearly detected, though, and was initially suggested by us a possible host at large offset \citep{KannGCN060502B}.}
\item{{\bf GRB 060801}: We obtained two epochs of imaging with the VLT and the FORS2 instrument in the $R_C$ band at 0.51 and 1.48 days after the GRB, obtaining $8\times300$ s and $11\times300$ s exposure time, respectively. The images clearly show the two galaxies near the XRT position \citep{Castro-TiradoGCN060801}. Photometry is calibrated against Landolt standards. Image subtraction using the second epoch as a template does not reveal any sources which vary by more than 0.1 mag to a limit of $R_C>25.3$. This is the deepest limit obtained for the afterglow of this GRB.}
\item{{\bf GRB 070714B}: We obtained one epoch of NIR imaging in the $J$ ($22\times2\times30$ s exposure time) and $K$ ($25\times4\times15$ s exposure time) bands one day after the GRB at high airmass. The field was calibrated against seven 2MASS stars, yielding zero point errors of 0.09 and 0.10 mag, respectively. No source is detected at the GRB position in the $J$ band, but we find a very faint source (only significant at $2\sigma$) in the $K$ band. Within errors, its magnitude is in agreement with that of the host galaxy \citep{Graham070714B}.}

\end{itemize}

\begin{deluxetable}{llcll}
\tablecolumns{5}
\tablewidth{0pt}
\tablecaption{Afterglow Photometry}
\tablehead{
\colhead{GRB}   &
\colhead{Days after trigger (midtime)} &
\colhead{Magnitude}  &
\colhead{Filter}  &
\colhead{Telescope}}
\startdata
050911	&	0.518530	& $ >	25.70			$ &	$V$	&	VLT	\\
	&	0.734070	& $ >	25.70			$ &	$V$	&	VLT	\\
	&	0.527130	& $ >	25.80			$ &	$R_C$	&	VLT	\\
	&	0.742490	& $ >	25.80			$ &	$R_C$	&	VLT	\\ \hline
051105A	&	0.476711	& $ >	23.00			$ &	$R_C$	&	TLS	\\
	&	0.563724	& $ >	23.70			$ &	$R_C$	&	TNG	\\
	&	1.457994	& $	\cdots			$ &	$R_C$	&	TLS	\\
	&	2.549481	& $	\cdots			$ &	$R_C$	&	TNG	\\ \hline
051210	&	1.807239	& $ >	25.20			$ &	$R_C$	&	VLT	\\
	&	1.807239	& $	24.00	\pm	0.1	$\tablenotemark{a} &	$R_C$	&	VLT	\\ \hline
060121	&	0.104525	& $	22.93	\pm	0.15	$ &	$R_C$	&	TNG	\\
	&	0.314757	& $	23.33	\pm	0.25	$ &	$R_C$	&	TNG	\\
	&	1.988715	& $ >	24.3			$ &	$R_C$	&	TNG	\\ \hline
060313	&	0.026663	& $ >	20.50			$ &	$B$	&	SMARTS	\\
	&	0.026663	& $ >	20.30			$ &	$V$	&	SMARTS	\\
	&	0.026663	& $	20.42	\pm	0.09	$ &	$R_C$	&	SMARTS	\\
	&	0.026663	& $	20.31	\pm	0.26	$ &	$I_C$	&	SMARTS	\\ \hline
060502B	&	0.115172	& $ >	20.30			$ &	$R_C$	&	TLS	\\ \hline
060801	&	0.514372	& $	23.30	\pm	0.05	$\tablenotemark{b} &	$R_C$	&	VLT	\\
	&	0.514372	& $	24.45	\pm	0.13	$\tablenotemark{c} &	$R_C$	&	VLT	\\
	&	0.514372	& $ >	25.30			$ &	$R_C$	&	VLT	\\
	&	1.477958	& $	\cdots			$ &	$R_C$	&	VLT	\\ \hline
070714B	&	0.973776	& $ >	21.2			$ &	$J$	&	TNG	\\
	&	0.996403	& $	20.63	\pm	0.52	$ &	$K$	&	TNG	
\enddata
\tablecomments{
VLT is the 8.2m Very Large Telescope at the European Southern Observatory Paranal Observatory, Chile. TLS is the 1.34m Schmidt telescope of the Th\"uringer Landessternwarte Tautenburg (Thuringia State Observatory) in Germany. TNG is the 3.6m Telescope Nazionale Galileo, on the island of La Palma, the Canary Islands, Spain (using the DOLoRes and OIG detectors in the optical and NICS in the NIR). SMARTS is the 1.3m Small and Moderate Aperture Research Telescope System telescope at the Cerro Tololo Interamerican Observatory in Chile.
}
\tablenotetext{a}{Probable host galaxy. \cite{BergerHighzSGRB} find $R_C=23.85\pm0.15$ for this galaxy, in agreement with our result \citep[see also][]{LeiblerBerger}.}
\tablenotetext{b}{Probable host galaxy, source ``B''. \cite{BergerHighzSGRB} find $R_C=23.03\pm0.11$ for this galaxy \citep[see also][]{LeiblerBerger}, which is brighter than our result.}
\tablenotetext{c}{Source ``D''. \cite{BergerHighzSGRB} find $R_C=23.95\pm0.3$ for this galaxy, which is brighter than our result.}
\label{tabPHOT}
\end{deluxetable}

\newpage\clearpage

\section{DETAILS ON TYPE I GRBS}
\label{AppA}

\paragraph{GRB 050509B.}This was a very faint and exceedingly short single-spiked \emph{Swift}-localized GRB, with $\TNT=0.04\pm0.004$ s, to which the satellite slewed immediately. \emph{Swift} detected an extremely faint X-ray afterglow which was undetectable after the first orbit \citep{Gehrels050509B}.
The proximity of the X-ray afterglow to a bright cD elliptical galaxy, 2MASX J12361286+285858026 in the galaxy cluster NSC J123610+28590131, was quickly noted, and it was proposed that this GRB was a merger event occurring in the halo of this galaxy, which lies at $z=0.2248\pm0.0002$ \citep[e.g.,][]{Gehrels050509B, Bloom050509B, Castro-Tirado050509B}. Still, the error circle is quite large and contains many faint and probably more distant galaxies, so the association with the elliptical galaxy is not absolutely secure \citep{FongHST}. Follow-up observations found no sign of SN emission down to very deep limits, supporting the merger hypothesis \citep{Hjorth050509B, BersierGCN050509B}. Upper limits are taken from \cite{RykoffGCN050509B}, \cite{WozniakGCN050509B}, \cite{Bloom050509B}, \cite{CenkoGCN050509B}, \cite{Hjorth050509B}, and \cite{BersierGCN050509B}, where we presume $R$-band observations for the last source. These upper limits are among the most constraining ever taken on a GRB afterglow.

\paragraph{GRB 050709.}This GRB was localized by \emph{HETE-2} \citep{Villasenor050709} and was the very first Type I GRB for which an optical afterglow was discovered \citep{Hjorth050709, Fox050709, Covino050709}, it was also marginally detected in the NIR \citep{Fox050709}. The subarcsecond localization was possible through the late detection of the X-ray afterglow with \emph{Chandra}. The burst was localized to a star-forming dwarf galaxy at $z=0.1606\pm0.0001$ \citep{Fox050709, Covino050709, ProchaskaSGRB}, immediately emphasizing the possibility of Type I GRBs to also occur in galaxies with ongoing star formation, just like Type Ia SNe. The burst itself had $\TNT=0.07\pm0.01$ s at high energies, was multispiked but had a low peak energy, and was followed by a bump of soft extended emission that had $\TNT=130\pm7$ s \citep{Villasenor050709}. Details on our construction of the afterglow light curve and SED, and the evidence for dust along the line of sight, can be found in \cite{Ferrero050813}.

\paragraph{GRB 050724.}This \emph{Swift}-localized GRB featured more ``firsts''. It was similar to GRB 050709 in the sense that it consisted of a short, hard spike of 0.25 s duration, some further emission \citep[$\TNT=3.0\pm1.0$ s,][]{Barthelmy050724} and a bump of soft extended emission \citep[total $\TNT=152.4\pm9.2$ s,][]{Campana050724}. The X-ray afterglow was bright and featured, for the first time, two late X-ray flares \citep{Campana050724}. Follow-up observations with \emph{Chandra} showed that the afterglow was not strongly collimated \citep{Grupe050724}. GRB 050724 was the first Type I GRB with a clearly detected NIR afterglow and a radio afterglow \citep{Berger050724}, and it was the first time that a GRB was securely associated with a galaxy that had no contemporaneous star formation \citep{Barthelmy050724, Berger050724, Gorosabel050724}, residing in a lone elliptical galaxy at $z=0.2576\pm0.0004$ \citep{Berger050724, ProchaskaSGRB}. Afterglow data is taken from \cite{Berger050724}, and \cite{Malesani050724}. 

\paragraph{GRB 050813.}This was a faint event localized by \emph{Swift}, which slewed immediately to the burst \citep{RetterGCN050813}. The X-ray afterglow was extremely faint, similar to that of GRB 050509B \citep{MorrisGCN050813}. The burst had $\TNT=0.6\pm0.1$ s and another faint peak at 1.3 s, but no extended emission \citep{SatoGCN050813}. Over time, the X-ray error circle has been revised and improved several times \citep[see Figure 1 of][]{Ferrero050813}. Observations of the error circle revealed a galaxy cluster at a redshift of $z\approx0.72$ \citep{ProchaskaSGRB}, which is believed to be associated with the GRB. We note that there is also photometric evidence for a background cluster \citep{Berger050813}. We use upper limits from \cite{LiGCN050813}, \cite{BlustinGCN050813} and \cite{Ferrero050813}.

\paragraph{GRB 050906.}This was a very faint single-spiked GRB with $\TNT=0.128\pm0.016$ s localized by \emph{Swift}, which slewed immediately to the burst, but did not detect any X-ray afterglow, leaving only the BAT error circle \citep{ParsonsGCN050906}.
\cite{Levan050906} report deep observations of the error circle, which includes, at its edge, the massive star-forming spiral galaxy IC 328 ($z=0.031$). Furthermore, a galaxy cluster at $z=0.43$ lies in the error circle. One possibility is that GRB 050906 is a SGR hyperflare from IC 328, but the softness of the spectrum speaks against this (Hurley et al., in preparation). Otherwise, the burst may be associated with the galaxy cluster, in which case it would energetically resemble GRB 050509B. A third possibility is a field galaxy at unknown redshift. In this work, we assume that this is a normal Type I GRB associated with the $z=0.43$ cluster. Upper limits are taken from \cite{FoxGCN050906} and \cite{Levan050906}.

\paragraph{GRB 050911.}This was a faint double-peaked burst followed by probably softer emission at 10 s after the trigger, it is $\TNT=16\pm2$ s. \emph{Swift} was unable to slew to the burst immediately and no X-ray afterglow was detected to a deep limit in observations starting 4.6 hr after the trigger, indicating a faint and steeply decaying afterglow. While there were some doubts that this was a Type I GRB, the presence of soft emission bumps has come to be seen as a typical sign of this type of GRB, and furthermore, the spectral lag was negligible. We find it to be a Type I GRB candidate according to the classification criteria of \cite{Zhang080913}. Information on the \emph{Swift} observations can be found in \cite{Page050911}. The BAT error circle coincides with the galaxy cluster EDCC 493 at $z=0.1646$. \cite{BergerClusterSGRB} find the significance of an association with GRB 050911 is $3.4\sigma$, thus we use this redshift in this work. We take upper limits on an optical afterglow from \cite{TristamGCN050911}, \cite{BergerGCN050911}, \cite{BreeveldGCN050911}, \cite{Page050911} as well as our own VLT data.

\paragraph{GRB 051105A.}This was a very faint and extremely short GRB, $\TNT=0.028\pm0.004$ s \citep{CummingsGCN051105A}. Although \emph{Swift} immediately slewed to it, no X-ray afterglow was detected at all \citep{MineoGCN051105A}. There are several faint and constant X-ray sources, one of them associated with a galaxy \citep{KloseGCN051105A} which does not show optical variability and may be an X-ray selected quasar \citep{HalpernGCN051105A}. We assume $z=0.5$. Upper limits are taken from \cite{BrownGCN051105A}, \cite{HalpernGCN051105A}, \cite{SharapovGCN051105A} as well as our own TLS and TNG data.

\paragraph{GRB 051210.}This was a faint two-spiked GRB localized by \emph{Swift}, which slewed immediately to the burst. The burst had $\TNT=1.27\pm0.05$ s and showed soft extended emission \citep[][but see \citealt{Norris2009}]{Laparola051210}. Near the XRT error circle, a host galaxy was discovered \citep{BloomGCN051210, BergerHighzSGRB}, spectroscopy of this galaxy reveals neither emission nor absorption lines. \cite{BergerHighzSGRB} argue that this implies $z\geq1.4$, we adopt $z=1.4$. \cite{LeiblerBerger} find $z=1.3\pm0.3$ from photometric redshift modeling of the host galaxy, in full agreement with the spectroscopic limit. No optical afterglow was discovered, we take upper limits from \cite{JelinekGCN051210}, \cite{BergerGCN051210}, \cite{BloomGCN051210}, \cite{BlustinGCN051210}, \cite{BergerHighzSGRB} as well as our own VLT data.

\paragraph{GRB 051211A.}This event was localized by \emph{HETE-2}, it was a multi-peaked burst with $\TNT=4.02\pm1.82$ s at high energies. The initial hard spike was detected by FREGATE and WXM, and the localization was derived from SXC which independently detected a soft extended bump. While $\TNT$ is high for a classical Type I burst, spectral analysis shows zero lag. \emph{Swift} XRT follow-up observations started half a day after the GRB and revealed no X-ray afterglow, so the best position is the $80\arcsec$ SXC error circle. Within this error circle, \cite{GuidorziGCN051211A} detected a candidate afterglow, but this is probably a star \citep{HalpernGCN051211A}. We take upper limits from \cite{GuidorziGCN051211A}, \cite{KlotzGCN051211A}, \cite{JelinekGCN051211A}, and \cite{UgarteGCN051211A}. As no host galaxy or redshift is known, we assume $z=0.5$.

\paragraph{GRB 051221A.}This was an intense, multi-spiked burst localized by \emph{Swift}, which slewed immediately to it \citep{ParsonsGCN051221A}. It was also detected by \emph{Konus-Wind} \citep{GolenetskiiGCN051221A}, \emph{Suzaku-WAM} \citep{EndoGCN051221A}, \emph{INTEGRAL SPI ACS}\footnote{Light curves for triggered GRBs can be found at the following sites. \emph{INTEGRAL SPI ACS}: http://isdc.unige.ch/cgi-bin/cgiwrap/$\sim$beck/ibas/spiacs/ibas\_acs\_web.cgi?month=20XX-YY with X = 02, 03, etc. and YY = 01, 02, ... 12. \emph{RHESSI}: http://grb.web.psi.ch/grb\_list\_200X.html with X = 2, 3, etc. (\emph{RHESSI} GRBs are only posted until 2009). \emph{Suzaku WAM}: http://www.astro.isas.ac.jp/suzaku/HXD-WAM/WAM-GRB/grb/trig/grb\_table.html for triggered events and http://www.astro.isas.ac.jp/suzaku/HXD-WAM/WAM-GRB/grb/untrig/grb\_table.html for untriggered events.} and \emph{RHESSI}. The burst has $\TNT=1.4\pm0.2$ s and shows no signs of extended emission \citep{CummingsGCN051221A}. \cite{BloomGCN051221A} discovered the afterglow in the NIR, and it was soon confirmed in the optical. \cite{Soderberg051221A} present a multi-color host-corrected light curve, we use their data and that of \cite{WrenGCN051221A}; for more details, see \cite{Ferrero050813}. Spectroscopy shows that the GRB happened in a star-forming galaxy at $z=0.5464$ \citep{Soderberg051221A}. The \emph{Swift} and \emph{Chandra} X-ray observations reveal what is probably a jet break at late times (between 2 and 9 days after the GRB), at a time when the afterglow is not detected anymore in the optical. The collimation-corrected energy is similar to that of the earlier Type I GRBs, indicating the possibility of a standard energy reservoir and processes that lead to collimation similar to Type II GRBs \citep{Burrows051221A}.

\paragraph{GRB 051227.}This was a multi-peaked burst localized by \emph{Swift}, which slewed immediately and flight-localized a bright X-ray afterglow \citep{BarbierGCN051227}. It was initially thought to be a Type II event, since it has $\TNT=8.0\pm0.2$ s \citep{HullingerGCN051227}, but spectral analysis showed negligible spectral lag, and together with a bump of soft emission starting several tens of seconds after the GRB, this indicated that GRB 051227 was a Type I event \citep{BarthelmyGCN051227}. Within the XRT error circle, an exceedingly faint afterglow was discovered \citep{MalesaniGCN051227a, D'AvanzoThree}. A nearby galaxy \citep{BloomGCN051227} was first thought to be the host galaxy, and was found to lie at $z=0.714$ \citep{FoleyGCN051227}. But follow-up observations \citep{MalesaniGCN051227c, D'AvanzoThree, BergerHighzSGRB} reveal a faint source directly underlying the afterglow position which does not fade in subsequent imaging and is thus the host galaxy. This is the one of the faintest host galaxies found so far for a Type I GRB, and suggests that it lies at an even higher redshift than the $z=0.714$ galaxy \citep{BergerHighzSGRB}. We thus adopt $z=1$ for this GRB \cite[see also the discussion in][]{D'AvanzoThree}. Afterglow data is taken from \cite{HalpernGCN051227}, \cite{RomingGCN051227}, \cite{KodakaGCN051227} (all upper limits), \cite{D'AvanzoThree}, and \cite{BergerHighzSGRB}, and has been host-corrected (we find $R_C=25.44\pm0.08$ from a fit, this is corrected for Galactic extinction). We find a very steep decay of $\alpha=2.78\pm1.31$ (highly unsure as it is only determined by two data points which lie close together in time), which may be indicative of a highly collimated event observed after the jet break, in agreement with \cite{D'AvanzoThree}. These authors report that the optical decay is clearly steeper than the X-ray decay.

\paragraph{GRB 060121.}This GRB was localized by \emph{HETE-2}, it is a double-peaked burst with $\TNT=1.60\pm0.07$ s at high energies, which is followed by several hundred seconds of faint soft emission \citep{DonaghyHETE}. It was also detected by \emph{Konus-Wind} \citep{GolenetskiiGCN060121}, \emph{Suzaku WAM} and \emph{RHESSI}. \emph{Swift} performed follow-up observations starting 3 hr after the GRB, and following the discovery of a bright X-ray afterglow \citep{ManganoGCN060121}, the faint optical afterglow was discovered \citep{MalesaniGCN060121}, finally leading to the discovery of an extremely faint host galaxy \citep{Levan060121, BergerHighzSGRB}. Analysis of the optical SED gives a best redshift solution of $z=4.6$ \citep{Ugarte060121}, which is supported by the host galaxy colors \citep{BergerHighzSGRB}, an alternate redshift solution is $z=1.7$. There is evidence of an overdensity of Extremely Red Objects in the field, possibly a high redshift cluster \citep{Levan060121, BergerHighzSGRB}. At such a high redshift, the energetics, both of the prompt emission and the afterglow, are extreme in comparison to other Type I GRBs, being more typical for Type II GRBs \citep{Ugarte060121}. This event is discussed in more detail in \kref{Hybrid}. Details on the light curve construction are given in \cite{Ferrero050813}, we additionally add our two TNG data points.

\paragraph{GRB 060313.}This event has been called ``a new paradigm for short-hard bursts''. It had the highest fluence, the highest observed peak energy and the second-highest hardness ratio (after GRB 060801) of all classical Type I GRBs observed by \emph{Swift} \citep{Roming060313} until GRB 090510 occurred. It is an intense, multi-spiked burst, $\TNT=0.7\pm0.1$ s, without any extended emission, the lower limit to the ratio of prompt to extended emission is also the largest in the \emph{Swift} sample. It was also detected by \emph{Konus-Wind} and \emph{INTEGRAL SPI ACS}. The optical afterglow was discovered by the VLT \citep{LevanGCN060313} after the XRT position was reported \citep{PaganiGCN060313}, with the earliest detections coming from \emph{Swift} UVOT \citep{Roming060313} and the Danish 1.54m \citep{ThoneGCN060313}. It remains roughly constant for several hours (Hjorth et al., in preparation) before fading rapidly. An extremely faint host galaxy is found at the afterglow position \citep[][Hjorth et al., in preparation]{BergerHighzSGRB, FongHST}, making the afterglow of GRB 060313 also the brightest afterglow in relation to its host galaxy for a Type I GRB. Spectral analysis of the UVOT data gives $z\leq1.1$ at 90\% confidence \citep{Roming060313}, and the faintness of the host galaxy suggests a redshift near this limit. As with GRB 051227, we adopt $z=1$. Afterglow data is taken from \cite{ThoneGCN060313}, \cite{NysewanderGCN060313}, \cite{BergerHighzSGRB}, as well as our own SMARTS data. We subtract the (small) host contribution. We do not use the UVOT data \citep{Roming060313}. Comparison with the early UVOT $U$ and $R_C$ data from the sources mentioned before indicates a very flat spectral slope.

\paragraph{GRB 060502B.}This event was detected by \emph{Swift} as a two-spiked GRB with $\TNT=0.09\pm0.02$ s, the satellite slewed immediately to it \citep{TrojaGCN060502B, SatoGCN060502B}. No extended emission is detected, and observations did not reveal an optical afterglow. The revised XRT position \citep{TrojaXRT060502B} lies close to a bright, early-type galaxy \citep[though it shows clear spiral structure in a deep HST image,][]{FongHST} which was first suggested as a possible host by \cite{KannGCN060502B}, a position put forth in more detail by \cite{Bloom060502B}, who find a redshift of $z=0.287$ and note the similarity to GRB 050509B and GRB 050724. While we use this redshift in this work, we note that \cite{BergerHighzSGRB} propose a far fainter galaxy within the XRT error circle (for which no redshift is known) as an alternate possible host, implying a much higher redshift \citep[see also][]{FongHST}. \cite{Church2011} note that if this GRB is associated with the elliptical galaxy \cite{Bloom060502B} propose, then it very likely occured in a globular cluster and was not ejected by a natal kick from the main galaxy. We take upper limits for our analysis from \cite{TrojaGCN060502B}, \cite{PooleGCN060502B}, \cite{ZhaiGCN060502B}, \cite{PriceGCN060502B}, \cite{RumyantsevGCN060502B}, \cite{BergerHighzSGRB}, and \cite{Bloom060502B}.

\paragraph{GRB 060505.}This burst was detected by \emph{Swift} as it entered the SAA, and the image significance was too low to lead to a flight-generated position. Ground analysis derived a higher significance and the burst was reported as a weak event lasting $\TNT=4\pm1$ s \citep{PalmerGCN060505}. \emph{Suzaku WAM} \citep{KrimmBATWAM} and \emph{RHESSI} also detected this GRB. Follow-up observations revealed a faint X-ray afterglow \citep{ConciatoreGCN060505}, and optical observations of the error circle lead to the discovery of an optical afterglow located in a spiral galaxy at $z=0.089$ \citep{Ofek060505}. Further observations revealed the absence of a SN down to very deep limits \citep{ThoneGCN060505, FynboNature, Ofek060505}, which posed the question if this was a Type II or a Type I GRB. The \emph{Swift} high-energy data does not allow any conclusions due to its low quality \citep{Schaefer060614} except that it does not follow the Amati relation \citep{Amati060614, KrimmBATWAM}, the recent joint \emph{Swift}-\emph{Suzaku WAM} spectral analysis places it well into the region of the Type I GRBs, though \citep{KrimmBATWAM}. Contrariwise, the burst originated in a young, massive star-forming region as is typical for Type II GRBs \citep{Ofek060505, Thoene060505, LeiblerBerger}. But the optical data show that there must be very little dust along the line of sight, excluding high extinction as a reason for the faintness of the afterglow and the absence of any SN emission \citep{Ofek060505, Thoene060505, Xu060505060614}. This event is discussed in more detail in \kref{Hybrid}. We take detections and upper limits from \cite{Ofek060505} and \cite{Xu060505060614}. 

\paragraph{GRB 060614.}This GRB, at first glance, seemed to be a classical high luminosity long GRB at moderate redshift \citep{Schaefer060614}. It was localized by \emph{Swift} \citep{ParsonsGCN060614}, which slewed immediately to the burst and also discovered the optical afterglow. Due to its brightness, it was also detected by \emph{Konus-Wind} \citep{GolenetskiiGCN060614} and \emph{RHESSI}. Initial spectroscopy \citep{FugazzaGCN060614} revealed no lines, but a redshift of $z=0.1254\pm0.0005$ was found several days later \citep{PriceGCN060614, DellavalleNature} from host galaxy lines. The host has very low mass (typical for Type II GRB hosts) but also a large mean stellar population age (much more typical for Type I GRB hosts, \citealt{LeiblerBerger}). Further observations revealed the absence of a SN \citep{FynboNature, GalyamNature, DellavalleNature}. A chance superposition with the faint, slightly star-forming galaxy \citep{Cobb060614} was ruled out strongly \citep{GalyamNature, GehrelsNature}, leading to the conclusion that this was a temporally clearly long GRB ($\TNT=102$ s) that was not associated with the typical bright, broad-lined Type Ic SN \citep{FynboNature, GalyamNature, DellavalleNature}. The observed optical afterglow of GRB 060614 is much brighter than that of any other Type I GRB, and we are able to create a high quality SED stretching from the UV to the NIR, taking data from \cite{FynboNature}, \cite{GalyamNature}, \cite{DellavalleNature}, \cite{Cobb060614}, \cite{Mangano060614}, \cite{YostDarkROTSE}, and \cite{Xu060505060614}. Our SED results are in agreement with \cite{DellavalleNature} and \cite{Xu060505060614}, the amount of dust along the line of sight is low and within the typical range found by K06 and Paper I for Type II GRBs. Still, \cite{Mangano060614} find, for SMC dust, an even lower value than we do, $\AV=0.05\pm0.02$. As the 2175 {\AA} feature is covered by the \emph{Swift} UVOT filters, we are able to derive a clear preference for SMC dust, as this feature is not visible. This is the first low-redshift GRB for which this has been possible in our analysis (see K06). For SMC dust, we find $\beta=0.41\pm0.09$, $\AV=0.28\pm0.07$ and a very good fit ($\chi^2_{dof}=0.72$). We caution that \cite{Xu060505060614} find $p\approx2.5$ from multi-band numerical modeling, which is significantly larger than our value ($p=1.82\pm0.18$, assuming $\nu_{opt}<\nu_c$). As \cite{DellavalleNature} and \cite{Xu060505060614}, we find that the late afterglow \citep[after the rising phase,][]{GalyamNature} can be fit with a broken power law. Using the precise host magnitude measurements from \cite{DellavalleNature}, we subtract the host galaxy from these data points and merge them with other, host-corrected sets. We derive the following light curve parameters: $m_k=20.32\pm0.05$ mag, $\alpha_1=1.05\pm0.04$, $\alpha_2=2.42\pm0.05$, $t_b=1.30\pm0.03$ days, and $n=10$ was fixed \citep[see][for a description of these parameters]{ZKK}. Again, this is the only Type I GRB afterglow where such an analysis could be performed. This event is discussed in more detail in \kref{Hybrid}.

\paragraph{GRB 060801.}This was a double-spiked burst (with a possible faint third peak) localized by \emph{Swift}, which slewed to the burst immediately \citep{RacusinGCN060801A}. It was also detected by \emph{Suzaku WAM}. No extended emission was detected, $\TNT=0.5\pm0.1$ s, and the burst is one of the hardest \emph{Swift} has ever detected \citep[][their Figure 1]{SatoGCN060801, HjorthMessenger}. Follow-up revealed no optical afterglow \citep[][this work]{BrownGCN060801, DuschaGCN060801}. There was initial confusion concerning the XRT error circle \citep{ButlerGCN060801, RacusinGCN060801B}, with two extended sources lying nearby, labeled ``B'' and ``D'' \citep{Castro-TiradoGCN060801}. Galaxy B, the brighter one, lies at a redshift of $z=1.1304$ \citep{BergerHighzSGRB}, no redshift is known for D. We find positions of RA = 14:12:01.26, Dec=+16:58:56.0 for B, RA = 14:12:01.43, Dec=+16:58:54.5 for D, with errors of 0\farcs3 for each. Comparing this with the final UVOT-enhanced XRT position (RA = 14:12:01.30, Dec=+16:58:54.0, error 1\farcs5) and the final (SDSS-calibrated) Butler position (RA = 14:12:01.35, Dec=+16:58:53.7, error 2\farcs4), we find that for the UVOT-enhanced position, the offset of both galaxies are almost identical (2\farcs$09\pm1$\farcs53, 2\farcs$01\pm1$\farcs53 for B and D, respectively), while for the Butler position, D is actually closer (2\farcs$67\pm2$\farcs42, 1\farcs$44\pm2$\farcs42 for B and D, respectively). Still, the much brighter magnitude of B makes it a significantly more likely host galaxy, making this the most distant Type I GRB with a reasonably secure redshift. We use the UVOT-enhanced position to determine the offset.

\paragraph{GRB 061006.}This event was initially detected by \emph{Swift} via a 64-s image trigger. IPN observations revealed this to be a bright Type I GRB with double-spiked emission lasting 0.5 s and occurring 22 s before the beginning of the \emph{Swift} trigger, which \emph{Swift} did not trigger on due to a preplanned slew \citep{HurleyGCN061006}. \emph{Swift} triggered on the extended emission of the GRB, $\TNT=130\pm10$ s, then slewed immediately. The \emph{Swift} observations are detailed in \cite{SchadyGCNR061006}. Initial observations \citep{SchadyGCN061006, PandeyGCN061006, MundellGCN061006} resulted in upper limits only, but VLT observations \citep{MalesaniGCN061006A, D'AvanzoThree} found a faint source that was subsequently found to fade, revealing the host galaxy \citep{MalesaniGCN061006B, BergerHighzSGRB, D'AvanzoThree}. \cite{BergerHighzSGRB} report a redshift of $z=0.4377\pm0.0002$ from several emission lines which indicate that this galaxy has moderate star-formation, which \cite{D'AvanzoThree} confirm. Relevant data have been taken from the sources aforementioned, with the afterglow data point being host-subtracted using data from \cite{D'AvanzoThree}. The host galaxy has blue colors consistent with a spiral galaxy \citep{D'AvanzoThree}, and \cite{FongHST} find it is well-fit by an exponential disk profile, confirming that it is an edge-on spiral galaxy. As there is only one detection which is clearly brighter than the host galaxy, the decay index is very unsure. Using all $I_C$ data, we find $\alpha=1.85\pm1.00$, in excellent agreement with \cite{D'AvanzoThree}. Using earlier upper limits, we find $\alpha\lesssim1$, but the afterglow contribution at 1.9 and 2.5 days is overestimated, indicating a break may have occurred.

\paragraph{GRB 061201.}This was a bright, multi-peaked GRB with $\TNT=0.8\pm0.1$ s localized by \emph{Swift} and also observed by \emph{Konus-Wind} \citep{GolenetskiiGCN061201}. \emph{Swift} slewed immediately to the burst which showed no sign of extended emission. The optical afterglow was discovered by UVOT and was seen to remain roughly constant for a long time, similar to the ones of GRB 060313 and GRB 090510. \emph{Swift} and VLT observations can be found in \cite{Stratta061201}. While it has been suggested that the GRB may be associated with the galaxy cluster Abell 995 \citep{BloomGCN061201} at a redshift of $z=0.0865$ \citep{BergerGCNz061201B}, the projected distance is very high (but note that \citealt{Salvaterra2010} show that even at this offset, the GRB may have occurred within an intra-cluster globular star cluster). A more likely association is a star-forming galaxy first noted by \cite{DAvanzoGCN061201} lying at a redshift of $z=0.111$ \citep{BergerGCNz061201A, Stratta061201, Berger2010}, at a distance of 34 kpc in projection. \cite{Berger2009} note this galaxy has properties which fit in well with the Type I GRB host galaxy population, but \cite{FongHST} present another possible host from HST observations, 1\farcs8 away at 25th mag, though they note that the chance probability is the same as for the $z=0.111$ galaxy. We use $z=0.111$ as the redshift in this work but note the similarity of this GRB to GRB 060313, which probably lies at $z\approx1$, and GRB 090510 at $z=0.903$. \cite{Stratta061201} find no host galaxy at the exact afterglow position down to $R>25.9$ (no host galaxy is detected in even deeper HST imaging either, see \citealt{Berger2010}); though note the host galaxy of GRB 070707 is even fainter \citep{Piranomonte070707}. Also, similar to GRB 060313 and GRB 090510, GRB 061201 is well detected in the \emph{Swift} UVOT \emph{uvw2} filter, and no sign of a Ly dropout is seen, indicating $z\lesssim1.3$ (the latter is of course no argument against the $z=0.111$ association). Relevant data on this GRB, detections and late upper limits, have been taken from \cite{MarshallGCNR061201}, \cite{HollandGCN061201B}, \cite{Berger2010}, and \cite{Stratta061201}, where we use the observation times detailed in \cite{HollandGCN061201B} to derive logarithmic mean times for the UVOT observations. We note that the afterglow of GRB 061201 is one of the faintest ever detected at early times and, assuming $z=0.111$ to be correct, is at early times the intrinsically faintest afterglow ever observed.

\paragraph{GRB 061210.}This was a very bright GRB consisting of triple-peaked emission lasting about 0.06 s but followed by a long tail of soft emission, $\TNT=85\pm5$ s. It was localized by \emph{Swift} \citep{CannizzoGCNR061210} and also detected by \emph{Suzaku-WAM} \citep{UrataGCN061210, KrimmBATWAM}. \emph{Swift} was unable to slew due to a Moon constraint. XRT observations that began 2.42 days after the GRB revealed two uncatalogued sources, one which was found to fade and was thus identified to be the X-ray afterglow \citep{GodetGCN061210, RacusinGCN061210}, indicating this was a very bright X-ray afterglow for a Type I GRB. While there are several sources within or near the error circle, the most likely host galaxy is a star-forming galaxy at $z=0.4095\pm0.0001$ \citep{BergerHighzSGRB}. Relevant data on this GRB, all upper limits, are taken from \cite{MirabalGCN061210}, \cite{CenkoGCN061210}, and \cite{MelandriGCN061210}.

\paragraph{GRB 061217.}This was a faint single-spiked GRB with $\TNT=0.212\pm0.041$ s localized by \emph{Swift}. \emph{Swift} slewed immediately to the burst which showed no sign of extended emission. The X-ray afterglow was very faint and initially wrongly localized. Details on the \emph{Swift} observations can be found in \cite{ZiaeepourGCNR061217}. Within the XRT error circle, a star-forming galaxy at redshift $z=0.8270\pm0.0001$ is found \citep{BergerHighzSGRB}. Relevant data on this GRB, all upper limits, are taken from \cite{SchaeferGCN061217}, \cite{JelinekGCN061217}, \cite{DAvanzoGCN061217}, and \cite{KlotzGCN061217}.

\paragraph{GRB 070209.}This was a faint single-spiked GRB with $\TNT=0.1\pm0.02$ s and no extended emission. It was detected by \emph{Swift}, which slewed to it immediately, but no fading X-ray source was detected at all. \emph{Swift} observations are detailed in \cite{SatoGCNR070209}. One constant X-ray source outside the BAT error circle is associated with an emission-line galaxy at $z=0.314$ \citep{BergerGCN070209}, possibly an AGN. We assume $z=0.5$. Relevant data on this GRB, all upper limits, are taken from \cite{SatoGCN070209} and \cite{JohnsonGCN070209}, the latter being the only ground-based follow-up reported. As there are only early and quite shallow upper limits reported, no significant magnitude limit can be derived, so this burst is not further included in the results.

\paragraph{GRB 070406.}This was a faint double-spiked GRB with $\TNT=0.7\pm0.2$ s detected by \emph{Swift}. The image significance was too low for a flight localization, and the ground analysis position was reported 19 hr after the GRB. \emph{Swift} ToO observations over the following 9 days revealed two faint X-ray sources, but both of them were found to be constant. One of these was associated with a bright blue galaxy visible in the SDSS data release \citep{CoolGCN070406}, found by \cite{KannGCN070406}, who proposed it as a host galaxy but also noted the colors and that it might be an AGN. The latter interpretation was confirmed spectroscopically by \cite{BergerGCN0704061}, who found it to be a quasar at $z=0.703$. The second X-ray source is possibly also associated with an AGN \citep{BergerGCN0704062}. No X-ray or optical afterglow was found, we assume $z=0.5$. The \emph{Swift} observations are reported in \cite{McBreenGCNR070406}. We take upper limits from \cite{ZhengGCN070406}, \cite{MalesaniGCN070406}, \cite{McBreenGCNR070406}, and \cite{BloomGCN070406}. Assuming $z=0.5$, these measurements span only a small time-span when shifted to $z=1$, 1.1 to 1.6 days. The deepest and last upper limit is $R_C>27$, we thus assume $R_C>26$ at one day for any reasonable afterglow evolution.

\paragraph{GRB 070429B.}This was a faint triple-spiked GRB with $\TNT=0.5\pm0.1$ s detected by \emph{Swift}. The satellite slew was delayed by 165 s due to Earth-limb constraint. After slewing, a faint X-ray afterglow was detected in ground analysis. The \emph{Swift} observations are reported in \cite{MarkwardtGCNR070429B}. Deep observations of the error circle revealed two faint galaxies of which one is probably the host \citep{CucchiaraGCN070429B, AntonelliGCN070429B}. For the brighter galaxy, \cite{PerleyGCN070429B} reported a redshift $z=0.904$, which is confirmed by \cite{Cenko070429B070714B}. Reanalysis of the \emph{Swift} UVOT data revealed a rising and then rapidly fading faint afterglow very close to the position of this galaxy \citep{HollandGCN070429B}, making this another secure association. We take detections from \cite{HollandGCN070429B} and upper limits from \cite{SchaeferGCN070429B}, \cite{MarkwardtGCN070429B},  \cite{AntonelliGCN070429B}, \cite{PerleyGCN070429B}, and \cite{HollandGCN070429B}.

\paragraph{GRB 070707.}This GRB was detected by \emph{INTEGRAL} and was first taken to be a temporally long event \citep{BeckmannGCN070707}, but was shortly afterward announced to be a Type I GRB of 1.1 s length, the first of its kind to be rapidly and accurately localized by \emph{INTEGRAL} \citep{GotzGCN070707}. The moderately bright, multi-spiked event was also detected by \emph{Konus-Wind} \citep{GolenetskiiGCN070707} and by \emph{Swift} BAT outside the coded FOV \citep{ParsonsGCNR070707}. The \emph{INTEGRAL} observations are described in \cite{McGlynn070707}. \emph{Swift} follow-up observations revealed an X-ray source that was subsequently seen to fade \citep{ParsonsGCNR070707}, confirming it as the X-ray afterglow. Optical observations with the VLT \citep{D'AvanzoGCN070707} revealed a single optical source in the X-ray error circle \citep{PiranomonteGCN070707}, which was seen to fade, confirming it to be the optical afterglow \citep{D'AvanzoGCN0707072}. Further very deep VLT follow-up revealed a break in the light curve, a steep decay, and an extremely faint host galaxy at $R_C\approx27.3$ \citep{Piranomonte070707}. No redshift is known \citep[deep spectroscopy 1.7 days after the GRB revealed only a low S/N continuum,][]{Piranomonte070707}, but we assume $z=1$ due to the extremely faint host \citep[see also][]{McGlynn070707}. We take upper limits from \cite{ParsonsGCNR070707} and detections from \cite{Piranomonte070707}. From a broken power-law fit with a smooth rollover ($n=1$ fixed), we find $\alpha_1=0.54\pm0.11$, $\alpha_2=5.44\pm0.89$, $t_b=1.98\pm0.18$ days and $m_h=27.04\pm0.09$ ($R_C$, corrected for extinction), in very good agreement with \cite{Piranomonte070707}.

\paragraph{GRB 070714B.}This was a very bright multi-spiked event lasting about 3 s which was followed by soft extended emission which lasted to probably 100 s. \emph{Swift} slewed immediately to the GRB and a bright X-ray afterglow was localized in flight. The \emph{Swift} observations are presented in \cite{RacusinGCNR070714B}. The GRB was not detected in spectral mode by \emph{Konus-Wind} (V. Pal'shin 2008, private communication), but a joint \emph{Swift} + \emph{Suzaku-WAM} spectral analysis yielded a hard spectrum and a high peak energy \citep{OhnoGCN070714B, KrimmBATWAM}. An optical afterglow was discovered just 10 minutes after the GRB by the Liverpool Telescope \citep{MelandriGCN070714B}, and was confirmed a day later by the WHT \citep{LevanGCN070714B}. Host galaxy observations reveal a redshift $z=0.92$ \citep{GrahamGCN070714B, Graham070714B}, confirmed by \cite{Cenko070429B070714B}. Late observations \citep{PerleyGCN070714B} show the source remains pointlike at $R\approx25.5$, and we note the similarity to GRB 051227 and GRB 060313, adding weight to the argument that these two GRBs lie at $z\approx1$. \cite{Graham070714B}, on the other hand, report $r^\prime=24.74\pm0.21$ (Vega mag, not extinction-corrected) from later Gemini host photometry; still, this is one of the faintest Type I GRB hosts detected so far.
We take upper limits from \cite{RacusinGCNR070714B} and detections from \cite{Graham070714B}, \cite{PerleyGCN070714B}, \cite{LandsmannGCN070714B} and our own TNG observation. The afterglow seems to show a rising evolution, or at least a plateau phase \citep{Graham070714B}.

\paragraph{GRB 070724A.}This was faint single-spiked GRB with $\TNT=0.4\pm0.04$ s localized by \emph{Swift}, which slewed to it immediately, flight-localizing the X-ray afterglow. While no extended emission was seen, \emph{Swift} detected two X-ray flares in the first 100 s, the first one rather bright, indicating that the GRB had extended emission, but it lay beneath the BAT detection threshold. The \emph{Swift} observations are detailed in \cite{ZiaeepourGCNR070724}. A comparison with the DSS lead to the discovery of what seemed to be a blue host galaxy \citep{BloomGCN0707241, BloomGCN0707242}. Spectroscopy of the object confirmed it to be a star-forming galaxy at $z=0.457$ \citep{CucchiaraGCN070724, CovinoGCN070724, Berger2009, Kocevski070724A}, and it is securely associated with the GRB due to the discovery of the optical/NIR afterglow \citep{Berger070724A}. \cite{Kocevski070724A} also present extensive and deep observations, but they do not reveal the afterglow as it is highly reddened \citep{Berger070724A}. We find a spectral slope of $\beta_0=2.1$, in full agreement with \cite{Berger070724A}. Assuming a fixed spectral slope ($\beta=0.6$ or $\beta=1.1$), we find $\AV\approx1.25-1.32$ (for different dust models) and $\AV\approx0.93-0.96$, respectively. Note that none of the dust models is able to fit the SED well. The X-ray afterglow has a hard spectral slope $\beta_X\approx0.7$ according to the \emph{Swift} XRT repository \citep{EvansXRT1, EvansXRT2}, so we use the values derived from $\beta=0.6$ in the optical \citep[though possibly this is influenced by an early flare,][]{Kocevski070724A}. We note that our analysis of the BAT data reveals this to be a quite soft event with an observed peak energy of just 41 keV and a photon index of $\Gamma=-2.2$, making this the first example of a ``short/soft GRB''. We take upper limits from \cite{ZiaeepourGCNR070724}, \cite{CenkoGCN070724}, and \cite{Kocevski070724A}, and detections from \cite{Berger070724A}.

\paragraph{GRB 070729.}This was faint triple-spiked GRB with $\TNT=0.9\pm0.1$ s localized by \emph{Swift}, which slewed to it immediately. The X-ray afterglow was faint and not flight-localized, similar to GRB 050509B and GRB 050813. The \emph{Swift} observations are detailed in \cite{GuidorziGCNR070729}. The GRB was also detected by the Konus-A experiment on Cosmos-2421 \citep{GolenetskiiGCN070729}. Magellan observations revealed an extended $K$-band source within the XRT error circle, which was presumed to be the host galaxy \citep{BergerGCN070729}, and was found to be red \citep{BergerMurphyGCN070729}. A later reanalysis\footnote{http://www.swift.ac.uk/xrt\_positions/00286373/image.php} of the UVOT-enhanced XRT position \citep[see][for the method]{GoadXRT} gave a strongly offset position, which contained another galaxy, measurements for this one are given in \cite{LeiblerBerger}. No spectroscopic redshift is known, but \cite{LeiblerBerger} find a photometric redshift of $z=0.8\pm0.1$, which we adapt. Upper limits are taken from \cite{GuidorziGCNR070729} and \cite{BergerGCN070729}. They do not reach late enough to allow for a derivation of a magnitude limit at one day.

\paragraph{GRB 070809.}This was a faint triple-spiked GRB with $\TNT=1.3\pm0.1$ s localized by \emph{Swift}, which slewed to it immediately. While not flight-localized, it had a moderately bright X-ray afterglow. The burst was relatively soft, which, combined with the relatively long duration, puts this burst in the realm intermediate between the clusters of ``short/hard'' and ``long/soft'' GRBs in the $\TNT$-hardness diagram. Further analysis places this GRB clearly into the Type I category \citep{BarthelmyGCN070809}. The \emph{Swift} observations are detailed in \cite{MarshallGCNR070809}. Keck observations revealed a faint source with $R\approx24$ at the edge of the XRT error circle \citep{PerleyGCN070809A}, which was found to have decayed significantly a day later, implying it to be the afterglow \citep{PerleyGCN070809B}. The afterglow is extremely red. Using the Keck data \citep{PerleyGCN070809B}, we find $\beta_0=3.97$, much steeper even than for GRB 070724A. At the low redshift (see below), no dust model can correctly represent this steep spectrum. Assuming a fixed spectral slope ($\beta=0.6$ or $\beta=1.1$), we find $\AV\approx1.38-1.56$ (for different dust models) and $\AV\approx1.29-1.45$, respectively. The closest approximation is gained using $\beta=1.1$ and MW dust ($\AV=1.45$), we will use this henceforth. A nearby galaxy that may be an edge-on spiral is a plausible host candidate; while an initial short spectrum showed no lines \citep{PerleyGCN070809B}, a second, longer integration showed it to lie at $z=0.2187$ \citep{PerleyGCN070809C}. We assume \citep[as is argued by][]{PerleyGCN070809C} that this is the host galaxy, and thus use this redshift. Note that \cite{Berger2010} also suggest a more distant (both in terms of $z$ as well as offset) elliptical galaxy at $z=0.473$ as the host. We take upper limits from \cite{RykoffGCN070809}, \cite{MarshallGCNR070809} and \cite{Berger2010}, and detections from \cite{PerleyGCN070809B}.

\paragraph{GRB 070810B.}This was a faint FRED GRB with $\TNT=0.08\pm0.01$ s localized by \emph{Swift}, which slewed to it immediately. No conclusive X-ray afterglow was detected. Deep XRT observations revealed several very faint X-ray sources. So far, none of these X-ray sources have been shown to fade. The \emph{Swift} observations are detailed in \cite{MarshallGCNR070810B}. Deep Keck observations linked one of the X-ray sources to a possibly interacting small cluster of galaxies, of which two seem to be strongly star-forming. The cluster lies at $z=0.49$ \citep{ThoeneGCN070810B}, we adopt this redshift. Furthermore, there is a large and nearby ($z=0.0385$) early type galaxy (LEDA 1354367) in the BAT error circle \citep{MarshallGCNR070810B, ThoeneGCN070810B}, but an association with this galaxy is unlikely due to the faintness of the GRB. A possible afterglow was reported in the $z=0.49$ cluster \citep{RumyantsevGCN070810B} but it was found to not be fading \citep{KocevskiGCN070810B}. Upper limits are taken from \cite{MarshallGCNR070810B}, \cite{XinGCN070810B}, \cite{GuidorziGCN070810B}, \cite{RumyantsevGCN070810B} and \cite{KocevskiGCN070810B}.

\paragraph{GRB 071112B}This was a faint double-peaked, very hard GRB with $\TNT=0.30\pm0.05$ s localized by \emph{Swift}. The satellite only started observing the burst after an hour due to Earth-limb constraint, and no X-ray afterglow was detected. The \emph{Swift} observations are detailed in \cite{PerriGCNR071112B}. The BAT error circle was observed by several large telescopes, but no variable sources were found via image subtraction. We assume $z=0.5$. Upper limits are taken from \cite{RykoffGCN071112B}, \cite{PerriGCNR071112B}, \cite{KocevskiGCN071112B}, and \cite{WiersemaGCN071112B}.

\paragraph{GRB 071227}This was a bright, multi-peaked, hard GRB with $\TNT=1.8\pm0.4$ s for the hard spike \citep{SatoGCN071227}, which was reported to have negligible spectral lag and possible extended emission by \cite{SakamotoGCN0712272}, the BAT ground-analysis page (http://gcn.gsfc.nasa.gov/gcn/notices\_s/299787/BA/) reports emission is detected up to 380 s. It was also detected by \emph{Konus-Wind} \citep{GolenetskiiGCN071227} and \emph{Suzaku-WAM} \citep{KrimmBATWAM}; the latter authors derive a high peak energy from a joint \emph{BAT-WAM} fit. \emph{Swift} slewed immediately and localized a bright X-ray afterglow, as expected from the extended emission \citep{BeardmoreGCN071227}. UVOT observations \citep{SakamotoGCN0712271, CucchiaraGCN071227} revealed only a single faint source near the XRT error circle, which was identified as a galaxy also visible in the DSS by \cite{BergerGCN0712271}. VLT \citep{D'AvanzoGCN0712271, D'AvanzoThree} and Magellan \citep{BergerGCN0712272} spectroscopy revealed it to lie at $z=0.381\pm0.001$ and further VLT follow-up revealed an optical afterglow \citep{D'AvanzoGCN0712272, D'AvanzoThree} at the tip of this edge-on spiral galaxy. Late-time follow-up revealed no sign of SN emission down to $R_C>24.9$ \citep{D'AvanzoThree}. We take photometry from \cite{SakamotoGCN0712271}, \cite{CucchiaraGCN071227} and \cite{D'AvanzoThree}. A decay rate cannot be determined except for a shallow limit due to a time gap of several days between the detection and the following observation \citep{D'AvanzoThree}.

\paragraph{GRB 080503}This was an extreme Type I GRB in several aspects, which has been extensively analyzed by \cite{Perley080503}. It consists of a typical short, multi-peaked spike ($\TNT=0.32\pm0.07$ s) followed by structured extended emission of a total duration of $\approx220$ s. The ratio between EE and IPC fluence is the largest for any Type I GRB in the \emph{Swift} era, over 30 \citep[only the \emph{BATSE} GRB 931222 has a higher value, with $\approx40$,][]{NorrisBonnell}. \emph{Swift} slewed immediately and detected a bright X-ray afterglow, which rapidly softens and then drops several orders of magnitude in flux, so that it is not detected at all anymore in the second orbit. This emission can be attributed to high-latitude emission of a possibly ``naked burst'' \citep{Genet080503, Perley080503}. The spectral lag of the IPC is negligible. Rapid, deep follow-up with Gemini revealed an extremely faint afterglow, $g^\prime=26.62\pm0.24$ (Vega mag, extinction-corrected) at about one hour after the GRB which seemed to be quickly fading. Deep observations a day later revealed the afterglow had significantly brightened, and in the following days it exhibited a slow rollover before fading again rapidly. While this was initially interpreted as a possible mini-SN signature, the contemporaneous detection in the X-rays by \emph{Chandra} indicates that it is synchrotron-powered afterglow emission \citep{Perley080503}. No host galaxy is detected in very deep \emph{HST} imaging down to $F606W>28.2$, the deepest limit ever found for a Type I GRB host galaxy. But there are two very faint nearby galaxies, one at $F606W\approx27$, 0\farcs8 distant, one at $F606W\approx26$, 2\farcs0 distant (all Vega mag, extinction-corrected). Both are host candidates (chance probability a few percent), unlike a more distant bright spiral galaxy at $z=0.561$ (chance probability close to unity, \citealt{Perley080503}). Due to the extreme faintness of any host, we assume $z=1$ similar to GRBs 051227, 060313 and 070707. Data are taken from \cite{MaoGCNR080503} and \cite{Perley080503}. As the afterglow is quite red \citep{Perley080503}, we use $\beta=1.1$ to shift the colors.

\paragraph{GRB 080905A}This event has been analyzed in detail by \cite{Rowlinson080905A}. It was a triple-peaked, hard \emph{Swift} GRB with $\TNT=1.0\pm0.1$ s and negligible lag. The X-ray afterglow was moderately bright for a Type I GRB, but decayed steeply and was not detected anymore in the second orbit. The afterglow is localized in one of the outer arms of a bright face-on star-forming spiral galaxy at $z=0.1218\pm0.0003$, making it the closest clear Type I GRB so far. Spatially resolved spectroscopy reveals no star-forming activity at the location of the GRB. High-energy emission has possibly been detected by \emph{Fermi} LAT for this event \citep{Akerlof2010}. Upper limits are taken from \cite{PaganiGCN080905A}, \cite{BrownGCN080905A}, \cite{TristamGCN080905A} and \cite{Rowlinson080905A}, and detections from the latter work.

\paragraph{GRB 090510}Even more than GRB 060313, GRB 090510 can be called ``a new paradigm for short-hard bursts''. Up to now (2011 February), it is the only GRB (Type I or Type II) which has been simultaneously localized by \emph{Swift} BAT and \emph{Fermi} LAT, allowing both rapid follow-up and broadband high-energy observations. It was also detected by \emph{AGILE} GRID at high energies \citep{Giuliani090510}, by \emph{Konus-Wind} \citep{GolenetskiiGCN090510} and by \emph{Suzaku-WAM} \citep{OhmoriGCN090510}. \emph{Swift} observations are detailed in \cite{DePasquale090510}, while information on the \emph{Fermi} observations can be found in \cite{Ackermann090510}, with time-resolved spectroscopy presented by \cite{Guiriec090510}. LAT detected gamma rays up to $\approx30$ GeV, which have been used to derive stringent lower limits on the quantum-gravity mass scale in certain theories \citep{Abdo090510, Xiao090510}. The GeV emission has a delayed onset and exhibits a power-law decay, being detected up until several 100 s after the GRB, which has been analyzed in the light of hadronic models \citep{Asano090510}, but finds a natural explanation as an external forward-shock synchrotron afterglow \citep{Ghirlanda090510, Gao090510, Kumar090510, DePasquale090510}. The GRB itself consists of a faint precursor pulse (which triggered GBM but not BAT) followed, after 0.5 s, by complex, intense, multi-spiked emission with $\TNT=0.3\pm0.1$ s \citep{HoverstenGCNR090510}. Recently, \cite{TrojaPrecursor} reported the existence of a significantly associated second precursor event 13 seconds before the main emission. The peak energy in the observer frame, $\approx4$ MeV, is the highest ever detected for any GRB \citep{Ackermann090510}, rising up to over 6 MeV on short timescales \citep{Guiriec090510}. BAT detects faint extended emission which connects directly to the XRT afterglow \citep{DePasquale090510}. The X-ray afterglow shows an early break to a steep decay, very probably a jet break \citep{DePasquale090510, Kumar090510}, whereas the UVOT afterglow shows a slow rise and decay (even beyond the X-ray break) initially \citep{DePasquale090510}, which \cite{Kumar090510} explain by the injection frequency $\nu_i$ being above the optical even at late times. The afterglow then turns over into a steep decay \citep{OlofssonGCN090510, McBreenLAT}, and is undetected down to deep limits one night later already, revealing a host galaxy at a small offset \citep{McBreenLAT}. Spectroscopy of this galaxy yields $z=0.903\pm0.001$ \citep{McBreenLAT}, making this one of the most distant spectroscopically confirmed Type I GRBs, which, together with the extreme broadband fluence, yields an immense isotropic energy release of $E_{iso}>10^{53}$ erg. Even in the bolometric frame, this is the highest energy release for any unambiguous (but see \citet{Panaitescu090510} for caveats) Type I GRB (only the controversial event GRB 060121 may be higher, though this is probably not a Type I GRB). From GROND \citep{McBreenLAT} and UVOT \citep{DePasquale090510} data, we derive a red spectral slope of $\beta=1.35\pm0.15$. Fixing the slope to the value from the X-ray afterglow \citep{EvansXRT1, EvansXRT2}, $\beta=0.79$, we find $\AV=0.24\pm0.07$. Note that while nominally lying blueward of Ly$\alpha$, the \emph{uvm2} and \emph{uvw2} detections are actually brighter than the more redward data, which would be expected from theory if the injection frequency $\nu_i$ still lies above the optical band at this time \citep{Kumar090510}. We take data from \cite{DePasquale090510}, \cite{OlofssonGCN090510} and \cite{McBreenLAT}.

\paragraph{GRB 090515}This event has been analyzed in detail by \cite{Rowlinson090515}. It was an extremely short ($\TNT=0.036\pm0.016$ s), soft, single-spiked \emph{Swift} FRED burst with a low peak energy and positive but almost negligible spectral lag. The X-ray afterglow was initially bright and almost flat, but then dropped precipitously and was not detected during \emph{Swift}'s second orbit, similar to GRB 080503. Despite the bright X-ray emission, which is typical for an extended emission component, none was discovered in the BAT data. \cite{Rowlinson090515} explain the X-ray emission as that of an unstable millisecond magnetar formed after a binary merger. Initial optical follow-up was rapid and quite deep, with an extremely faint detection after just 0.08 days, even fainter than GRB 080503 at a similar time, making this the faintest-ever detected GRB afterglow. There is no host galaxy down to very deep limits (\citealt{Rowlinson090515}, comparable to the magnitude of the GRB 070707 host galaxy, \citealt{Piranomonte070707}) under the afterglow position. \cite{Berger2010} present spectroscopy of several nearby galaxies and suggest an early-type galaxy at $z=0.403$ (part of a galaxy cluster) as the host, we adopt this in terms of redshift and offset determination. Upper limits are taken from \cite{RujopakarnGCN090515}, \cite{SiegelGCN090515}, \cite{PaceGCN090515}, and \cite{PerleyGCN090515}, and two detections from \cite{Rowlinson090515}.

\paragraph{GRB 091109B}This \emph{Swift} event \citep{OatesGCN091109B} had a short ($\TNT=0.3\pm0.03$ s) symmetrical peak and no sign of extended emission \citep{MarkwardtGCN091109B}. It was also detected by \emph{Suzaku-WAM} which detected a high peak energy \citep{OhnoGCN091109B}. A faint, rapidly decaying afterglow was discovered with the VLT \citep{LevanGCN091109B, MalesaniGCN091109B}. No host galaxy candidate has been reported, and no redshift is known, we assume $z=0.5$. Next to the detections by the aforementioned authors, we add upper limits from \cite{SchaeferGCN091109B}, \cite{JelinekGCN091109B}, and \cite{OatesGCN091109BB}.

\paragraph{GRB 100117A}This \emph{Swift} burst featured a short ($\TNT=0.3\pm0.05$ s) single peak and no sign of extended emission \citep{MarkwardtGCN100117A}, the burst was also detected by \emph{FERMI} GBM \citep{PaciesasGCN100117A}. \cite{Fong100117A} report a detailed analysis of the optical properties of the event, they present a faint afterglow which is clearly associated with an elliptical host galaxy at $z=0.915$, thus being only the second burst next to GRB 050724 \citep{Barthelmy050724, Berger050724, Gorosabel050724} to be unambiguously associated with an early-type galaxy. We use the detection and upper limits from \cite{Fong100117A}, as well as upper limits from \cite{XuGCN100117A}, \cite{CenkoGCN100117A}, and \cite{DePasqualeGCN100117A}.

\paragraph{Type I (and other short) GRBs not in the sample.}The following Type I GRBs are not in the sample:
\begin{itemize}
\item GRBs 050112, 060303, 060425, 060427B and 060429 were all localized by the IPN to large error boxes only.
\item GRBs 081007B, 081122B, 081204B, 081209, 081216, 081223, 081229, 090108A, 090108B, 090126C, 	090219, 090227B, 090228, 090305C, 090328B, 090412, 090418C, 090520B, 090617, 090717B, 090802A, 090814C, 090902A, 091126, 091126B have 
been detected by the GBM on \emph{Fermi}, but in all cases, the error circles are too large to yield any follow-up observations.
\item GRBs 051103 and 070201 are thought to be extragalactic SGR hyperflares in M81 and M31, respectively \citep{Ofek051103, Frederiks051103, PerleyGCN070201, GolenetskiiGCN070201, HurleyGCN070201, Abbott070201, Ofek070201}, they were localized by the IPN only.
\item The short but soft GRB 050925 is thought to be a small flare from an otherwise unknown Galactic SGR \citep{HollandGCN050925, MarkwardtGCN050925, Sakamoto050925}, though a background Type I GRB cannot be excluded with the available data.
\item The short INTEGRAL GRB 071017 is probably associated with a known Galactic X-ray source \citep{MereghettiGCN071017, EvansGCN071017}.
\item GRBs 050202 \citep{TuellerGCN050202}, 070923 \citep{StrohGCN070923}, 090417A \citep{ManganoGCN090417A} and 090715A \citep{RacusinGCN090715A} were all very close to the Sun, so \emph{Swift} did not slew, and ground-based follow-up is minimal. Note that GRB 090417A has an early-type 2MASS galaxy at $z=0.088$ at the edge of the BAT error circle \citep{FoxGCN090417A, O'BrienGCN090417A, BloomGCN090417A}.
\item GRBs 051114 \citep{CummingsGCN051114} and GRB 080121 \citep{CummingsGCN080121} were ground-localized and not reported until many hours after they occurred.
\item GRB 080123 \citep{UkwattaGCN080123} only had very limited optical follow-up with UVOT.
\item GRB 080702A \citep{dePasqualeGCN080702A} only had very limited optical follow-up.
\item GRB 080919 \citep{PregerGCN080919} occurred near the Galactic plane, in a crowded field with high extinction, and had a constant source in the XRT error circle.
\item GRB 081024B \citep{Abdo081024B, ConnaughtonGCN081024B} was the first Type I GRB detected by \emph{Fermi} LAT at GeV energies. While bright, it was not reported until 17 hr after the event, and multiple observations by \emph{Swift} XRT \citep{GuidorziGCN081024BA, GuidorziGCN081024BB, GuidorziGCN081024BC, GuidorziGCN081024BD} as well as several ground-based facilities \citep{CenkoGCN081024B, KannGCN081024B, FatkhullinGCN081024B} failed to reveal any credible afterglow.
\item GRB 081211B \citep{CopeteGCN081211B} was detected in the BAT slew survey, and comparison with \emph{Konus-Wind} data showed it is very probably a Type I GRB, with a short spike (which did not trigger \emph{Swift}, as it was almost outside the coded FOV) followed by extended emission, which BAT detected while slewing \citep{GolenetskiiGCN081211B}. XRT observations \citep{PageGCN081211BA, PageGCN081211BB} revealed an X-ray afterglow. While optical follow-up is minimal, \cite{PerleyGCN081211B} report that the GRB, which shows no host galaxy down to very deep limits, may be associated with a galaxy cluster at $z=0.216$.
\item GRB 081226A \citep{GodetGCN081226A} had an afterglow candidate \citep{AfonsoGCN081226A, BergerGCN081226AA} which was seen to neither fade \citep{BergerGCN081226AC} nor show any spectral lines \citep{BergerGCN081226AB}. These observations do not report any magnitude limits and thus cannot be used.
\item GRB 081226B \citep{MereghettiGCN081226B} was detected by \emph{INTEGRAL} IBIS as well as \emph{Fermi} GBM \citep{BissaldiGCN081226B}. Follow-up with \emph{Swift} revealed neither an XRT nor an optical afterglow \citep{EvansGCN081226B}. No further observations are reported.
\item GRB 090305 \citep{BeardmoreGCN090305A} is a case where a faint optical afterglow was discovered by the rapid reaction of large ground-based telescopes \citep[A. Updike 2009, private communication]{CenkoGCN090305, BergerGCN090305}, while the X-ray afterglow was almost undetected due to pointing constraints \citep{BeardmoreGCN090305B}. \cite{Berger2010} report that no host galaxy is detected down to deep limits at the afterglow position. They do not report redshifts of surrounding galaxies. An inclusion in our sample awaits the publication of the optical observations.
\item GRB 090607 \citep{MarshallGCN090607} was announced with a 30 minute delay and has hardly any optical follow-up.
\item GRB 090621B \citep{CurranGCN090621B}, which was also detected by \emph{Fermi} GBM \citep{GoldsteinGCN090621B}, occurred at low Galactic latitude, with high foreground extinction and field crowding. While optical follow-up was given, extinction-corrected limits are shallow, and the single object in the XRT error circle turned out to be an M dwarf \citep{BergerGCN090621B}.
\item GRB 090916 \citep{TrojaGCN090916} was Moon-constrained, and a late slew revealed no X-ray afterglow \citep{BeardmoreGCN090916}, ground follow-up was minimal.
\item GRB 091117 \citep{CummingsGCN091117} was a bright event not flight-localized by \emph{Swift}, it was reported over a day later after the derivation of an IPN localization led to a ground-analysis BAT position. X-ray follow-up by \emph{Swift} \citep{BergerGCN091117, FoxGCN091117B, D'EliaGCN091117A, D'EliaGCN091117B, D'EliaGCN091117C} and \emph{Chandra} \citep{FoxGCN091117} did not reveal any afterglow candidates, but two X-ray emitting AGNs \citep{ChornockGCN091117}.
\end{itemize}
\section{CONSTRAINTS ON EXTRA LIGHT IN TYPE I GRB AFTERGLOWS: METHODS}
\label{AppC}

\paragraph{SN1998bw light}
The procedure for calculating and redshifting a SN~1998bw component is explained in detail in \cite{Zeh2004}.

\paragraph{The mini-SN model and the macronova model}
Following LP98ab we assumed the ejecta consists of a variety of nuclei with a very broad range of decay times. This leads to their power-law decay model. The equations given in LP98ab provide the bolometric luminosity, $L_{\rm
bol}$, and the time-dependent effective temperature, $T_{\rm eff}$, of a mini-SN assuming black body radiation. We are interested in the $R$-band luminosity, $L_R$, which we write as $L_R(t) = y(t) \, L_{\rm bol}(t)$. Hence,
\begin{equation}
y(t) = \frac{\int_{\lambda_1}^{\lambda_2}\ S_R(\lambda)\ F^{\rm bb}_\lambda
       (T_{\rm eff}(t'), \lambda';z)\ d\lambda}
       {\int_0^{\infty}\ F^{\rm bb}_\lambda (T_{\rm eff}(t'),
        \lambda';z)\ d\lambda}\,,
\label{y}
\end{equation}
where $S_R(\lambda$) is the filter response function in the $R$ band, $F^{\rm bb}_\lambda$ is the Planck function, $t=t_{\rm obs}, t'=t_{\rm host} = t/(1+z)$, $\lambda' = \lambda/(1+z)$, and $z$ is the redshift.  The radius $R$
of the non-relativistically expanding SN is $R(t')=R_0 + vt'$, where $v$ is the velocity of the ejecta which is assumed to be independent of time; $R_0$ is negligible. The effective temperature of the SN at the time $t$ in the observer frame, needed to calculate $y(t)$, follows from $L_{\rm bol}(t).$

The above formalism assumes (following LP98ab, and K05 as well) that the spherical ejecta is optically thick, so that its radiation can be described by a black body.  This assumption holds up to a critical time, $t_c$ (LP98ab,
their Eq. 7), when the ejecta becomes optically thin:
\begin{equation}
t_c = 1.13\ \mbox{day} \ \left(\frac{M_{\rm ej}}{0.01}\right)^{1/2}
\left(\frac{c}{3v}\right)\ \left(\frac{\kappa}{\kappa_e}\right)^{1/2}\,.
\end{equation}
Here $M_{\rm ej}$ is the ejected mass  in units of solar masses, $\kappa$ is the opacity of a gas, and the index $e$ stands for electron scattering. While at $t>t_c$ the time evolution of the bolometric luminosity of the ejecta is approximately calculable (LP98ab), the  fraction of luminosity that goes into the optical bands cannot be calculated anymore adopting a black body.  Therefore, for $t \ge t_c$ our results are less reliable.

\section{UPDATES TO THE TYPE II GRB AFTERGLOW SAMPLE}
\label{PaperI}
Since the publication of Paper I, we have expanded the Type II GRB afterglow sample with several additional GRBs, as well as revised the analysis of GRBs within the sample with new data. Energetics for the three completely new GRBs are given in Table \ref{tabTypeIISample} analogous to Table 2 of Paper I.

\paragraph{GRB 060729, $z=0.5428$.}This GRB was already included in Paper I. In addition to the data used in Paper I, we add the data published in \cite{Cano060729}. The addition of high S/N $R_C$ data as well as HST NIR data allows us to construct a very broad SED ($U_3U_2U_1\,F300W\,UBg^\prime VR_CI_Cz^\prime F160W\, F222M$) and place this GRB within the Golden Sample. We find: $\beta_0=0.67\pm0.07$ ($\chi^2_\nu=1.98$) for no dust; $\beta=0.22\pm0.19$, $\AV=0.36\pm0.15$ ($\chi^2_\nu=1.45$) for MW dust; $\beta=0.16\pm0.31$, $\AV=0.32\pm0.19$ ($\chi^2_\nu=1.89$) for LMC dust; and $\beta=0.67\pm0.20$, $\AV=-0.00\pm0.10$ ($\chi^2_\nu=2.18$) for SMC dust. While MW (and to a lesser degree LMC) dust fits the SED better, the intrinsic slope is very flat, and the existence of a 2175 {\AA} bump is ruled out by the \emph{Swift} $U$ data. We therefore adopt the fit without any dust, which is also in good agreement with the X-ray spectral slope from the \emph{Swift} XRT repository, $\Gamma_X = 2.096^{+0.040}_{-0.039}$ \citep{EvansXRT1,EvansXRT2} if one assumes a cooling break between the two bands. \cite{Schady2010} also find evidence for a very low amount of dust. \cite{Cano060729} determine the additional light from the associated supernova by using the HST $F300W$ data at late times, which is expected to contain almost no SN light. We pursue a similar approach, creating a composite light curve which contains only UV data at $>9$ d. From a broken power-law fit, we find: $\alpha_1=1.26\pm0.01$, $\alpha_2=2.42\pm0.21$, $t_b=8.72\pm0.28$ days ($n=10$ fixed). The last $F300W$ data point indicates a slight upward inflection of the light curve, a fit leaving the host galaxy free results in a host galaxy magnitude $F300W_{Host}=25.13\pm0.48$ (AB magnitude, not corrected for Galactic extinction), in agreement with the late non-detection in this filter \citep{Cano060729}. Using $R_C$ data and fixing the underlying afterglow evolution, we find evidence for a late supernova bump in full agreement with \cite{Cano060729}, with $k=1.04\pm0.03$, $s=0.86\pm0.02$ (note that these errors are underestimated, as they do not include the uncertainties of the underlying afterglow evolution). These values are in good agreement with those derived by \cite{Cano060729}. Furthermore, we find $dRc=1.55\pm0.02$. At 86.4 s in the $z=1$ frame, the afterglow has $R_C=19.8\pm0.5$ (statistical error only, an unsure back-extrapolation), putting it clearly in the faint category (see Figure 6 of Paper I). At 1 d after the GRB [in the $z=1$ system]), we find $R_C=19.59\pm0.09$ ($M_B=-23.23\pm0.10$), and it is $R_C=21.51\pm0.20$ at 4 d ($M_B=-21.31\pm0.20$).

\paragraph{GRB 080413B, $z=1.1014$.}We constructed the light curve and the SED with data from the following sources: \cite{Filgas080413B}, \cite{OatesGCN080413B}, and \cite{GombocGCN080413B}. The redshift is from \cite{FynboSpectra}. A detailed analysis can be found in \cite{Filgas080413B}, who find a strong color change in the afterglow between the first epoch (up to 0.06 d) and later observations ($>1$ d), and model a strong rebrightening with a double-jet model. We undertake separate broken power-law fits for the early and late data. There is marginal evidence for flattening at very early times ($<0.003$ d), but the data is too sparse, therefore an additional early break does not improve the fit significantly. For the data up to 0.06 d, we find from a simultaneous multi-color fit: $\alpha_1=0.84\pm0.03$, $\alpha_2=0.52\pm0.01$, $t_b=0.0099\pm0.0011$ d, $n=-10$ fixed. For the late data ($>0.36$ d) we find from a simultaneous multi-color fit: $\alpha_3=0.84\pm0.02$ (identical to $\alpha_1$), $\alpha_4=2.56\pm0.16$, $t_b=3.55\pm0.15$ d, $n=10$ fixed, and $r_G=25.13\pm0.14$ mag for the host galaxy (Vega mag, corrected for foreground extinction). The color change is clearly visible when deriving the SEDs from these multi-color fits (there is no significant evidence of chromatic evolution within the two epochs themselves). The early SED ($UBg_GVr_GR_Ci_Gz_GJ_GH_GK_G$, $U_3U_2U_1$ are suppressed by Lyman absorption) is very blue, similar to that of GRB 080319B (Paper I), and there is no evidence for dust: We find $\beta_0=0.25\pm0.06$ ($\chi^2_\nu=0.74$) for no dust; $\beta=0.28\pm0.16$, $\AV=-0.02\pm0.10$ ($\chi^2_\nu=0.83$) for MW dust; $\beta=0.28\pm0.22$, $\AV=-0.02\pm0.15$ ($\chi^2_\nu=0.83$) for LMC dust; and $\beta=0.25\pm0.24$, $\AV=0.00\pm0.16$ ($\chi^2_\nu=0.83$) for SMC dust. The late SED ($g_Gr_Gi_Gz_GJ_GH_GK_G$) is much steeper (and more typical), but still, there is no significant evidence for any local extinction. We find $\beta_0=0.74\pm0.04$ ($\chi^2_\nu=0.11$) for no dust; $\beta=0.72\pm0.12$, $\AV=0.02\pm0.08$ ($\chi^2_\nu=0.12$) for MW dust; $\beta=0.71\pm0.16$, $\AV=0.03\pm0.11$ ($\chi^2_\nu=0.12$) for LMC dust; and $\beta=0.70\pm0.18$, $\AV=0.03\pm0.12$ ($\chi^2_\nu=0.12$) for SMC dust. We shift the two epochs individually using the two dust-free fits. We find $dRc=-0.22\pm0.00$ and $dRc=-0.25\pm0.00$ for the early and late epochs, respectively. At 86.4 s in the $z=1$ frame, the afterglow has $R_C=15.88\pm0.20$ (statistical error only), putting it clearly in the faint category (see Figure 6 of Paper I). The afterglow is already decaying at $t=0.0008427$ d, with $R_C=15.73\pm0.19$ (in the $z=1$ frame), this places it in category 4 (see Chapter 3.3.4 of Paper I). At 1 d after the GRB (in the $z=1$ system), we find $R_C=19.50\pm0.11$ ($M_B=-23.34\pm0.11$), and it is $R_C=21.12\pm0.15$ at 4 d ($M_B=-21.72\pm0.15$).

\paragraph{GRB 080603A, $z=1.68742$.} This GRB was localized by \emph{INTEGRAL}. The Faulkes Telescope North obtained early observations, finding a faint optical flare contemporaneous with the prompt emission, followed by a slow rise and rollover comparable to the afterglow evolution of GRB 080710 \citep{Kruehler080710}. Observations and detailed analysis are given in \cite{Guidorzi080603A}. We use data from \cite{Guidorzi080603A}, \cite{MilneGCN080603A}, and \cite{SbarufattiGCNR080603A}. We confirm the SED results of \cite{Guidorzi080603A}, and find a red SED ($Bg^\prime Vr^\prime R_Ci^\prime I_CJHK_S$) which is excellently fit by LMC dust: $\beta_0=2.29\pm0.05$ ($\chi^2_\nu=3.71$) for no dust; $\beta=2.08\pm0.10$, $\AV=0.19\pm0.08$ ($\chi^2_\nu=3.26$) for MW dust; $\beta=0.85\pm0.31$, $\AV=0.90\pm0.19$ ($\chi^2_\nu=0.86$) for LMC dust; $\beta=2.11\pm0.18$, $\AV=0.13\pm0.08$ ($\chi^2_\nu=3.83$) for SMC dust. This is among the highest extinction values derived in Paper I, only GRB 070802 has a higher value among the (secure) Golden Sample GRBs. The only two other cases in the sample of Paper I where LMC dust is strongly preferred are GRB 061007 and GRB 070802. We find $dRc=-3.64^{+0.39}_{-0.40}$. At 86.4 s in the $z=1$ frame, the afterglow (just rising to the prompt emission flare, \citealt{Guidorzi080603A}) has $R_C=16.85\pm0.20$ (statistical error only), putting it clearly in the faint category (see Figure 6 of Paper I). The afterglow reaches a late peak at $t=0.0137517$ d, with $R_C=14.31\pm0.03$ (in the $z=1$ frame), this places it in category 3 (see Chapter 3.3.4 of Paper I). At 1 d after the GRB [in the $z=1$ system]), we find $R_C=18.21\pm0.42$ ($M_B=-24.66\pm0.43$), and it is $R_C=21.00\pm0.47$ at 4 d ($M_B=-21.87\pm0.47$).

\paragraph{GRB 080607, $z=3.0363\pm0.0003$.} This was an extremely energetic GRB at moderately high redshift, having one of the fighest isotropic energy releases and the highest peak luminosity ever measured for a GRB \citep{GolenetskiiGCN080607}. Rapid Keck spectroscopy revealed the most distant known 2175 {\AA} absorption bump as well as the first detection of molecular lines in a GRB afterglow \citep{Prochaska080607, Sheffer080607}. It also revealed an extremely red afterglow, despite it being initially very bright. Afterglow observations and dust modeling are described in detail in \cite{Perley080607}. The dust model that yields a good fit to the afterglow is different from the local universe dust models used in Paper I \citep{Perley080607}, we confirm this and find that while MW dust gives a marginally acceptable fit, the intrinsic spectral slope is still much redder than theoretically allowed. From the $VR_Ci^\prime I_Cz^\prime JHK_S$ SED, we find: $\beta_0=2.99\pm0.05$ ($\chi^2_\nu=20.11$) for no dust; $\beta=1.88\pm0.13$, $\AV=0.77\pm0.09$ ($\chi^2_\nu=4.89$) for MW dust; $\beta=1.54\pm0.33$, $\AV=0.80\pm0.19$ ($\chi^2_\nu=19.76$) for LMC dust; $\beta=4.30\pm0.21$, $\AV=-0.57\pm0.08$ ($\chi^2_\nu=15.69$) for SMC dust, the latter yields an obviously unphysical result and is strongly ruled out. Therefore, we use the photometry from their paper as well as their extinction correction (5.76 mag from dust, 0.21 mag from absorption lines in the observed $R_C$ band) and the assumption $\beta_0=0.7$ to find $dRc=8.73$ (the error is hard to estimate, but $\approx0.5$). The extinction \cite{Perley080607} find, $A_V=3.26\pm0.35$ (or $A_V=3.07\pm0.32$ for a different intrinsic spectral slope) is much higher than any found in the sample of Paper I, showing that this a ``Rosetta Stone'' for the study of dark GRBs -- an extremely extinguished GRB afterglow which was luminous enough to allow detailed measurements nonetheless. Correcting for this, we confirm the early afterglow of GRB 080607 is the second most luminous ever observed, we find $R_C=5.83\pm0.04$ (statistical error only) at $t=0.00013977$ d (12 s after the trigger) in the $z=1$ frame. The afterglow is already decaying at this point, placing it in category 4 (see Chapter 3.3.4 of Paper I). At 86.4 s in the $z=1$ frame, the afterglow is the 4th brightest, just 0.1 mag fainter than that of GRB 061007 (which peaks around this time, whereas the afterglow of GRB 080607 has already decayed by over 2 mag). Until it goes over into a plateau at 0.005 d, it is almost identical in luminosity and evolution to that of GRB 061007. Afterglow observations extend only to 0.043 days ($z=1$ frame), at this time, only the afterglows of GRB 060210 and GRB 090313 are (possibly, see Paper I) more luminous. A composite light curve can be fit with four power-laws, we find $\alpha_1=0.77\pm0.07$, $\alpha_2=1.54\pm0.02$, $\alpha_3=0.40\pm0.05$, and $\alpha_4=1.92\pm0.05$, with breaks at $t_{b1}=(3.85\pm0.298)\times10^{-4}$, $t_{b2}=(4.0\pm0.14)\times10^{-3}$, and $t_{b3}=(18\pm0.44)\times10^{-3}$ d. Extrapolating the last, well-constrained decay, we find $R_C=19.60\pm0.52$ ($M_B=-23.24\pm0.52$) at 1 d after the GRB (in the $z=1$ system), dominated by the error of $dRc$.

\paragraph{GRB 090926A, $z=2.1071$.}This GRB was already included in Paper I. \cite{D'Elia090926A} find $z=2.1071$ from VLT/X-Shooter observations. The light curve now also includes the UVOT photometry from \cite{Swenson090926A}.

\section{ERRATUM TO PAPER I}
\label{Erratum}
In Table 5 of Paper I, due to an unfortunate error, several \texttt{tablenotemark} commands are associated with the incorrect \texttt{tablenotetext}. The following corrections need to be made:
\begin{itemize}
\item GRB 071025 needs to have the \texttt{tablenotemark} $^\mathrm{m}$.
\item GRB 080210 needs to have the \texttt{tablenotemark} $^\mathrm{n}$ instead of $^\mathrm{p}$.
\item GRB 050730 needs to have the \texttt{tablenotemark} $^\mathrm{p}$ instead of $^\mathrm{q}$.
\item GRB 080319C needs to have the \texttt{tablenotemark} $^\mathrm{q}$ instead of $^\mathrm{r}$.
\item GRB 081203A needs to have the \texttt{tablenotemark} $^\mathrm{r}$ instead of $^\mathrm{s}$.
\item GRB 050820A needs to have the \texttt{tablenotemark} $^\mathrm{s}$ instead of $^\mathrm{t}$.
\item GRB 060210 needs to have the \texttt{tablenotemark} $^\mathrm{t}$ instead of $^\mathrm{u}$.
\item GRB 080928 needs to have the \texttt{tablenotemark} $^\mathrm{n}$ removed. Furthermore, the final data from \cite{Rossi080928} gives $t=0.0182117$ d for the peak, and $R_C=14.51\pm0.06$.
\item GRB 060906 needs to have the \texttt{tablenotemark} $^\mathrm{x}$ instead of $^\mathrm{y}$.
\end{itemize}



\newpage\clearpage

\LongTables

\begin{landscape}
\begin{deluxetable}{lcccccccclr}
\tablecolumns{11}
\tabletypesize{\tiny}
\tablewidth{0pt}
\tablecaption{Selection criteria for the Type I GRB Sample according to the flowchart of \cite{Zhang080913}}
\tablehead{
\colhead{GRB} &
\colhead{$\TNT>2$s?\tablenotemark{a}} &
\colhead{IPC+ESEC?} &
\colhead{Elliptical host?\tablenotemark{c}} &
\colhead{Low SSFR?} &
\colhead{Large Offset?\tablenotemark{d}} &
\colhead{Off Amati or lag-lum?\tablenotemark{e}} &
\colhead{Low $E_\gamma$,$E_K$?} &
\colhead{} &
\colhead{Classification} &
\colhead{\cite{Lü2010}\tablenotemark{f}}
}
\startdata
050509B	&	N	&	$\cdots$	&	Y	&	$\rightarrow$	&		&		&		&		&	Type I	&	Type I	\\
050709	&	Y	&	Y	&	N	&	Y	&	$\rightarrow$	&		&		&		&	Type I	&	Type I	\\
050724	&	Y	&	Y	&	Y	&	$\rightarrow$	&		&		&		&		&	Type I	&	Type I	\\
050813	&	N	&	$\cdots$	&	Y?	&	Y?	&	Y?	&	N/N	&	Y	&	$\rightarrow$	&	Type I Candidate	&	Type I	\\
050906	&	N	&	$\cdots$	&	$\cdots$	&	$\cdots$	&	$\cdots$	&	N/?	&	Y	&	$\rightarrow$	&	Type I Candidate	&	$\sim$ Type I	\\
050911	&	Y	&	Y	&	$\cdots$	&	$\cdots$	&	$\cdots$	&	N/N	&	Y	&	$\rightarrow$	&	Type I Candidate	&	$\sim$ Type I	\\
051105A	&	N	&	$\cdots$	&	$\cdots$	&	$\cdots$	&	$\cdots$	&	N/?	&	Y	&	$\rightarrow$	&	Type I Candidate	&	$\sim$ Type I	\\
051210	&	N	&	Y	&	unclear	&	Y?	&	Y	&	$\rightarrow$	&		&		&	Type I	&	$\sim$ Type I	\\
051211A	&	Y	&	Y	&	$\cdots$	&	$\cdots$	&	$\cdots$	&	N/N	&	Y	&	$\rightarrow$	&	Type I Candidate	&	$\sim$ Type I	\\
051221A	&	N	&	$\cdots$	&	N	&	Y	&	$\rightarrow$	&		&		&		&	Type I	&	Type I	\\
051227	&	Y	&	Y	&	unclear	&	unclear	&	N	&	N/N	&	Y	&	$\rightarrow$	&	Type I Candidate	&	$\sim$ Type I	\\
060121	&	N	&	Y	&	unclear	&	unclear	&	N	&	Y/Y	&	N	&	$\rightarrow$	&	Type II Candidate	&	$\sim$ Type II	\\
060313	&	N	&	$\cdots$	&	unclear	&	unclear	&	Y	&	N/N	&	Y	&	$\rightarrow$	&	Type I Candidate	&	$\sim$ Type I	\\
060502B	&	N	&	$\cdots$	&	Y?	&	Y?	&	Y?	&	N/?	&	Y	&	$\rightarrow$	&	Type I Candidate	&	Type I	\\
060505	&	Y	&	N\tablenotemark{b}	&	N	&	N	&	N	&	N/N?	&	Y	&	$\rightarrow$	&	Type I Candidate	&	Type I	\\
060614	&	Y	&	Y	&	N	&	Y	&	$\rightarrow$	&		&		&		&	Type I	&	Type I	\\
060801	&	N	&	$\cdots$	&	N	&	Y	&	$\rightarrow$	&		&		&		&	Type I	&	Type I	\\
061006	&	Y	&	Y	&	N	&	Y	&	$\rightarrow$	&		&		&		&	Type I	&	Type I	\\
061201	&	N	&	$\cdots$	&	N?	&	Y?	&	Y	&	$\rightarrow$	&		&		&	Type I	&	Type I	\\
061210	&	Y	&	Y	&	N	&	Y	&	$\rightarrow$	&		&		&		&	Type I	&	Type I	\\
061217	&	N	&	$\cdots$	&	N	&	Y	&	$\rightarrow$	&		&		&		&	Type I	&	Type I	\\
070209	&	N	&	$\cdots$	&	$\cdots$	&	$\cdots$	&	$\cdots$	&	N/?	&	Y	&	$\rightarrow$	&	Type I Candidate	&	$\sim$ Type I	\\
070406	&	N	&	$\cdots$	&	$\cdots$	&	$\cdots$	&	$\cdots$	&	N/?	&	Y	&	$\rightarrow$	&	Type I Candidate	&	$\sim$ Type I	\\
070429B	&	N	&	$\cdots$	&	N	&	Y	&	$\rightarrow$	&		&		&		&	Type I	&	Type I	\\
070707	&	N	&	$\cdots$	&	unclear	&	unclear	&	N	&	N/N	&	Y	&	$\rightarrow$	&	Type I Candidate	&	$\sim$ Type I	\\
070714B	&	Y	&	Y	&	N	&	Y	&	$\rightarrow$	&		&		&		&	Type I	&	Type I	\\
070724A	&	N	&	$\cdots$	&	N	&	Y	&	$\rightarrow$	&		&		&		&	Type I	&	Type I	\\
070729	&	N	&	$\cdots$	&	unclear	&	unclear	&	unclear	&	N/?	&	Y	&	$\rightarrow$	&	Type I Candidate	&	$\sim$ Type I	\\
070809	&	N	&	$\cdots$	&	Y or N	&	Y	&	$\rightarrow$	&		&		&		&	Type I	&	$\sim$ Type I	\\
070810B	&	N	&	$\cdots$	&	$\cdots$	&	$\cdots$	&	$\cdots$	&	N/?	&	Y	&	$\rightarrow$	&	Type I Candidate	&	$\sim$ Type I	\\
071112B	&	N	&	$\cdots$	&	$\cdots$	&	$\cdots$	&	$\cdots$	&	N/?	&	Y	&	$\rightarrow$	&	Type I Candidate	&	$\sim$ Type I	\\
071227	&	Y	&	Y	&	N	&	Y	&	$\rightarrow$	&		&		&		&	Type I	&	Type I	\\
080503	&	Y	&	Y	&	$\cdots$	&	$\cdots$	&	$\cdots$	&	N/N	&	Y	&	$\rightarrow$	&	Type I Candidate	&	$\sim$ Type I	\\
080905A	&	N	&	$\cdots$	&	N	&	Y	&	$\rightarrow$	&		&		&		&	Type I	&	$\sim$ Type I	\\
090510	&	N	&	Y	&	N	&	Y	&	$\rightarrow$	&		&		&		&	Type I	&	Type I	\\
090515	&	N	&	$\cdots$	&	Y?	&	Y?	&	Y?	&	N/N	&	Y	&	$\rightarrow$	&	Type I Candidate	&	$\sim$ Type I	\\
091109B	&	N	&	$\cdots$	&	$\cdots$	&	$\cdots$	&	$\cdots$	&	N/?	&	Y	&	$\rightarrow$	&	Type I Candidate	&	$\sim$ Type I	\\
100117A	&	N	&	$\cdots$	&	Y	&	$\rightarrow$	&		&		&		&		&	Type I	&	$\sim$ Type I	
\enddata
\tablecomments{
\tablenotetext{a}{In all cases where $\TNT>2$s, $\TNT/(1+z)>2$s was also the case, therefore we do not list this step here.}
\tablenotetext{b}{GRB 060505 is the only GRB with $\TNT>2$s which does not show the IPC+ESEC light curve spectral shape. It, along will all others, does not have SN emission associated with it \citep{FynboNature, Ofek060505}, therefore we do not list this step here.}
\tablenotetext{c}{``$\cdots$'' implies that no host (candidate) is known. ``unclear'' implies that the host is so faint that no classification can be done, e.g., Sersic profiles with both $n=1$ or $n=4$ fit the host light profile equally well \citep{FongHST}. For GRB 050813, several elliptical galaxies lie in or close to the error circle \citep{ProchaskaSGRB, Ferrero050813}, but their association with the GRB is not secure \cite{Berger050813}. The host of GRB 051210 shows a featureless spectrum \citep{BergerHighzSGRB} and no clear solution can be determined from multicolor photometry either \citep{LeiblerBerger}. GRB 060502B is possibly associated with a large elliptical galaxy at high offset \citep{Bloom060502B, Church2011}, but there are other faint galaxies within the error circle \citep{BergerHighzSGRB, FongHST}. GRB 061201 is most likely associated with a spiral galaxy at large offset \citep{Stratta061201}, but there is a faint galaxy of unknown classification nearby \citep{FongHST}. The host galaxy of GRB 070809 could either be a spiral galaxy \citep{PerleyGCN070809C} or an elliptical galaxy at larger offset \citep{Berger2010}. GRB 090515 does not have a host galaxy beneath its optical afterglow to deep limits \citep{Rowlinson090515} but is statistically associated with an elliptical galaxy at large offset \citep{Berger2010}.}
\tablenotetext{d}{For GRB 050813, the offset is large if the elliptical galaxies discussed in \cite{ProchaskaSGRB} are associated with the GRB, but it is unclear if the host candidate of \cite{Berger050813} is correct (it lies within the rather large X-ray error circle). For GRB 060502B, the offset is large if the \cite{Bloom060502B} host candidate is correct, and unclear otherwise. Note we skip the next two selection criteria, as none of the GRBs in our sample shows evidence for a wind medium, and there is no significant constraining evidence for the external medium density $n$.}
\tablenotetext{e}{GRB 060505 has a significant lag \citep{McBreen060505}, and yet it does not lie on the lag-luminosity correlation \citep{Zhang080913}. Instead, it occupies an area of very low-energy GRBs with significant but not extremely large lag, together with the SN-associated GRBs 031203 and 980425 \citep{Zhang080913}. Therefore, it shows neither the almost always negligible lags of Type I GRBs, nor does it exhibit a lag large enough to satisfy the correlation for its energy release.}
\tablenotetext{f}{``Type I'' denotes GRBs that are Type I according to the log $\varepsilon$ classification of \cite{Lü2010}, with results given in that paper (essentially, only GRBs with a secure or reasonably secure redshift). ``$\sim$ Type I'' are Type I GRBs as derived in this work, using the redshifts we assume \citep[GRB 100117A has a secure redshift,][]{Fong100117A}. GRB 060121 is a marginal Type II GRB using the $z\approx4.6$ solution and a Type I GRB using the less likely $z\approx1.7$ solution.}
}
\label{tabTypeISelection}
\end{deluxetable}
\clearpage
\end{landscape}

\newpage\clearpage

\tabletypesize{\tiny}

\begin{landscape}
\begin{deluxetable}{lccccccccccccc}
\tablecolumns{13}
\tablecaption{Properties of the Type I GRB Sample}
\tablehead{\colhead{GRB} & \colhead{Redshift $z$} & \colhead{$\TNT$ (s)\tablenotemark{a}} & \colhead{Fluence} & \colhead{Band (keV)} & \colhead{$\tn{E}_{\tn{iso,bol}}$} & \colhead{$\tn{E}_{\tn{p,rest}}$} & \colhead{Photon} & \colhead{Spikes} & \colhead{ESEC} & \colhead{$\TNT$ (s)\tablenotemark{c}} & \colhead{Afterglow} & \colhead{Host offset}\\
\multicolumn{2}{c}{} & \colhead{(IPC)} & \colhead{($10^{-7}\;\tn{erg}\;\tn{cm}^{-2}$)} & \colhead{(Satellite)\tablenotemark{l}} & \colhead{$(10^{50}\;\tn{erg})$} & \colhead{(keV)} & \colhead{Index $\Gamma$\tablenotemark{b}} & \colhead{in IPC} & \colhead{(Bump)} & \colhead{(total)} & \colhead{X/O/R?} & \colhead{(kpc)}}
\startdata
050509B	&	$0.2248\pm0.0002$	&	$0.04\pm0.004$	&	$0.2\pm0.09$	&	$15-350$ (\emph{S})	& $48.38^{+0.45}_{-0.23}$	&	$100.4^{+748.4}_{-98.0}$	&	$-1.7\pm0.7$	&	1	&	N	&	$\cdots$	&	Y/N/N	&	$63.7\pm12.2$	\\
050709	&	$0.1606\pm0.0001$	&	$0.07\pm0.01$	&	$14.93\pm1.81$	&	$2-400$ (\emph{H})	& $49.93^{+0.05}_{-0.06}$	&	$100.4^{+18.6}_{-12.8}$	&	$-0.82^{-0.13}_{+0.14}$	&	$\sim3$	&	Y	&	$130\pm7$\tablenotemark{d}	&	Y/Y/N	&	$3.74\pm0.005$	\\
050724	&	$0.2576\pm0.0004$	&	$3.0\pm1.0$\tablenotemark{e}	&	$15\pm2$	&	$15-350$ (\emph{S})	&	$50.38^{+0.38}_{-0.02}$	&	$138.4^{+503.3}_{-56.6}$	&	$-1.6\pm0.2$\tablenotemark{f}	&	1\tablenotemark{g}	&	Y	&	$152.4\pm9.2$	&	Y/Y/Y	&	$2.76\pm0.024$	\\
050813	&	$\sim0.72$	&	$0.6\pm0.1$	&	$1.0^{+0.3}_{-0.6}$	&	$15-350$ (\emph{S})	&	$50.18^{+0.43}_{-0.33}$	&	$361.2^{+1221}_{-223.6}$	&	$-1.2\pm0.5$	&	1\tablenotemark{h}	&	N	&	$\cdots$	&	Y/N?/N	&	$\cdots$	\\
050906	&	0.43?	&	$0.128\pm0.016$	&	$0.6^{+0.4}_{-0.3}$	&	$15-350$ (\emph{S})	& $49.49^{+0.42}_{-0.59}$	&	$517.4^{+2282}_{-492.1}$	&	$-0.94^{-0.1.05}_{+0.99}$	&	1	&	N	&	$\cdots$	&	N/N/N	&	$\cdots$	\\
050911	&	0.1646?	&	$\sim1.5$	&	$4.4^{+0.9}_{-1.3}$	&	$15-350$ (\emph{S})	& $49.43^{+0.39}_{-0.10}$	&	$63.9^{+419.9}_{-60.9}$	&	$-1.83^{-0.33}_{+0.36}$	&	2	&	Y	&	$16\pm2$	&	N/N/N	&	$\cdots$	\\
051105A	&	$\cdots$	&	$0.028\pm0.004$	&	$0.5\pm0.2$	&	$15-350$ (\emph{S})	&	$49.68^{+0.34}_{-0.39}$\tablenotemark{i}	&	$450.6^{+1036}_{-327.6}$\tablenotemark{i}	&	$-1.30^{-0.42}_{+0.43}$	&	1	&	N	&	$\cdots$	&	N/N/N	&	$\cdots$	\\
051210	&	$\geq1.4$	&	$1.27\pm0.05$	&	$2.2^{+0.5}_{-0.8}$	&	$15-350$ (\emph{S})	& $51.41^{+0.39}_{-0.48}$	&	$721.0^{+2104.9}_{-557.8}$	&	$-1.00^{-0.33}_{+0.31}$	&	2	&	Y	&	$\sim40$	&	Y/N/N	&	$35.7\pm14.4$	\\
051211A	&	$\cdots$	&	$4.02\pm1.82$	&	$8.0\pm1.29$	&	$2-400$ (\emph{H})	& $50.70^{+0.16}_{-0.21}$\tablenotemark{i}	&	$187.5^{+44.0}_{-36.2}$\tablenotemark{i}	&	$-0.12^{-0.54}_{+0.21}$	&	$\sim4$	&	Y	&	$\sim40$	&	N/N/N	&	$\cdots$	\\
051221A	&	0.5464	&	$1.4\pm0.2$	&	$32^{+1}_{-17}$	&	$20-2000$ (\emph{K})	& $51.41^{+0.02}_{-0.33}$	&	$621.7^{+143.8}_{-111.3}$	&	$-1.08^{-0.13}_{+0.14}$	&	8	&	N	&	$\cdots$	&	Y/Y/Y	&	$1.92\pm0.18$	\\
051227	&	$\cdots$	&	$8.0\pm0.2$	&	$8.0\pm1.0$	&	$15-350$ (\emph{S})	& $51.44^{+0.37}_{-0.20}$\tablenotemark{j}	&	$270.0^{+1167.1}_{-128.2}$\tablenotemark{j}	&	$-1.09\pm0.23$	&	$\sim3$	&	Y	&	$\sim100$	&	Y/Y/N	&	$0.40\pm0.16$\tablenotemark{j}	\\
060121	&	4.6	&	$1.60\pm0.07$	&	$47.7\pm2.8$	&	$2-400$ (\emph{H})	& $53.34\pm0.05$	&	$627.2^{+79.5}_{-68.9}$	&	$-0.78^{-0.12}_{+0.11}$	&	3	&	Y	&	$\sim120$	&	Y/Y/N	&	$0.79\pm0.31$	\\
060121	&	1.7	&		&		&		& $52.61\pm0.06$	&	$302.4^{+38.3}_{-33.2}$	&		&		&		&		&		&	$1.02\pm0.39$	\\
060313	&	$\cdots$	&	$0.7\pm0.1$	&	$129^{+15}_{-31}$	&	$15-3000$ (\emph{S+K})	& $52.54^{+0.05}_{-0.12}$\tablenotemark{j}	&	$1894^{+448}_{-346}$\tablenotemark{j}	&	$-0.61^{-0.09}_{+0.11}$	&	$>20$	&	N	&	$\cdots$	&	Y/Y/N	&	$2.28\pm0.50$\tablenotemark{j}	\\
060502B	&	0.287?	&	$0.09\pm0.02$	&	$1.2^{+0.2}_{-0.6}$	&	$15-350$ (\emph{S})	& $49.48^{+0.43}_{-0.48}$	&	$437.6^{+926.6}_{-244.5}$	&	$-1.0\pm0.2$	&	2	&	N	&	$\cdots$	&	Y/N/N	&	$71.0\pm15.9$	\\
060505	&	0.0889	&	$4\pm1$	&	$23^{+10.8}_{-6.6}$	&	$15-2000$ (\emph{S+Sz})	& $49.64^{+0.17}_{-0.15}$	&	$482.4^{+524.8}_{-167.7}$	&	$-1.23\pm0.33$	&	1	&	N	&	$\cdots$	&	Y/Y/N	&	$7.06\pm0.33$	\\
060614	&	$0.1254\pm0.0005$	&	$\sim5$	&	$409^{+18}_{-34}$	&	$20-2000$ (\emph{K})	& $51.40^{+0.05}_{-0.04}$	&	$438.9^{+922.8}_{-281.4}$	&	$-1.90\pm0.04$	&	6\tablenotemark{k}	&	Y	&	$102\pm5$	&	Y/Y/N	&	$1.11\pm0.22$	\\
060801	&	$1.1304\pm0.0001$	&	$0.5\pm0.1$	&	$9.0^{+0.8}_{-1.0}$	&	$15-2000$ (\emph{S+Sz})	& $51.49\pm0.05$	&	$1400^{+864.9}_{-449.5}$	&	$-0.44\pm0.32$	&	2	&	N	&	$\cdots$	&	Y/N/N	&	$17.3\pm12.6$	\\
061006	&	$0.4377\pm0.0002$	&	$\sim0.5$	&	$35.7^{+3.1}_{-19.2}$	&	$20-2000$ (\emph{K})	& $51.24^{+0.04}_{-0.34}$	&	$954.6^{+326.4}_{-207.0}$	&	$-0.62^{-0.18}_{+0.21}$	&	2	&	Y	&	$130\pm10$	&	Y/Y/N	&	$1.30\pm0.24$	\\
061201	&	0.111?	&	$0.8\pm0.1$	&	$53.3^{+7.0}_{-44.4}$	&	$20-3000$ (\emph{K})	& $50.15^{+0.06}_{-0.78}$	&	$969.9^{+508.8}_{-315.5}$	&	$-0.36^{-0.40}_{+0.65}$	&	5	&	N	&	$\cdots$	&	Y/Y/N	&	$34.0\pm2.0$	\\
061210	&	$0.4095\pm0.0001$	&	$\sim0.06$	&	$26.0^{+4.7}_{-15.9}$	&	$15-2000$ (\emph{S+Sz})	& $51.06^{+0.08}_{-0.41}$	&	$1012^{+451.0}_{-286.1}$	&	$-0.72\pm0.2$	&	3	&	Y	&	$85\pm5$	&	Y/N/N	&	$10.9\pm9.9$	\\
061217	&	$0.8720\pm0.0001$	&	$0.212\pm0.041$	&	$1.2^{+0.3}_{-0.4}$	&	$15-350$ (\emph{S})	& $50.48^{+0.37}_{-0.48}$	&	$730.8^{+1480}_{-475}$	&	$-0.96\pm0.28$	&	1	&	N	&	$\cdots$	&	Y/N/N	&	$58.2\pm29.5$	\\
070209	&	$\cdots$	&	$0.10\pm0.02$	&	$0.7^{+0.2}_{-0.3}$	&	$15-350$ (\emph{S})	& $49.73^{+0.38}_{-0.39}$\tablenotemark{i}	&	$498.0^{+875.0}_{-341.3}$\tablenotemark{i}	&	$-1.02\pm0.33$	&	1	&	N	&	$\cdots$	&	N/N/N	&	$\cdots$	\\
070406	&	$\cdots$	&	$0.7\pm0.2$	&	$0.45\pm0.10$	&	$15-150$ (\emph{S})	& $50.32^{+0.70}_{-1.29}$\tablenotemark{i}	&	$1263^{+5072}_{-1011}$\tablenotemark{i}	&	$-0.9\pm0.4$	&	2	&	N	&	$\cdots$	&	N/N/N	&	$\cdots$	\\
070429B	&	$0.9023\pm0.0003$	&	$0.5\pm0.1$	&	$0.6\pm0.2$	&	$15-350$ (\emph{S})	& $50.13^{+0.42}_{-0.20}$	&	$228.3^{+1419}_{-125.6}$	&	$-1.53^{-0.38}_{+0.40}$	&	3	&	N	&	$\cdots$	&	Y/Y/N	&	$8.27\pm5.24$	\\
070707	&	$\cdots$	&	$\sim1.1$	&	$14.1^{+1.6}_{-10.7}$	&	$20-2000$ (\emph{K})	& $51.58^{+0.35}_{-0.99}$\tablenotemark{j}	&	$854^{+748}_{-288}$\tablenotemark{j}	&	$-0.57^{-0.43}_{+0.59}$	&	$\sim12$	&	N	&	$\cdots$	&	Y/Y/N	&	$\cdots$	\\
070714B	&	$0.9224\pm0.0001$	&	$\sim3$	&	$39^{+5.4}_{-8.7}$	&	$15-2000$ (\emph{S+Sz})	& $52.04^{+0.08}_{-0.12}$	&	$2470^{+988.1}_{-688.2}$	&	$-1.00\pm0.09$	&	$>4$	&	Y	&	$\sim100$	&	Y/Y/N	&	$3.15\pm0.63$	\\
070724A	&	$0.4571\pm0.0003$	&	$0.4\pm0.04$	&	$0.44^{+0.11}_{-0.20}$	&	$15-350$ (\emph{S})	& $49.39^{+0.36}_{-0.15}$	&	$59.2^{+147.5}_{-54.2}$	&	$-2.18^{-0.24}_{+0.26}$	&	$1$	&	N	&	$\cdots$	&	Y/Y/N	&	$4.76\pm0.06$	\\
070729	&	$0.8\pm0.1$	&	$0.9\pm0.1$	&	$5.59^{+0.05}_{-4.38}$	&	$20-1000$ (\emph{KA})	&	$51.05^{+0.16}_{-0.68}$	&	$840.6^{+1526}_{-351.0}$	&	$-1.08^{-0.28}_{+0.36}$	&	$3$	&	N	&	$\cdots$	&	Y/N/N	&	$3.21\pm18.8$\tablenotemark{n}	\\
070809	&	0.2187?	&	$1.3\pm0.1$	&	$0.87^{+0.40}_{-0.17}$	&	$15-350$ (\emph{S})	&	$49.12^{+0.34}_{-0.10}$	&	$75.6^{+12.6}_{-13.9}$	&	$-0.80^{-0.94}_{+0.13}$	&	$3$	&	N	&	$\cdots$	&	Y/Y/N	&	$21.1\pm1.98$	\\
070810B	&	0.49?	&	$0.08\pm0.01$	&	$0.70^{+0.37}_{-0.43}$	&	$15-350$ (\emph{S})	& $49.71^{+0.43}_{-0.52}$	&	$494.7^{+1372}_{-385.8}$	&	$-1.01^{-0.73}_{+0.64}$	&	$1$	&	N	&	$\cdots$	&	N/N/N	&	$\cdots$	\\
071112B	&	$\cdots$	&	$0.30\pm0.05$	&	$1.81^{+0.60}_{-0.92}$	&	$15-350$ (\emph{S})	& $50.24^{+0.34}_{-0.51}$\tablenotemark{i}	&	$672.3^{+1271}_{-441.3}$\tablenotemark{i}	&	$-0.83^{-0.50}_{+0.44}$	&	$2$	&	N	&	$\cdots$	&	N/N/N	&	$\cdots$	\\
071227	&	$0.381\pm0.001$	&	$1.8\pm0.4$	&	$11^{+29.2}_{-1.8}$	&	$15-2000$ (\emph{S+Sz})	& $50.75^{+0.56}_{-0.12}$	&	$2251^{+1019}_{-665.6}$	&	$-0.71\pm0.22$	&	$\sim10$	&	Y	&	$\gtrsim300$	&	Y/Y/N	&	$14.8\pm0.34$	\\
080503	&	$\cdots$	&	$0.32\pm0.07$	&	$27.5^{+2.0}_{-2.1}$	&	$15-350$ (\emph{S})	& $52.22^{+0.07}_{-0.08}$	&	$99.3^{+35.9}_{-97.3}$	&	$-1.94\pm0.14$	&	$\sim5$	&	Y	&	$\sim220$	&	Y/Y/N	&	$\cdots$	\\
080905A	&	$0.1218\pm0.0003$	&	$1.0\pm0.1$	&	$4.0^{+0.68}_{-0.60}$	&	$15-350$ (\emph{S})	& $49.36^{+0.46}_{-0.35}$	&	$502.8^{+950.6}_{-280.5}$	&	$-0.86^{+0.26}_{-0.24}$	&	$3$	&	N	&	$\cdots$	&	Y/Y/N	&	$18.5\pm0.5$	\\
090510\tablenotemark{o}	&	$0.903\pm0.001$	&	$0.3\pm0.1$	&	$320\pm20$	&	$8-40000$ (\emph{F})	& $52.61\pm0.04$	&	$7490^{+532.8}_{-494.8}$	&	$-0.58^{+0.06}_{-0.05}$	&	$7$	&	N	&	$\sim100$	&	Y/Y/N	&	$9.38\pm3.91$	\\
090515	&	$0.403?$	&	$0.036\pm0.016$	&	$0.22\pm0.08$	&	$15-350$ (\emph{S})	&	$48.97^{+0.35}_{-0.39}$	&	$90.1^{+47.4}_{-16.8}$	&	$+0.05^{-1.18}_{+1.54}$	&	$1$	&	N	&	$\cdots$	&	Y/Y/N	&	$75.2\pm0.54$	\\
091109B	&	$\cdots$	&	$0.3\pm0.03$	&	$9.98^{+0.72}_{-3.69}$	&	$100-1000$ (\emph{Sz})	&	$51.14^{+0.20}_{-0.47}$\tablenotemark{i}	&	$1195^{+1680}_{-915.0}$\tablenotemark{i}	&	$-0.91^{-0.42}_{+0.78}$	&	$4$	&	N	&	$\cdots$	&	Y/Y/N	&	$\cdots$	\\
100117A	&	$0.915$	&	$0.3\pm0.05$	&	$4.10\pm0.5$	&	$8-1000$ (\emph{F})	&	$50.96^{+0.05}_{-0.06}$	&	$549.6^{+141.7}_{-95.8}$	&	$+0.14^{-0.33}_{+0.27}$	&	$1$	&	N	&	$\cdots$	&	Y/Y/N	&	$0.471\pm0.314$	
\enddata
	\tablecomments{References for $z$ and $\TNT$ can be found in \kref{AppA}. Further references for energetics (Fluence, Band, $E_p$, Photon Index):
GRB 050509B: \cite{ButlerBAT};
GRB 050709: \cite{Villasenor050709};
GRB 050724: \cite{Barthelmy050724, Campana050724, ButlerBAT};
GRB 050813: \cite{SatoGCN050813, ButlerBAT};
GRB 050906: \cite{ParsonsGCN050906, ButlerBAT}, this work;
GRB 050911: \cite{Page050911, ButlerBAT}, this work;
GRB 051105A: \cite{BarbierGCN051105A, ButlerBAT}, this work;
GRB 051210: \cite{Laparola051210, ButlerBAT}, this work;
GRB 051211A: \cite{DonaghyHETE};
GRB 051221A: \cite{GolenetskiiGCN051221A};
GRB 051227: \cite{HullingerGCN051227, SakamotoGCN051227, ButlerBAT}, this work;
GRB 060121: \cite{DonaghyHETE, GolenetskiiGCN060121};
GRB 060313: \cite{Roming060313};
GRB 060502B: \cite{SatoGCN060502B, ButlerBAT};
GRB 060505: \cite{HullingerGCN060505, KrimmBATWAM};
GRB 060614: \cite{Mangano060614, GolenetskiiGCN060614, ButlerBAT};
GRB 060801: \cite{SatoGCN060801, KrimmBATWAM};
GRB 061006: \cite{SchadyGCNR061006, GolenetskiiGCN061006};
GRB 061201: \cite{MarshallGCNR061201, GolenetskiiGCN061201};
GRB 061210: \cite{CannizzoGCNR061210, KrimmBATWAM};
GRB 061217: \cite{ZiaeepourGCNR061217, ButlerBAT};
GRB 070209: \cite{SatoGCNR070209, ButlerBAT}, this work;
GRB 070406: \cite{McBreenGCNR070406, Liang2007b};
GRB 070429B: \cite{MarkwardtGCNR070429B, ButlerBAT}, this work;
GRB 070707: \cite{GolenetskiiGCN070707};
GRB 070714B: \cite{RacusinGCNR070714B, KrimmBATWAM};
GRB 070724A: this work;
GRB 070729: \cite{GolenetskiiGCN070729};
GRB 070809: this work;
GRB 070810B: this work;
GRB 071112B: this work;
GRB 071227: \cite{KrimmBATWAM};
GRB 080503: this work;
GRB 080905A: this work;
GRB 090510: \cite{GuiriecGCN090510, Abdo090510SUB};
GRB 090515: \cite{Rowlinson090515};
GRB 091109B: \cite{OhnoGCN091109B};
GRB 100117A: \cite{PaciesasGCN100117A}.
\\
References for host galaxy offset:
In all cases where only an X-ray afterglow was detected, we used the XRT position from \cite{ButlerX} and the associated Web page (http://astro.berkeley.edu/$\sim$nat/swift/xrt\_pos.html) for newer GRBs (GRB 061210: v8.6; GRB 061217: v2.7; GRB 070429B: v0.7), except for GRB 060801 and GRB 070729, where we used XRT positions enhanced by UVOT astrometry \citep{GoadXRT} taken from the associated Web page (http://www.swift.ac.uk/xrt\_positions/index.php). We note that in several cases where we could compare (GRB 050509B, GRB 051210, GRB 060502B) discrepancies on the $\approx2\sigma$ level exist between Butler and Goad positions \citep[see also][]{FongHST}, whereas in other cases (GRB 060801) the error circles overlap well. X-ray error circles are given at 90\% confidence level, therefore, the offset errors are larger than $1\sigma$ confidence. Optical afterglow positions and host galaxy positions (or direct offsets) are taken from:
GRB 050509B, GRB 050709, GRB 050724, GRB 051210, GRB 060121, GRB 060313, GRB 061006, GRB 071227: all \cite{FongHST};
GRB 050813: \cite{Ferrero050813};
GRB 051221A: \cite{Soderberg051221A};
GRB 051227: \cite{BergerHighzSGRB};
GRB 060502B: \cite{Bloom060502B};
GRB 060505: \cite{Ofek060505};
GRB 060614: \cite{GalyamNature};
GRB 060801: \cite{PiranomonteGCN060801A};
GRB 061201: \cite{DAvanzoGCN061201};
GRB 061210: \cite{BergerHighzSGRB};
GRB 061217: \cite{BergerHighzSGRB};
GRB 070429B: \cite{Cenko070429B070714B};
GRB 070714B: \cite{Graham070714B}, J. Graham, private communication;
GRB 070724: \cite{Berger070724A};
GRB 070729: \cite{LeiblerBerger};
GRB 070809: \cite{PerleyGCN070809A};
GRB 080905A: \cite{Rowlinson080905A}
GRB 090510: \cite{McBreenLAT};
GRB 090515: \cite{Berger2010};
GRB 100117A: \cite{Fong100117A}.\\
\tablenotetext{a}{Duration of the Initial Pulse Complex (IPC), identical to the complete GRB if no extended emission is detected.}
\tablenotetext{b}{Identical to the low energy spectral index $\alpha$ in case of a fit with a power law plus exponential cutoff or a Band function fit.}
\tablenotetext{c}{Total duration in case extended emission is observed at low energies. \cite{Norris2009}, using Bayesian Blocks analysis, give the following durations for the ESEC only (of \emph{Swift} GRBs, in s): 050724: 104.3; 050911: 105.1; 051227: 119.1; 060614: 168.8; 061006: 157.1; 061210: 89.6; 070714B: 64.9; 071227: 106.6; 080503: 245.4. The find no ESEC for 051210 and 090510.}
\tablenotetext{d}{Extended emission only, total about 10 s more.}
\tablenotetext{e}{Dominated by a short, hard spike of 0.25 s duration.}
\tablenotetext{f}{For the short spike only; \cite{Barthelmy050724} give $\Gamma=1.38\pm0.13$ for the short spike, \cite{Campana050724} give $\Gamma=1.75\pm0.16$ for the short spike, this softens to $\Gamma=2.5\pm0.2$ afterward.}
\tablenotetext{g}{Excluding low-level emission.}
\tablenotetext{h}{Overlaid on a broader peak.}
\tablenotetext{i}{Assuming $z=0.5$.}
\tablenotetext{j}{Assuming $z=1$.}
\tablenotetext{k}{Initial Pulse Complex only, the extended emission has several significant peaks too.}
\tablenotetext{l}{Satellite: \emph{S} = \emph{Swift}, \emph{K} = \emph{Konus-Wind}, \emph{H} = \emph{HETE-2}, \emph{Sz} = \emph{Suzaku HXD-WAM}, \emph{KA} = \emph{Konus-A (Cosmos-2421)}, \emph{F} = \emph{Fermi}}
\tablenotetext{n}{The radius of the XRT error circle is larger than the offset between the center of the error circle and the host galaxy center.}
\tablenotetext{o}{The fluence is taken from \cite{GuiriecGCN090510}, the other spectral information from the submitted version of the Nature paper \citep[][see also \citealt{Ackermann090510}]{Abdo090510SUB}, it has been completely removed from the final published version \citep{Abdo090510}. The spectrum is best described by a Band function with low-energy power law $\alpha=-0.58^{+0.06}_{-0.05}$, high-energy power law $\beta=-2.83^{+0.14}_{-0.20}$, and an additional rising power-law at even higher energies which describes the LAT emission but which we ignore when computing the bolometric energy release. Concerning the extended emission, Figure 1 of \cite{DePasquale090510} shows BAT detections until 60 s, whereas \cite{HoverstenGCNR090510} state extended emission is visible from 110 to 170 s, and \cite{Norris2009} find no extended emission at all.}
}
\label{tabTypeISample}
\end{deluxetable}
\clearpage
\end{landscape}

\newpage\clearpage

\tabletypesize{\footnotesize}

\begin{deluxetable}{llcccccc}
\tablecolumns{8}
\tablecaption{Results on Type I GRBs}
\tablehead{
\colhead{GRB} &
\colhead{$\beta$\tablenotemark{a}} &
\colhead{$\AV$\tablenotemark{b}} &
\colhead{$dRc$\tablenotemark{c}} &
\colhead{mag\tablenotemark{d}} &
\colhead{$M_B(AG)$\tablenotemark{e}} &
\colhead{$k$\tablenotemark{f}} &
\colhead{$M_R(SN)$\tablenotemark{g}}
}
\startdata
050509B	&	0.6	&	0	&	$+3.68$	&	$>28.95$	&	$>-13.9$	&	$<2.5\times10^{-3}$	&	$>-12.7$	\\
050709	&	1.12	&	0.67	&	$+4.15$	&	$25.3\pm0.2$	&	$-17.6\pm0.2$	&	$<1.5\times10^{-3}$	&	$>-12.1$	\\
050724	&	0.76	&	0	&	$+3.43$	&	$23.9\pm0.1$	&	$-18.95\pm0.13$	&	$<0.06$	&	$>-16.1$	\\
050813	&	0.6	&	0	&	$+0.81$	&	$>24.3$	&	$>-18.5$	&	$<0.29$	&	$>-17.8$	\\
050906	&	0.6	&	0	&	$+2.10$	&	$>28.0$	&	$>-14.8$	&	$<0.08$	&	$>-16.4$	\\
050911	&	0.6	&	0	&	$+4.41$	&	$>30.19$	&	$>-12.6$	&	$\cdots$	&	$\cdots$	\\
051105A\tablenotemark{h}	&	0.6	&	0	&	$+1.73$	&	$>25.35$	&	$>-17.5$	&	$\cdots$	&	$\cdots$	\\
051210	&	0.6	&	0	&	$-0.83$	&	$>23.70$	&	$>-19.1$	&	$\cdots$	&	$\cdots$	\\
051211A\tablenotemark{h}	&	0.6	&	0	&	$+1.73$	&	$>23.43$	&	$>-19.8$	&	$\cdots$	&	$\cdots$	\\
051221A	&	0.62	&	0	&	$+1.52$	&	$23.82\pm0.2$	&	$-19.0\pm0.2$	&	$<0.60$	&	$>-18.6$	\\
051227\tablenotemark{h}	&	0.6	&	0	&	$+0.00$	&	$27.17\pm1.03$	&	$-15.6\pm1.0$	&	$\cdots$	&	$\cdots$	\\
060121\tablenotemark{i}	&	0.6	&	0.5	&	$-6.67$	&	$18.5\pm0.5$	&	$-24.3\pm0.5$	&	$\cdots$	&	$\cdots$	\\
060121\tablenotemark{i}	&	0.6	&	1.1	&	$-4.11$	&	$21.0\pm0.3$	&	$-21.8\pm0.3$	&	$\cdots$	&	$\cdots$	\\
060313\tablenotemark{h}	&	0.6	&	0	&	$+0.00$	&	$22.72\pm0.07$	&	$-20.08\pm0.07$	&	$\cdots$	&	$\cdots$	\\
060502B	&	0.6	&	0	&	$+3.09$	&	$>27.28$	&	$>-15.5$	&	$<3.8\times10^{-3}$	&	$>-13.1$	\\
060505	&	1.1	&	0	&	$+6.16$	&	$26.6\pm0.3$	&	$-16.35\pm0.3$	&	$<3.3\times10^{-4}$	&	$>-10.5$	\\
060614	&	0.41	&	0.28	&	$+4.67$	&	$24.04\pm0.05$	&	$-18.71\pm0.05$	&	$<6.0\times10^{-3}$	&	$>-13.6$	\\
060801	&	0.6	&	0	&	$-0.31$	&	$>24.94$	&	$>-17.9$	&	$\cdots$	&	$\cdots$	\\
061006	&	0.6	&	0	&	$+2.06$	&	$25.32\pm0.2$	&	$-17.5\pm0.2$	&	$\cdots$	&	$\cdots$	\\
061201	&	0.6	&	0	&	$+5.32$	&	$28.9\pm0.4$	&	$-13.9\pm0.4$	&	$<3.3\times10^{-3}$	&	$>-13.0$	\\
061210	&	0.6	&	0	&	$+2.22$	&	$>25.6$	&	$>-17.2$	&	$\cdots$	&	$\cdots$	\\
061217	&	0.6	&	0	&	$+0.47$	&	$>22.5$	&	$>-20.3$	&	$\cdots$	&	$\cdots$	\\
070209\tablenotemark{h}	&	0.6	&	0	&	$+1.73$	&	$\cdots$	&	$\cdots$	&	$\cdots$	&	$\cdots$	\\
070406\tablenotemark{h}	&	0.6	&	0	&	$+1.73$	&	$>26.0$	&	$>-16.8$	&	$\cdots$	&	$\cdots$	\\
070429B	&	0.6	&	0	&	$+0.26$	&	$>25.1$	&	$>-17.7$	&	$\cdots$	&	$\cdots$	\\
070707\tablenotemark{h}	&	0.6	&	0	&	$+0.00$	&	$23.46\pm0.05$	&	$-19.34\pm0.05$	&	$\cdots$	&	$\cdots$	\\
070714B	&	0.6	&	0	&	$+0.21$	&	$23.95\pm0.21$	&	$-18.85\pm0.21$	&	$\cdots$	&	$\cdots$	\\
070724A	&	0.6	&	1.29	&	$+0.31$	&	$>25.4$	&	$>-17.4$	&	$<0.06$	&	$>-16.1$	\\
070729	&	0.6	&	0	&	$+0.56$	&	$\cdots$	&	$\cdots$	&	$\cdots$	&	$\cdots$	\\
070809	&	1.1	&	1.45	&	$+2.55$	&	$26.62\pm0.25$	&	$-16.3\pm0.25$	&	$\cdots$	&	$\cdots$	\\
070810B	&	0.6	&	0	&	$+1.78$	&	$>27.0$	&	$>-15.8$	&	$\cdots$	&	$\cdots$	\\
071112B\tablenotemark{h}	&	0.6	&	0	&	$+1.73$	&	$>26.6$	&	$>-16.2$	&	$\cdots$	&	$\cdots$	\\
071227	&	0.6	&	0	&	$+2.40$	&	$26.20\pm0.30$	&	$-16.6\pm0.3$	&	$\cdots$	&	$\cdots$	\\
080503\tablenotemark{h}	&	1.1	&	0	&	$+0.00$	&	$25.06\pm0.18$	&	$-17.89\pm0.18$	&	$\cdots$	&	$\cdots$	\\
080905A	&	0.6	&	0	&	$+5.11$	&	$29.32\pm0.30$	&	$-13.48\pm0.30$	&	$<5.5\times10^{-3}$	&	$>-13.6$	\\
090510	&	0.79	&	0.24	&	$-0.14$	&	$25.5\pm0.5$	&	$-17.4\pm0.5$	&	$\cdots$	&	$\cdots$	\\
090515	&	0.6	&	0	&	$+2.26$	&	$28.68\pm0.25$	&	$-14.12\pm0.25$	&	$\cdots$	&	$\cdots$	\\
0901109B\tablenotemark{h}	&	0.6	&	0	&	$+1.73$	&	$28.4\pm0.5$	&	$-14.4\pm0.5$	&	$\cdots$	&	$\cdots$	\\
100117A	&	0.6	&	0	&	$+0.22$	&	$>26.2$	&	$>-16.6$	&	$\cdots$	&	$\cdots$	\\
\enddata
\tablenotetext{a}{Excepting GRB 060121, if the slope is $\beta=0.6$, then this is the assumed value.}
\tablenotetext{b}{If the table gives extinction in the host frame as $\AV=0$, then this is the assumed value, except for GRB 050724 and GRB 060505, where no extinction is found in the SED.}
\tablenotetext{c}{The magnitude shift to $z=1$.}
\tablenotetext{d}{The $R_C$ magnitude of the afterglow (or upper limit thereon) at 1 day after the GRB in the $z=1$ frame.}
\tablenotetext{e}{The absolute $B$-band magnitude of the afterglow at one day after the burst (for the $z=1$ frame).}
\tablenotetext{f}{The upper limit on a SN contribution. This has only been obtained for GRBs with deep late detections or upper limits (see Figure \ref{SNfig}). For the definition of $k$, see \cite{Zeh2004}.}
\tablenotetext{g}{The limit on the absolute $R_C$-band luminosity of a contributing SN at peak.}
\tablenotetext{h}{No redshift known. A shift $dRc=+1.73$ implies that we assume $z=0.5$, a shift $dRc=+0.00$ implies we assume $z=1$.}
\tablenotetext{i}{For $z=4.6$ (upper line) and $z=1.7$ (lower line)}
\label{tabTypeI}
\end{deluxetable}

\newpage\clearpage
\begin{deluxetable}{lccccccccl}
\tablecolumns{9}
\tablecaption{Energetics of the Type II GRB Sample expansion added in this work}
\tablehead{
\colhead{GRB} &
\colhead{Redshift $z$} &
\colhead{Fluence ($10^{-7}\;$} &
\colhead{Band} &
\colhead{Low-energy} &
\colhead{High-energy} &
\colhead{$E_{\tn{p,rest}}$} &
\colhead{log $E_{\tn{iso,bol}}$} &
\colhead{References} \\
\colhead{} &
\colhead{} &
\colhead{$\rm erg\;\rm cm^{-2}$)} &
\colhead{(keV)} &
\colhead{index $\alpha_B$} &
\colhead{index $\beta_B$} &
\colhead{(keV)} &
\colhead{(erg)} &
\colhead{(for Energetics)}
}
\startdata
080413B & $1.1014$ & $37.7\pm4.93$ & $15-350$ & ${-1.22}^{-0.29}_{+0.27}$ & ${-3.02}^{-0.36}_{+0.32}$ & ${151.7}^{+39.7}_{-18.8}$ & ${52.20}^{+0.08}_{-0.10}$ &1\\
080603A & $1.68742$ & $11\pm2$ & $20-200$ & $-1.63\pm0.17$ & $\cdots$ & ${160}^{+920}_{-130}$ & ${52.34}^{+0.13}_{-0.20}$ &2\\
080607 & $3.0363\pm0.0003$ & ${893}^{+52}_{-47}$ & $20-4000$ & ${-1.08}^{-0.06}_{+0.07}$ & $\cdots$ & ${1691}^{+185.7}_{-153.4}$ & ${54.28}^{+0.03}_{-0.02}$ &3\\
\enddata
\tablecomments{
References for $z$:
GRB 080413B: \cite{FynboSpectra},
GRB 080603A: \cite{Guidorzi080603A},
GRB 080607: \citep{Prochaska080607}.
References for energetics (Fluence, Band, $E_p$, Band function parameters):
(1) \cite{BarthelmyGCN080413B}, \cite{EnotoGCN080413B}, BAT refined analysis page: http://gcn.gsfc.nasa.gov/gcn/notices\_s/309111/BA/,
(2) \cite{Guidorzi080603A},
(3) \cite{GolenetskiiGCN080607}.
}
\label{tabTypeIISample}
\end{deluxetable}



\end{document}